\documentclass[apj,twocolumn]{openjournal} 
\usepackage{lineno} 
\usepackage{xcolor} 
\usepackage{multirow} 
\usepackage{longtable} 
\usepackage{booktabs} 
\usepackage{color} 

\usepackage{tikz} 
\usetikzlibrary{arrows.meta,positioning} 
\newcommand{\cm}[1]{} 

\newcounter{magicrownumbers} 
 
\shorttitle{The Mass Accretion Rates in Novae}  
\shortauthors{Godon et al.}
\graphicspath{{./}{figures/}}

\begin{document}


\title{ 
The Instantaneous Mass Accretion Rate of Novae in Quiescence: \\
- an Archival Ultraviolet Optical Spectral Analysis 
}


\author{Patrick Godon, Edward M. Sion}  
\affiliation{Department of Physics and Planetary Science, Villanova University, Villanova, PA 19085,  USA} 
\email{patrick.godon@villanova.edu} 
%
%

%
\author{Tim Naylor} 
\affiliation{Department of Physics and Astronomy, University of Exeter, Exeter, EX4 4QL, UK} 
\author{Frederick A. Ringwald}
\affiliation{Department of Physics, California State University, Fresno, CA 93740, USA} 
%

\begin{abstract}

We present the first results of our archival spectral analysis 
to derive the quiescent instantaneous mass transfer rates $\dot{M}$ in novae, 
using synthetic disk spectra generated with \textsc{tlusty},   
white dwarf (WD) masses estimates from the literature, Gaia parallax-derived distances, 
and updated color excess values. 
Our results for nine novae, based on ultraviolet spectra and on a number of optical spectra,
yield mass accretion rates that are higher than those derived from simple integration of the UV and optical luminosity. 

Twenty one years after its eruption, HR Del, with a mass transfer rate of $4 \times 10^{-7}M_\odot$/yr,   
must have been burning the H-rich material as it accreted onto the WD.   
V842 Cen, which was previously thought to be a low mass transfer system, happened to be on 
the other extreme with $\dot{M}$ of the order of $10^{-7}M_\odot$/yr.   
For both novae, the enhanced mass transfer rate must have been self-sustained by a feedback loop.   
	RR Pic, with $\dot{M} \sim 5 \times 10^{-8}M_\odot$/yr, is better fitted with an accretion
disk in which the outer disk is heated up to a temperature of 12,000~K, in agreement with 
H and He emission lines coming from the outer disk and a large emission region on the leading side 
of the disk. Such a disk, augmented with a heated outer disk, also gives a good fit to the spectra of 
CP Lac, and DI Lac with $\dot{M} \sim4.5$ and $9 \times 10^{-9} M_\odot$/yr.  
V1974 Cyg and V533 Her, with an accretion rate of $\sim 3 \times 10^{-9}M_\odot$/yr, 
have an even flatter spectrum. 
As to V446 Her and BK Lyn, they were caught in a state of low accretion and have  
the lowest mass accretion rates in our sample: $\dot{M} \sim 10^{-9}$ and $\sim 10^{-10}M_\odot$/yr respectively. 

We find the higher mass transfer rate systems, with $\dot{M} \approx \sim 10^{-7}M_\odot$/yr, agree with the standard disk
model, while the remaining systems, with $\dot{M}$ of the order of $10^{-8}M_\odot$/yr and lower,  
are better fitted when the outer disk is heated to $\sim 12,000$~K. 
We suggest that irradiation from the heated WD and inner disk, together with tidal interaction, the bright spot, 
and material overflowing the disk edge and falling back onto the disk near phase 0.5,     
can increase the temperature of the outer disk to $\sim 12,000$~K. The relative contribution of this component  
to the UV and optical spectra increases with decreasing accretion rate.

\end{abstract}


\section{Introduction}  

Cataclysmic variables (CVs) are close interacting binary stars in which 
matter is transferred from a Roche lobe filling donor star (the secondary)  
to a white dwarf (WD) star (the primary). 
The binary orbital period ranges  from 
a fraction of an hour to about a day,  
with a gap in the orbital period distribution between 2 hours 
and 3 hours, where very few CVs are found - the {\it period gap}. 
CVs are believed to evolve from long 
to short orbital period as they lose binary 
angular momentum. Angular momentum loss (AML) is due primarily
to magnetic braking 
\citep[from the magnetic wind of the donor star,][]{ver81}  
above the period gap, which is shut off around 
$P\sim 3$~hr as the donor becomes fully convective and mass
transfer stops. At that point gravitational radiation becomes the
only remaining AML 
\citep{pac67,kin88} 
and the binary evolves as a detached system
until the donor re-makes contact with its Roche lobe, which 
occurs around $P \sim 2$~hr. 
   
In many systems the magnetic field of the WD is negligible 
({\it non-magnetic} CVs),  
and the gas (hereafter ``matter'') is transferred by means of an accretion disk 
reaching all the way to the WD surface. Disk systems are divided mainly into 
dwarf nova (DN) and nova-like (NL) systems. 
DNe are found mostly in a state of low mass accretion rate (quiescence) and 
interrupted every week to months by phases of intense accretion (outburst) lasting
days to weeks. 
As to the NLs, they are found in a state of high mass accretion rate, though 
the VY Sculptoris (VY Scl) subclass of NLs can drop into a temporary state of low mass
accretion rate (``anti dwarf novae'').  
In some systems, however,  the WD magnetic field cannot be ignored.  
{\it Magnetic} CVs are divided into intermediate polars (IPs) where the
magnetic field truncates the inner disk and channel the matter along the magnetic field lines
to the WD poles, and polars where the field is so strong that it prevents the formation
of an accretion disk.  For a review see \citep{war95}. 
      
With time, as the accreted matter piles up onto the WD, 
gravity compresses the accreted envelope until a critical pressure is reached at the base of the accreted 
layer above the electron-degenerate core, triggering  
a brief and powerful thermonuclear runaway (TNR) of the accreted 
H-rich material on the WD surface: the nova eruption \citep{sta76}. 
A shell of gas is ejected and expands at hundreds to thousands of km/s, 
the system quickly peaks in the optical 
and the visible luminosity increases by up to eight orders of magnitude \citep{sha09}.
As the ejecta dilute (and thin) they become transparent to shorter wavelengths
and the nova shifts to the ultraviolet. 
As the photosphere recedes the WD is seen as a luminous supersoft
X-ray source (SSS) due to residual quasi-static hydrogen nuclear 
burning of the remaining accreted layer on its surface.
The SSS phase can last from weeks to years \citep[for a review see][]{cho21}.  

A system that has been (directly or indirectly) observed to undergo 
a nova eruption is referred to as 
a {\it classical nova (CN),} and often just referred to as a `nova'.
After the nova eruption, accretion resumes and matter, again, accumulates on the WD surface.  
Consequently, a CV WD can potentially undergo such a nova eruption every few millennia.  

{\it Recurrent novae (RNe)} are nova systems that are known to have erupted at least twice 
\citep{sch10}, and  may account for about 1/3 of all nova eruptions
\citep{pag14,sha15,wil16}.   
Overall, however, the majority of known nova systems are classical novae.
The observed recurrence periods are the order of $\sim 1$ year
\citep{dar14} to $\sim 100$ years \citep{pag09}, 
but these are certainly limited by selection effects 
\citep{kat14,hil16,sha17} since all novae are likely to be recurrent.
The inter-eruption period depends upon the WD mass ($M_{\rm wd}$) 
and mass accretion rate\footnote{
Throughout this paper we use mass `transfer' rate and mass `accretion' rate interchangeably,
assuming negligible mass loss, so the two are equal.} ($\dot{M}$): 
RNe with a large $M_{\rm wd}$ and high $\dot{M}$ 
exhibit the shortest recurrence periods \citep{sch10} as the pressure
at the base of the accreted layer increases more rapidly due to 
the larger WD gravity and/or the larger accumulating mass. 
The high mass transfer rates in RNe are due to    
the elevated mass-loss rates from their donor star. 

During classical nova eruptions, the WD loses mass, and between 
recurrent nova eruptions it gains mass through accretion;  
whether a nova WD mass overall increases or decreases over its life-time  
is, however, still a matter of debate \citep{hil20,sta20}. 
Accreting WDs in CVs, and in particular novae, could reach the
Chandrasekhar mass limit and explode as Type Ia Supernovae:
they are candidates for SN Ia progenitors
through the single degenerate (SD) channel \citep[e.g.][]{liv11}. 

Furthermore, the long-term evolution of CVs, including their nova phase, 
is regulated by AML, which, by reducing the orbital
binary separation, shrinks the donor's Roche lobe, thereby controlling
the mass transfer rate. Deriving the average mass transfer rate is therefore
essential for the theories of evolution of CVs and novae. 

However, the CNe and RNe during their {\it quiescence} 
(i.e. before, after, or between their nova eruptions) 
are poorly understood,  
mostly because the observational coverage focusses primarily 
on the nova explosion itself, the ejecta emission lines, and
their corresponding chemical abundances. 
Nova studies, \citep[e.g.][]{cas02},  
only rarely cover the transition into quiescence and beyond 
in the ultraviolet (UV), and very often for no more than a few months 
to a couple of years.  
\citet{sel13} have stressed how poorly understood the state of CNe are, 
far  into quiescence, and \citet{pag15} emphasized how little is
known about the long-term secular behavior of novae. 
This is unfortunate since their outburst
characteristics, such as energetics, duration, and nucleosynthetic yields, 
all depend on the details of the accretion process in quiescence, 
including their quiescent mass accretion rate \citep{muk05}.  
   
A large UV-optical analysis of 18 novae in quiescence 
was carried out by \citet {sel13,sel19}, who found an average mass accretion
rate of $\dot{M} \approx 3 \times 10^{-9} M_\odot$/yr,   
using Gaia parallax-derived distances and based on optical and UV luminosities, 
but without using realistic accretion disk models.  
In fact, there has been very little realistic accretion disk modeling and/or 
WD photosphere modeling of novae in the UV following their post nova eruption 
decline into quiescence or optical light minimum. 

A limited number of objects were studied in detail in their 
quiescent phases with realistic accretion disk models in the UV.
So far, the most comprehensive and self-consistent study of novae in their quiescent phase 
with realistic accretion disk models is that of \citet{pue07}, in which ten novae were included 
in a statistical analysis using a multi-parameter optimization method 
to model International Ultraviolet Explorer (IUE) spectra of 33 CVs. 
They found that novae have a mass transfer rate of $1.3 \times 10^{-8}M_\odot$/yr, 
on average four times larger than that derived by \citet{sel19}  
by integration of the observed flux.  
It is to be noted that, in their the pre-Gaia analysis, \citet{pue07} adopted accurate distances for 
some CNe (CP Pup, V533 Her, and BT Mon).

In this paper we present the first results from our archival spectroscopic  
analysis of ultraviolet (and to some extent optical) spectra for nine novae in quiescence, 
with an emphasize on using realistic accretion disk models generated with \textsc{tlusty/synspec} 
\citep{hub17a,hub17b,hub17c}, 
Gaia DR3 parallax-derived distances \citep[applying the method of][]{sch18}, 
and WD mass estimates \citep{sha18,hac19}.   
Our analysis is similar to the one carried out for the RNe  M31N 2008-12a \citep{dar17} and T Pyxidis \citep{god18}.

\section{The Mass Transfer Rate in Novae} 
   
The high mass transfer rate derived from observations of nova binaries 
in quiescence, $\dot{M} \sim 10^{-9}- 10^{-8} M_{\odot}$/yr
\citep{pat84,war95,sel19}, is much larger than predicted by the theory,  
$\dot{M} \sim 10^{-10}M_\odot$/yr \citep{sha18}. 

For some systems, such as HR Del \citep{sel19} and T Pyx \citep{god24}, 
$\dot{M}$ reaches $\sim 10^{-7}M_\odot$/yr. 
Such a high accretion rate may result in steady nuclear burning on the WD surface 
: the matter `burns' as it is accreted, no flash, 
or a very weak flash with no nova eruption \citep{kut80,pri82,fuj82}.  
Such a high mass transfer rate for a short-period RN such as T Pyx (and similarly for IM Nor) 
also conflicts with CV evolution theories, since mass transfer rate is expected to decrease 
with decreasing orbital period \citep{pac81,rap82,spr83}. 
%

While it is clear that the mass transfer rates derived from observations are 
the instantaneous values of $\dot{M}$'s, and those inferred from the theory are the secular
(long-term average) values of $< \dot{M}>$, one has still to explain why years after
the eruption $\dot{M}$ is so high, and how and when $\dot{M}$ decreases, 
and by how much. 

To compensate for these discrepancies, intermittent cycles 
have been proposed as explained here below. 

\subsection{Hibernation Theory.} 

The original hibernation theory \citep{sha86} advances that the 
high mass transfer rate (which is more characteristic of NLs) 
observed before and after the nova eruption lasts only a few centuries. 
It is suggested that after a nova eruption 
the high mass transfer rate decreases over hundreds of years  
(reaching values as low as DNe in quiescence). 
The secondary star loses contact with its Roche lobe, 
the system becomes `detached' and mass transfer stops completely.  
The system stays in this hibernation
phase for thousands of years, until binary gravitational radiation 
(and/or magnetic braking) 
drives the two stars closer. Mass transfer resumes, increasing back
over centuries to a high rate before the next nova eruption.  
Namely, the assumption is that
novae spent most of their lives hibernating between
nova eruptions, i.e. the system returns periodically (every thousands of years)
into a detached phase of zero mass transfer.

The hibernation theory, however, has remained a theory, and has faced
significant skepticism for more than 3 decades, primarily due to conflicting observational data 
\citep[see e.g.][]{nay92,sch23} 

\subsection{Self-sustained Enhanced Mass Transfer Rate.} 

\begin{figure}[h!] 
\begin{tikzpicture}[>=Stealth, thick, every node/.style={inner sep=3pt, align=center}]
  \node at         (5.5,2.5) {{\color{red}{Nova Eruption}}} ; 
  \node (step0) at (5.5,2.25) {{\color{red}{SSS phase}}}; 
  \node at (0.0,2.50) {Steady H-burning}; 
  \node (step1) at (0.0,2.25) {or}; 
  \node at (0.0,2.00)  {Hot Inner Disk/WD};        
  \node at (3.00,2.50)     {Irradiating};      
  \node (step2) at (3.00,2.25) {Donor};        
  \node (step3) at (3.00,-1.0)  {Heated Donor};        
  \node at (3.00,-1.30) {Inflated Radius};  
  \node at (3.00,-1.60) {/Strong Wind};  
  \node (step4) at (0.0,-1.0)  {High Mass Transfer};        
  \node at (0.0,-1.30) {$\dot{M}\sim 10^{-7}M_\odot$/yr}; 
%
  \draw[->] (step0) -- (step2);                
  \draw[->] (step1) -- (step2);                
  \draw[->] (step2) -- (step3);                
  \draw[->] (step3) -- (step4);                
  \draw[->] (step4) to[out=90,in=-90] (step1); 
	\node (cycle1) at (0.5,1.5) {}; 
	\node (cycle2) at (2.5,1.5) {}; 
	\node (cycle3) at (2.5,-0.3) {}; 
	\node (cycle4) at (0.5,-0.3) {}; 
	{\color{blue}{ 
  \node at (1.5,1.10) {Feedback};
  \node at (1.5,0.75) {Loop};
  \node at (1.5,0.40) {Sustained}; 
  \node at (1.5,0.05) {High $\dot{M}$}; 
	\draw[->] (cycle1) -- (cycle2);     
	\draw[->] (cycle2) -- (cycle3);     
	\draw[->] (cycle3) -- (cycle4);     
	\draw[->] (cycle4) -- (cycle1);     
        } } 
\end{tikzpicture}
	\caption{
{\bf{ Self-sustained  Mass Transfer.}} 
	Schematic Self-sustained $\dot{M}$ feedback loop where a nova remains in a state of high mass 
accretion rate (with or without nuclear burning of H on its WD) for hundreds of years
or more. See text for details. 
\label{loop}}
\end{figure}
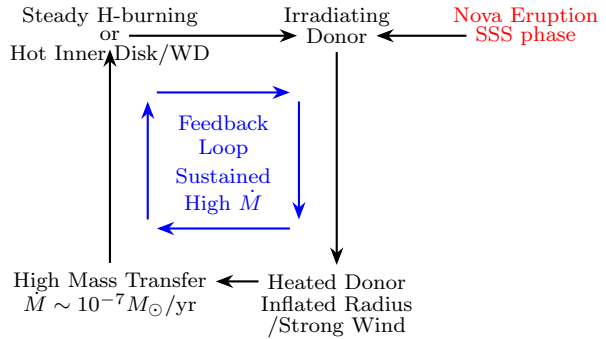

In order to explain the high mass transfer  rate 
observed in some nova systems (such as e.g. T Pyx),  
a self-sustained enhancing mass transfer rate 
was proposed \citep{kni00}. In this scenario, the hot WD star    
(and/or inner disk) {\it irradiates} the donor star, inducing a strong wind from the latter
(or possibly inflating its radius), thereby 
inducing a large mass transfer rate. 
As the matter accretes onto the WD, its temperature increases, 
and the hot inner disk and WD, in turn, further irradiate the secondary and thereby  
make the process self-sustained (see Fig.\ref{loop}).  
In such a scenario the self-sustained enhanced mass transfer rate
could start due to an unusually large burst of mass transfer from the secondary.

On the other hand, the original trigger of such a process might also be the  
residual H nuclear burning on the WD surface (SSS phase) after a nova eruption, 
where the hot luminous WD irradiates the donor, heating it and inflating its radius, 
causing it to transfer matter at a high rate. 
As the transferred mass is accreted onto the WD surface it undergoes 
stable nuclear burning and `closes the loop'. 
Such a self-sustained high $\dot{M}$ loop would significantly increase the duration
of the SSS phase of a nova. The enhanced mass transfer loop could also continue
as $\dot{M}$ decreases below the threshold for steady nuclear burning, 
provided $\dot{M}$ is large enough to heat up the WD and inner disk 
which in turn continue to irradiate and heat up the secondary
(Fig.\ref{loop}).

\citet{gin21} calculate (both analytically and numerically) the donor's 
heating and expansion and find that irradiation drives enhanced mass transfer 
from the donor at a rate $\dot{M} \propto T_{\rm irr}^{5/3}$, which reaches 
$10^{-6}M_\odot$/yr at the peak of the nova eruption and drops to lower values 
as the nova subsides.  They show that under certain circumstances the mass transfer 
persists at a self-sustained  rate of $\dot{M}\approx 10^{-7}M_\odot$/yr, 
close to the WD's stable H burning limit, for up to $\approx 10^3$~yr after the eruption.  
Irradiation of the donor by the WD is sufficient to drive the mass transfer. 
This mechanism can simultaneously explain RNe with orbital periods $\sim 2$~hr 
with high mass transfer rate such as T Pyx, IM Nor, CI Aql, as well as long-lived supersoft 
X-ray sources with periods $\sim 4$~hr such as RX J0537.7-7034, 1E 0035.4-7230.  
Whether a system reaches this self-sustained  $\dot{M}$ feedback loop or not depends on the donor's 
chromosphere structure and the orbital period change during nova eruptions \citep{gin21}.  
The timescale underlying the enhanced accretion rates induced 
by nova irradiation of the donor star is highly uncertain \citep{sch10b}.

T Pyx is suspected to have undergone a highly energetic nova eruption in 1866 (more powerful than
the subsequent nova outbursts) that increased its 
mass transfer rate through the self-sustained $\dot{M}$ feedback loop mentioned above. 
Since 1866, however, it has faded by 2 mag \citep{sch10b}, 
and more recent UV data indicates that it has been decreasing steadily before and 
after its 2011 outburst \citep{god24}: 
its accretion rate, currently of the order of $10^{-7}M_{\odot}$/yr 
\citep{pat17} continues to decline.
This seems to indicate that the self-sustained $\dot{M}$ feedback loop in T Pyx
is weakening. 
It has even been suggested \citep{pat22} that the self-sustained
enhanced mass transfer in the short-period RNe IM Nor and T Pyx 
might ``evaporate'' the donor star in a {\it paroxysm} of 
classical nova eruptions.  Namely, that these two stars
might represent a final stage of nova - and CV - 
evolution.

\section{The Systems and their Archival Data} 

\begin{table*}[t!] 
\begin{minipage}[t]{0.50\textwidth} 
\begin{tabular}{lrlccl}
	{\bf Table 1.} & {\bf System} & {\bf Parameters.}   \\  [4pt]   
\hline  
\noalign{\vskip 1.0ex}   
System     & $P_{\rm orb}$ &   d            & $E(B-V)$        & $M_{\rm wd}$ & $i$  \\   [2pt] 
Name       &  (hr)         &       (pc)     &                 & $(M_\odot)$  & (deg) \\   [2pt] 
\hline
\noalign{\vskip 1.0ex}   
BK Lyn     & 1.7995  & $527 ^{+9}  _{-8}$   & 0.018$\pm0.005$ & ....      & 20-60 \\  
HR Del     & 5.13996 & $896 ^{+18} _{-16}$  & 0.180$\pm0.030$ & 0.6-0.84  & 42$\pm2$ \\  
RR Pic     & 3.4806  & $501^{+6}   _{-5}$   & 0.034$\pm0.020$ & 0.6-0.95  & 65-68 \\ 
CP Lac     & 3.48343 & $1163^{+45} _{-36}$  & 0.300$\pm0.070$ & 1.0-1.24  & 60$\pm5$ \\ 
DI Lac    & 13.05055 & $1731^{+57} _{-47}$  & 0.240$\pm0.030$ & 0.91      & $15\pm10$ \\ 
V533 Her   & 3.528   & $1248^{+130}_{-80}$  & 0.025$\pm0.004$ & 0.95-1.07 & 57-62 \\ 
V446 Her   & 4.968   & $1272^{+140}_{-83}$  & 0.430$\pm0.050$ & 1.11      & $40\pm20$ \\ 
V1974 Cyg  & 1.95024 & $1617^{+186}_{-109}$ & 0.320$\pm0.020$ & $>1.0$    & 39$\pm2$ \\ 
V842 Cen   & 3.555   & $1376^{+105}_{-72}$  & 0.800$\pm0.100$ & 0.88-1.02 & low  \\ [3pt]  
\hline   
\end{tabular} 
\end{minipage} 
%
%
\hfill 
\begin{minipage}[t]{0.50\textwidth} 
\hspace{1.3cm} 
\begin{tabular}{llc}
	{\bf Table 3.} & {\bf Optical  Spectra} \\ [5pt]   
\hline  
\noalign{\vskip 1.0ex}   
	System     &  Telescope & Observation  \\  [2pt] 
	Name       &            & Year      \\  [2pt] 
\hline 
\noalign{\vskip 1.0ex}   
HR Del & Shane 3-m (at Lick)       & 1990  \\  
CP Lac & Shane 3-m (at Lick)       & 1990  \\  
DI Lac & Shane 3-m (at Lick)       & 1990  \\  
	& Isaac Newton (at La Palma)   & 1991  \\  [3pt]  
\hline  
\noalign{\vskip 1.0ex}   
\end{tabular} 
\end{minipage} 
Note: The references for the orbital periods, WD masses and inclinations are given in the text.   
\end{table*}

In an attempt to further elucidate the problem of the mass accretion rate
in novae, we have been carrying out an UV-optical spectral 
analysis of archival data of nova in quiescence. 
In Table 1 we list nine novae from our on-going archival study which includes 
a much larger number of systems. 
The novae were chosen for their relatively good quality UV spectra 
obtained late enough after the nova eruption to
ensure they were all disk-dominated at the time of the observations. 
They all have distances derived from Gaia DR3 parallaxes, known extinction, 
and (almost all) known WD masses and orbital inclination.

\begin{table*} 
\centering         
	{\bf Table 2. UV Spectra}  \\   [2pt] 
\begin{small}  
\begin{tabular}{llrllccccl}
\hline  
\noalign{\vskip 1.0ex}   
	System     &  Nova &  Telescope & Observation & Data ID                              \\  [2pt] 
	Name       &  Year &            & Year        &                                      \\  [2pt] 
\hline 
\noalign{\vskip 1.0ex}   
BK Lyn     & 101 AD&  HST STIS  & 2003        & o6li34010                                      \\  
HR Del     &  1967 &  IUE       & 1979        & SWP05757 LWR04993 SWP05918 LWR05161                \\  
           &       &  IUE       & 1988        & SWP33402 LWP13136 SWP33403 LWP13137                \\  
           &       &  IUE       & 1992        & SWP45481 SWP45482                                  \\  
RR Pic     &  1925 &  IUE       & 1979        & SWP05774 LWR05010 SWP06625 LWR05687                \\ 
           &       &  IUE       & 1981        & SWP15632/3 LWR12071/2 SWP15635/7 LWR12074/5        \\ 
           &       &  IUE       & 1989        & SWP35522 LWP15003 SWP36558 LWP15758                \\ 
           &       &  IUE       & 1996        & SWP57075 LWP32292                                \\ 
CP Lac     &  1936 &  IUE       & 1991        & SWP42000 LWP27044                                \\ 
DI Lac     &  1910 &  IUE       & 1986        & SWP29325 LWP09208   \\  
V533 Her   &  1963 &  IUE       & 1980        & SWP10250 LWR08915                                \\ 
           &       &  IUE       & 1992        & SWP44805 LWP23205                                \\ 
V446 Her   &  1960 &  HST STIS  & 2003        & o8n001010 o8n001020 o8n001030 o8n001040        \\ 
V1974 Cyg  &  1992 &  HST FOS   & 1995        & y30o010ct y30o010at y30o0109t y30o0107t y30o0105t y30o0104t  \\ 
V842 Cen   &  1986 &  HST COS   & 2010        & lb1a03010                                      \\ 
	   &       &  IUE       & 1987        & SWP31020 LWP10801 (used for dereddening)         \\ 
\hline  
\end{tabular} 
\\
\vspace{0.2cm} 
Note: We list the spectra used for the spectroscopic analysis, namely to derive the
mass accretion rates for each system. For RR Pic with use additional IUE SWP spectra
to generate an IUE light curve which we display in Fig.\ref{rrpiciuelc}. 
\vspace{0.3cm} 
\end{small}
\end{table*} 

We use the Data Release 3 from Gaia \citep{lur18,gai21} to derive the distance
following the procedure described by \citet{sch18}. 
This distance is then used to derive a first assessment of the color excess $E(B-V)$ using the 
tridimentional map of the local interstellar medium (ISM) {\it STructuring by
Inversion the Local Interstellar Medium} \citep[Stilism\footnote{We downloaded the {\sc stilism} package
from \url{github.com} and extracted color excess values $E(B-V)$ using a {\texttt{Python}} scritp.};][]{lal14,lal18,cap17}.
When possible, we use near-UV (NUV) spectra to derive $E(B-V)$ more accurately
(see next section). For most systems, the values we obtain for $E(B-V)$ are consistent with those found
in the literature (e.g. \citet{sel13}; within error bars). 
The largest discrepancy is for V842 Cen, where Stilism gave $E(B-V)=0.289$, while our direct dereddening
using its IUE spectra gave $0.8 \pm 0.1$ (see \S4.3). We further checked using Galextin\footnote{Galextin.org.}
\citep{amo05} models A \& S from \citet{amo21}, and found a value $E(B-V) \approx 1.0-1.1$
consistent with our result.  

The inclinations are taken from the literature and are given in the following subsections. 
The WD masses are from \citet{sha18}, except for DI Lac for which we use the value given in \citet{sel19}.

In Table 2 we list the UV spectra (for the systems listed in Table 1) that we use for our spectroscopic analysis.  
The IUE spectra cover both the far-UV (FUV) and the NUV 
with the IUE SWP segment from $\sim$1150~\AA\ to $\sim$2000~\AA , 
and IUE LWP (or LWR) segment from $\sim$1850~\AA\ to $\sim$3200~\AA . 
When the spectra have nearly the same continuum flux level and 
shape, we combine them together to reduce the signal to noise ratio, even if they were obtained
years apart. 

The HST Space Telescope Imaging Spectrograph (STIS) spectrum of BK Lyn has a spectral coverage of [1150\AA -1715~\AA ], 
that of V446 Her: [1148~\AA - 3170~\AA ]; 
the HST Faint Object Spectrograph (FOS) spectrum of V1974 Cyg covers the spectral region [1137~\AA - 2508~\AA ] , 
and that of V842 Cen covers [1120~\AA - 2049~\AA ] .

Three of these systems, HR Del, DP Lac, \& DI Lac, were observed in the optical
\citep{rin96} 
and have optical spectra that can be used with the UV spectra for our analysis. 
These are listed in Table 3. Their spectral coverage is as follows:
[3945~\AA - 8295~\AA ] for the spectra obtained with the Cassegrain CCD Spectrograph
on the 3-m Shane telescope at Lick Observatory in 1990,
and [3408~\AA - 9784\AA ] for the spectrum of DI Lac obtained 
with the Faint-Object Spectrograph (FOS-1) on the 2.5 m Isaac Newton Telescope
(INT) on La Palma.  

Due to their high accretion rates, all these systems are disk dominated.
Their spectra are expected to be consistent with synthetic spectra of accretion disk models. 

We review here below each of the nine novae separately and justify the 
choice of inclinations and WD masses we make. 
It is to be noted that our grid of accretion disk models was generated for discrete values 
of the WD mass (0.35, 0.55, 0.81, 1.03, and 1.21$M_\odot$), inclination 
(18, 41, 50 and 75~deg) and accretion rate. As a consequence, we adopt the grid values that are close
or in agreement with those derived from observations (which are given in Table 1). 
See \S4.2 for full details. 

\subsection{BK Lyncis - Nova Lyncis 101 AD} 

BK Lyn was the first known NL below the CV period gap with a period of 
107.97~min \citep[=1.7995~hr;][]{rin96a},  
and was identified with a Chinese-recorded nova from 101 AD \citep{pat13}.  
Its accretion rate has apparently remained very high up to the present. 
It exhibits positive superhumps indicative of an eccentric accretion disk
\citep[see e.g.][and references therein]{bru23a}. 
It appears to have transitioned from a NL in a permanent high state/$\dot{M}$ to a DN with 
outburst/quiescence cycles and with a relatively lower $\dot{M}$
than a NL \citep{pat13}. Its photometric light curve is virtually identical 
to the light curves of ER UMa systems, i.e. DN systems with  
higher than normal DNe accretion rates \citep{guz19}, suggesting that   
ER UMa systems may still be cooling from recent (but unrecorded) nova explosions. 
The spectral analysis of the UV spectrum of BK Lyn 
allows us to measure the accretion rate of a CN 
${\sim}$2000 years after the nova explosion. 
 
\citet{pat13} estimated an inclination between $20^\circ$ and $60^\circ$ based on the absence of eclipses, 
width of emission lines, and weak orbital modulation.  
\citet{dob92} derived $i=32^\circ \pm 12^\circ$ and a rather small WD mass of $0.3^{+0.35}_{-0.12} M_\odot$. 
We note that no WD mass was derived for BK Lyn in \citet{sha18}. 
From its Gaia distance of $527^{+9}_{-8}$~pc we find a reddening towards the source of $E(B-V)=0.018\pm0.005$ 
using Stilism. Based on \citet{dob92} and \citet{pat13} the inclination is between $20^\circ$ and $60^\circ$,
therefore, for our analysis we assume inclinations $i=18^\circ,41^\circ,60^\circ$, which 
are the discrete values of our grid of models. 
However, we leave the WD mass as a free parameter.  

BK Lyn has only one STIS spectrum from 2003 \citep{zel09}, but   
it has optical spectra from 3 decades ago \citep{dob92,szk92,rin96a}  
showing that the continuum flux level varies by a factor of two,
a possible indication that it was already on its way to transiting from NL to DN. 
The optical spectra have a steeper slope than the UV spectrum.
From the AAVSO light curve data and the fact that the low state in BK Lyn 
seems to last about 2 weeks \citep{pat13}, we find that the optical spectra were obtained  
when $V=14.6$, while the STIS spectrum was collected when $V\approx 15.7$. 
Namely, the optical spectra do not match the state and brightness of the STIS spectrum of BK Lyn. 
Since the disk emission peaks in the UV, we do not use optical spectra alone to derive $\dot{M}$,    
and, therefore, we carry out the analysis on the STIS spectrum alone.  
\\

\subsection{HR Delphini - Nova Delphini 1967}    

HR Del erupted in 1967 as a very slow nova which remains exceptionally 
bright in the optical and UV.  
A WD mass of $0.6M_\odot$ was derived by \citet{kue88} 
who also found an orbital period of 5.13996~hr. 
However, \citet{sha18} derived a slightly higher value of $0.84 M_\odot$; 
\citet{bru82} found an inclination $i=42^\circ \pm 2^\circ$.

\begin{figure} 
\includegraphics[scale=0.33,trim= 30 40 0 10]{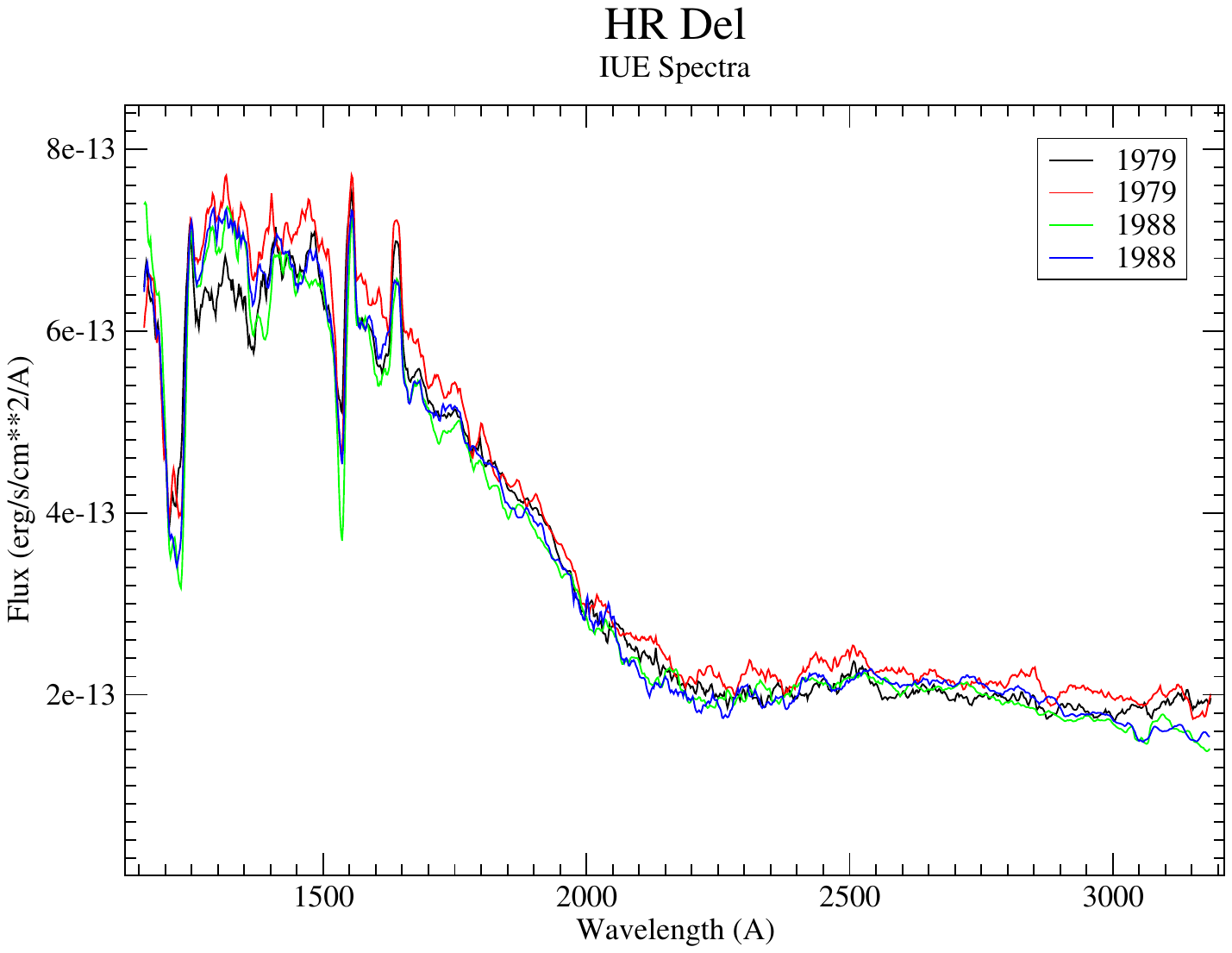}   
\caption{Four IUE spectra of HR Del obtained in 1979 and 1988.
Even though they were obtained 9 years apart, they look remarkably similar.  
The `elbow' shape is due to extinction with the prominent depression
near 2175~\AA. 
	\label{hrdeliue}}  
	\vspace{0.3cm} 
\end{figure}

HR Del was observed with IUE between 1979 and 1992;  
we find the FUV flux (in the SWP channel) 
decreased by not more than 20\% over that period of time,
but the overall shape of the spectrum remains the same. 
In Fig.\ref{hrdeliue} we display two pairs of spectra,
one from 1979 and one from 1988. In spite of the 9 years
difference, the spectra look remarkably similar. 
Based the prominent 2175~\AA\ ISM absorption feature in the IUE spectra,  
we derive a color excess $E(B-V)=0.18\pm0.03$. 

HR Del has an optical spectrum from \citet{rin96}, 
which exhibits some weak nebular emission lines of C\,{\sc iii}, N\,{\sc iii} and O\,{\sc iii} 
flanking the H$\beta$ (Balmer) emission line ($\lambda$4861~\AA).  
Its IUE UV spectra reveal a variable 
P Cygni profile in C\,{\sc iv} ($\lambda 1550$\AA), with an outflow velocity of the 
order of 5000~km/s.  All this suggests that continuing weak steady TN burning 
was still taking place on the WD surface when these spectra were collected, and kept on ionizing 
the expanding shell \citep{mor09,fri10}. 

Therefore, for our analysis, we use the optical spectrum of HR Del obtained 
1990, as well as the co-added IUE spectra obtained between 1988 and 1992,
since no IUE spectrum was collected in 1990 when the optical spectrum 
was obtained. 
The 1992 spectra only include the shorter segment SWP, no LWP or LWR 
data were collected.

\subsection{RR Pictoris - Nova Pictoris 1925}

RR Pic has been extensively studied, including its nova shell \citep[see][and references therein]{cel24}.   
The system exhibits a partial {\it eclipse} \citep{war96,hae91}, 
as the light curve of RR Pic, 
folded onto its orbital period \citep[of 3.4806~hr;][]{hae91}, 
exhibits a consistent gradual 
drop in flux (reaching 30\%) from phase $\sim 0.0$ to $\sim 0.5$ (the phase of the {\it eclipse}). 
This drop cannot be explained with the {\it usual} stream-disk overflow, nor with 
partial eclipse of the disk: it has been attributed to the
occultation of a rather large emission region on the leading side of the disk \citep[i.e. it is not 
an eclipse per se;][]{sch03}. 
The disk has He\,{\sc i} and H$ \alpha$ emission lines coming from its outer region. 
All this indicates that the outer region of the disk is heated up, likely due 
to a combination of tidally induced spiral waves, the bright spot 
(the region where the L1 stream of matter hits the outer edge of the disk near orbital phase 0.9), 
stream disk overflow falling back onto the disk, 
and irradiation from the hot inner disk and/or WD. 
\citet{fue18} emphasized that RR Pic has a magnitude that has been steadily decreasing since
its eruption in 1925 (based on AAVSO light curve) and presented evidence of positive superhumps.
A third body has even been invoked to explain the O-C deviations of its linear ephemeris \citep{vog17}.

The reddening toward the source is not very large. We find a maximum of 0.05 from the ISM Galactic reddening map 
\citep[][]{sch98,sch11}.  At a Gaia distance of $510^{+6}_{-5}$~pc, Stilism gives $E(B-V)=0.034\pm0.020$, which is consistent with 
its IUE spectra in which it is difficult to find even a hint of the 2175~\AA\ ISM absorption feature. 
The WD mass was estimated to be $0.95 \pm 0.1 M_\odot$ by \citet{hae82} and \citet{sha18},
but values as low as $0.62\pm0.2M\odot$ and $0.66\pm0.15M_\odot$ have also been derived by \citet{sel19} and \citet{sch25}, respectively. 
Its inclination has been put near $65^\circ-68^\circ$ \citep{hae82,sel19}, consistent
with its disk self-occulting ({\it partial eclipses}). 
Hence, we carry out disk models assuming 
$M_{\rm wd} = 0.55 M_\odot$ and $1.03 M_\odot$, $i=60^\circ$ and $75^\circ$, and $E(B-V)=0.034\pm0.020$.  

\begin{figure}[h!] 
\includegraphics[scale=0.33,trim= 30 20 0 0]{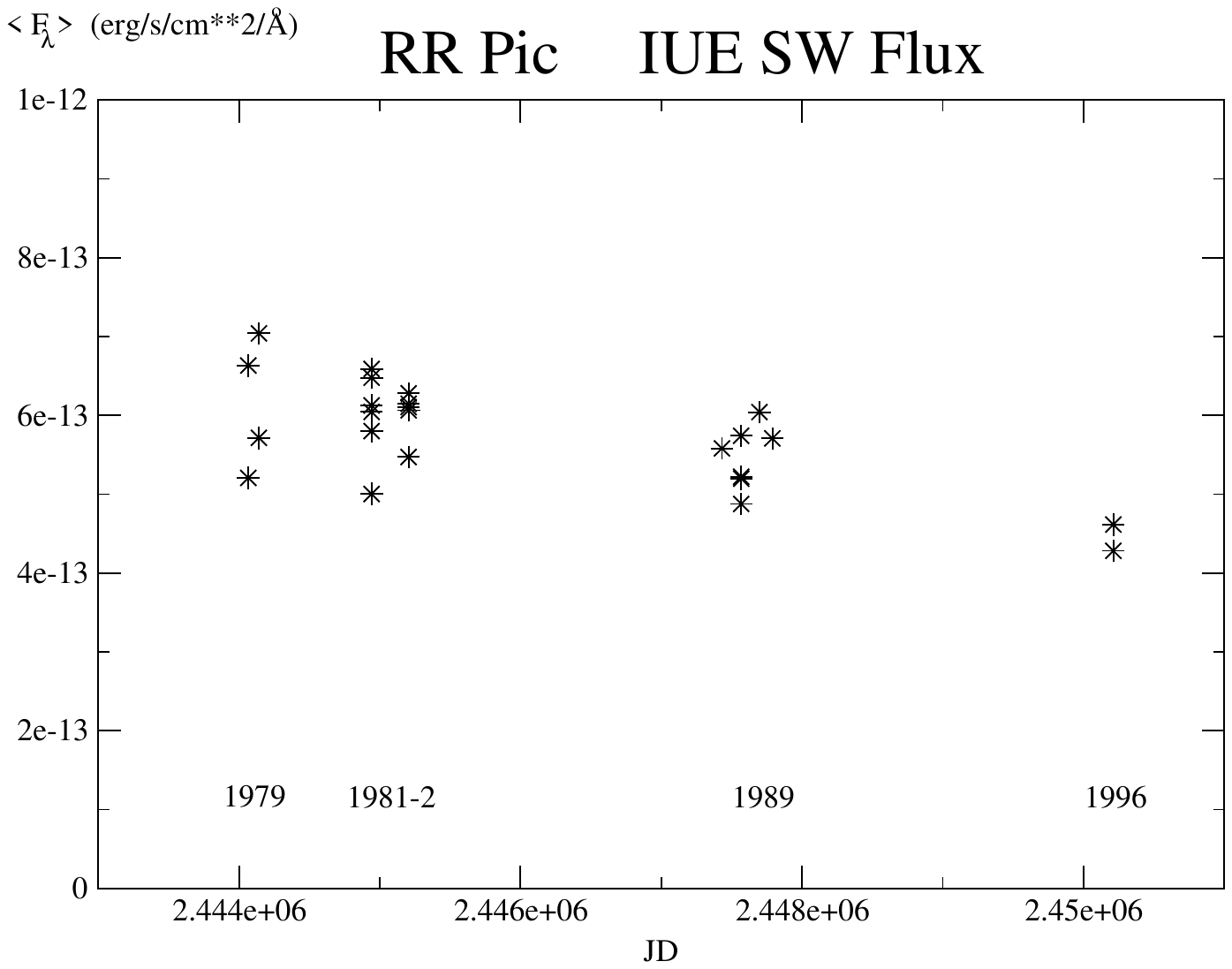}
\caption{The IUE SWP light curve. The average fluxes of the IUE SWP spectra
have been drawn as a function of time, years have been indicated for convenience. 
The UV flux exhibits strong short-period modulation ($\pm A$) reaching a maximum amplitude
of $A=$17\%. The amplitude seems to decrease as the flux declines.   
\label{rrpiciuelc} 
}
\end{figure}

We coadded the IUE spectra and combined them together to produce a spectrum extending from 1150~\AA\ to 3200~\AA . 
In Fig.\ref{rrpiciuelc} we present the IUE light curve generated with the average SWP fluxes
($\frac{1}{\lambda_2 - \lambda_1} \int_{\lambda_1}^{\lambda_2} F_\lambda d\lambda$) 
obtained between 1979 and 1996, reflecting both the orbital modulation as well as the decline between 
1979 (first IUE spectrum) and 1996 (last IUE spectrum).

\begin{figure}[h!]  
\includegraphics[scale=0.33,trim= 30 20 0 0]{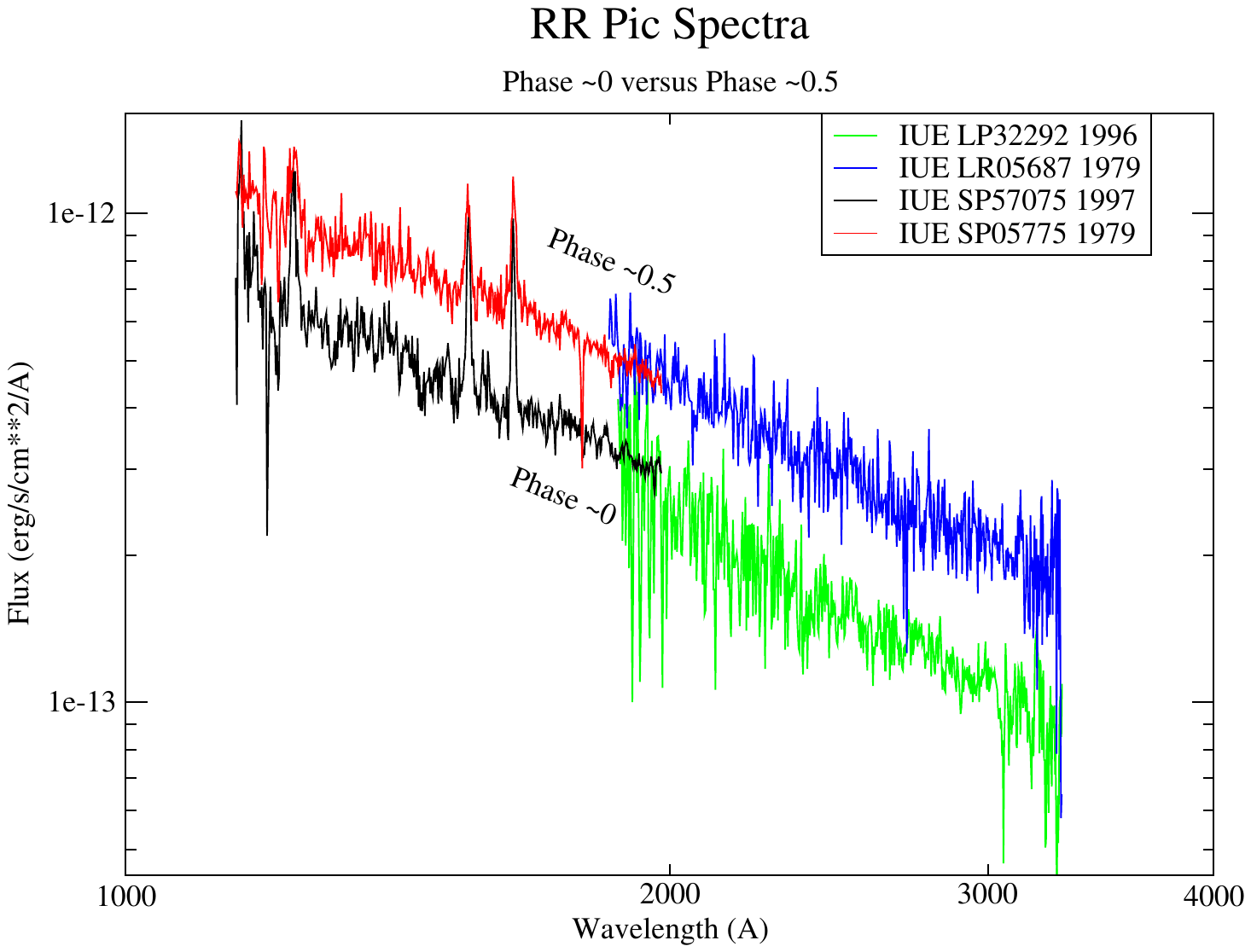}
\caption{IUE Spectra at phase 0.0 and 0.5. While the flux of these
spectra have different continuum flux levels; they do have the same
continuum shape and same continuum slope. 
\label{rrpiciuephase} 
}  
\end{figure}

In Fig.\ref{rrpiciuephase} we display the spectra exhibiting the largest discrepancy:
the spectrum (SWP+LWR) obtained in 1979 near phase 0.5 with the highest flux versus the
spectrum (SWP+LWP) obtained in 1996 near phase 0.0 with the lowest flux.  
While the IUE spectra of RR Pic obtained over the years have different continuum flux levels, 
they do, however, have the same continuum shape and same continuum slope. 
Thus, we coadd the spectra listed in Table 4 for our spectral analysis.  
The coadded spectrum is an average and one has to take into account that
the flux level fluctuates by $\pm 17$\% as a function of the orbital phase. 

\subsection{CP Lacertae - Nova Lacertae 1936} 

CP Lac is a very fast nova ($t_2=5.3$ days, $t_3$=9.8 days), 
showing small-amplitude outbursts of dwarf nova-type (though their amplitude is smaller
than typical DN outbursts). It has tentatively been classified as a SW Sex type nova-like \citep{rod05}, 
and has negative superhumps with a period near 3.8~hr \citep[][indicative of a tilted accretion disk]{bru23b}.  
It is, however, a poorly studied nova remnant \citep[see e.g.][and references therein]{rod05}.    
It has an orbital period of 3.48343~hr \citep{pet06}. 

With a Gaia distance of 1163$^{+45}_{-36}$pc, we find a color excess of $E(B-V)=0.30\pm0.07$
from Stilism. 
Its WD mass is rather large: \citet{sha18} derived $M_{\rm wd}=1.24M_\odot$. 
It has no reliable inclination, but it might be exhibiting shallow eclipses \citep{rod05},
namely $i$ cannot be too high. 
\citet{sel19} derived $i=60\pm5^\circ$, using $K_1=100$~km/s from \citet{pet06}, 
and taking $M_{\rm wd}=1 M_\odot$, $M_2=0.26 M_\odot$. 
We adopt this value of the inclination for our analysis
as well disk models with a WD mass of $1.03 M_\odot$ and of $1.21 M_\odot$. 

When choosing from our grid of disk models a $1.03 M_\odot$ and a $1.21 M_\odot$ accreting WD, this is equivalent 
to considering a $\sim 1.1 M_\odot$ accreting WD with an uncertainty of $\sim 0.1 M_\odot$ in the WD mass.
Namely, we average the accretion rates we obtained for the $1.03 M_\odot$ and $1.21M_\odot$ models
$$ \dot{M}_{1.1 M_\odot} = \frac{1}{2} \left( \dot{M}_{1.03 M_\odot} + \dot{M}_{1.21 M_\odot} \right). $$ 
The uncertainty (error) in $\dot{M}$ due to an uncertainty of $0.1M_\odot$ in the WD mass (as is the case here for CP Lac)
is given by                
$$ \Delta \dot{M}_{1.1 M_\odot} = \frac{1}{2} \left( \dot{M}_{1.03 M_\odot} - \dot{M}_{1.21 M_\odot} \right). $$ 
These values are for CP Lac and for each system, depending on the assumptions on the WD mass, 
we compute $\dot{M}$ and its error due to the uncertainty in the WD mass. 

CP Lac has only one IUE spectrum made of LWP20744 and SWP42000 (collected in 1991), 
which is very noisy and exhibits emission lines. We remove the short wavelength region ($< 2,300$~\AA ) 
of the LWP spectrum as there is no signal.  
It has optical spectra from \citet{rin96} and \citet{hon98}.  
For our analysis we use the IUE spectrum 
and the optical spectrum from \citet{rin96}.

\subsection{DI Lacertae - Nova Lacertae 1910} 

DI Lac showed a decrease in its brightness for about
four decades after its 1910 outburst, followed by an apparent constant
magnitude for several decades, then increasing in the 1990s. 
This might be an indication that the donor is being slowly heated
via irradiation, increasing the mass transfer rate \citep{hoa00}. 
Its sharp emission lines indicate a relatively low inclination \citep{kra64}, 
and it shows striking P Cygni wind outflow \citep{sel13} also implying 
a low to moderate inclination (based on the depth of the absorption 
part of the P Cygni profile). 
It also exhibits optical flare-ups on a 35-day  times-scale \citep{hon95}. 
It has a relatively long orbital period of 13.05055~hr \citep[listed in the Ritter and Kolb catalog
of cataclysmic variables;][which they obtained from their private communication with Webbink, R.F. in 1982]{rit98}. 

It has a single IUE spectrum obtained in 1986; 
it has also an HST STIS spectrum with a continuum flux level 30\% higher than its IUE spectrum,
but it extends only from 1150~\AA\ to 1715~\AA\ \citep{moy03} . Therefore, we use the IUE spectrum 
for our spectral analysis, from which we derived a color excess $E(B-V)$ of $0.24\pm0.03$.  
Since the continuum flux level is 30\% higher in the STIS spectrum than in the IUE spectrum,
the mass accretion rate in DI Lac at the time the STIS spectrum was obtained 
was about 30\% higher than when the IUE spectrum was obtained.  
\citet{sha18} did not derive a WD mass for DI Lac, but \citet{sel19} derived $M_{\rm wd}= 0.91 \pm 0.2 M_\odot$
and an inclination $i=15^\circ \pm 10^\circ$.  
In the present work we assume a low inclination $i=18^\circ$ and a WD mass of $1.03 M_\odot$ with an uncertainty of $0.1M_\odot$. 

As explained in \S3.4 for CP Lac, an uncertainty
of $0.1M_\odot$ in the $1.03 M_\odot$ WD mass is taken into account by considering disk models 
from our grid with $M_{\rm wd}=0.8M_\odot$ and $M_{\rm wd}=1.21 M_\odot$. 
Explicitly for V533 Her, where $\Delta M_{\rm wd} = 0.1 M_\odot$ (i.e. $M_{\rm wd} \approx 1.0 M_\odot \pm 0.1 M_\odot$), 
we have     
$$ \Delta \dot{M}_{1.0 M_\odot} = \frac{1}{2} \left( \Delta \dot{M}_{1.1 M_\odot} + \Delta \dot{M}_{0.9 M_\odot} \right), $$ 
since our grid model is in steps of about $0.2 M_\odot$ in the WD mass. 

Furthermore, we use two optical spectra obtained in 1990 (Lick) \& 1991 (La Palma).

\subsection{V533 Herculis - Nova Herculis 1963} 

With a Gaia parallax-derived distance of $1248^{+130}_{-80}$pc, we find for V533 Her a   
reddening of about $E(B-V)=0.025\pm0.025$ from the online Stilism extinction 3D map.   
With an orbital period of 3.528~hr \citep{tho00}, 
it is expected to have 
a secondary mass near $0.3 M_\odot$, and, with the presence of positive superhumps, 
\citet{lei22} infers a rather large WD mass of the order of $1 M_\odot$, since the outer disk edge    
is to reach the 3:1 resonance radius.  
V533 Her has also been suspected to have a moderately magnetized WD, making
it an IP, based on the detection of coherent oscillations, dwarf nova oscillations \citep[DNOs][]{pat79,pat94}. 
However, this has not been confirmed so far, since the DNOs have not been detected since. 
\citet{rod02} derived an inclination of $62^\circ$ and a WD mass of $0.95M_\odot$.
\citet{sel19} adopted an inclination of $57^\circ\pm7^\circ$ and derived a WD mass of $0.95\pm0.15M_\odot$.  
\citet{sha18} gives it a WD mass of $1.07M_\odot$, therefore we adopt a WD mass of $1.03M_\odot$ (from our
grid) with an uncertainty of $0.1M_\odot$. 

\citet{tho00}, based on a phase-resolved spectroscopic analysis showing phase 0.5 absorption, 
suggest V533 Her is an SW Sextantis star with an inclination low enough to avoid eclipses, 
consistent with $i\approx 60^\circ$. This is the value we adopt here.   
As such, V533 Herculis sits in a rather unusual place at the intersection of two  
classifications: SW Sextantis stars and IPs.  

We use a combined and co-added IUE spectrum made of SWP10250 (29/09/80) and LWR08915 (30/09/80), 
together with SWP44805 (29/05/92) and LWP23205 (29/05/92). Though these two sets were obtained 12 years
apart, they have nearly the same flux level and the same continuum shape, making them
almost indistinguishable. The IUE spectrum is consistent with a rather low extinction 
as no obvious 2175~\AA\ absorption feature is detected. We therefore adopt $E(B-V)=0.025\pm0.025$
from the Stilism map for a Gaia parallax-derived distance of $1248^{+130}_{-80}$~pc.   
It has one more IUE SWP spectrum from 1979, but with an extremely low flux level
which we discarded. 


Optical spectra of V533 Her obtained between 1996 and 1998 
reveal that the optical flux varied by a factor 3 \citep{tho00}. 
We therefore use only the combined co-added IUE spectrum for the analysis.  

\subsection{V446 Herculis - Nova Herculis 1960} 

V446 Her is one of a few post-novae exhibiting dwarf nova behavior \citep{hon95b}.
Therefore, its {\it average} mass accretion rate should be below $3 \times 10^{-9} M_\odot$/yr. 
One can expect $\dot{M} \sim 10^{-9}-10^{-8}M_\odot$/yr in outburst
and of the order of $\sim 10^{-10}M_\odot$ or lower in quiescence.  
\citet{tho00} derived an orbital period of 4.968~hr, and   
from its light curve they derived a moderate inclination; therefore, we will assume $i=40^\circ\pm20^\circ$. 
From observation of its decline from the nova eruption, \citet{sha18} derived $M_{\rm wd}=1.11 M_\odot$.
It is to be remarked that, while \citet{sha18} do not list uncertainties for the WD masses in their Table 3, 
they note that "{\it ... the results have significant errors (we estimate $0.1 M_\odot$ or so, ..)}".  
We therefore adopt an uncertainty of $0.1 M_\odot$ in the WD masses from \citet{sha18}. 

V446 Her has only one SWP and one LWP IUE spectrum (collected in 1992).
However, it has one STIS spectrum (made of 4 exposures) obtained in 2003 with a spectral range from 1150~\AA\ to 3150~\AA , 
thereby covering both the Ly$\alpha$ and the 2175~\AA\ ISM bump.
From the STIS spectrum we derived a color excess $E(B-V)=0.43\pm0.05$. 
Optical spectra of V446 Her were obtained in June/July 1997, but 
only the average spectrum was presented which makes it impossible to use for comparison
\citep{tho00}. 

Therefore, for the analysis we use accretion disk models with $M_{\rm wd}=1.03$ and $1.21 M_\odot$, 
$i=18^\circ, 41^\circ,$ and $60^\circ$, and $E(B-V)=0.43\pm0.05$. 

The STIS spectrum of V446 Her was obtained on MJD 52,862.47 corresponding to JD=2,452,862.97. 
We carried out a meticulous visual inspection of the long-term light curve given
in \citet{hon11} near time 4,863 (=JD-2448000) in their Fig.6 upper panel, and found that,    
at the time of the HST/STIS observation, the system was at V=16.5, in `decline' from a magnitude 15.3
reaching eventually a magnitude of 17.0. Namely, it seems that the system might have been reaching a quiescent state
(as originally planned for the observation).

\subsection{V1974 Cygni - Nova Cygni 1992} 

V1974 Cyg was the most thoroughly analyzed nova at the time     
of (and following) its outburst, observed across the entire electromagnetic spectrum 
(from $\gamma$ rays to radio), and with an SSS phase lasting at least 500 days \citep[e.g.][]{sho93,sho96,aus96}. 
As a consequence, it has many UV spectra, but they were collected no more than 3 years after the outburst.  
It was discovered with binoculars on February 19, 1992 by Peter Collins 
\citep{col92} and IUE spectra were already collected on the night of February 20-21. 

At a Gaia distance of 1617~pc, V1974 Cyg has a color excess of $0.32\pm 0.02$ \citep{van05,hac05,hac16} 
which we confirm by using the 2175~\AA\ ISM feature in its FOS spectrum. 
The inclination of the polar ejecta with respect to the observer was derived,  
giving a likely disk inclination $i=39^\circ \pm 2^\circ$ \citep{cho97}. 
It has been shown that its ONeMg WD is strongly magnetic \citep{cho97}, 
and its mass has been estimated to be near or above solar; 
in the present work we assume values of $1.03 M_\odot$ and $1.21M_\odot$, in general
agreement with \citet{aus96,sal05,hac05,hac16,sha18}.  
The system also exhibits positive superhumps \citep{ret96,ole02,bru23a}. 
It has a binary orbital period of 1.95024~hr \citep{shu02}.  

\begin{figure}[h!] 
\includegraphics[scale=0.33,trim= 20 30 0 30]{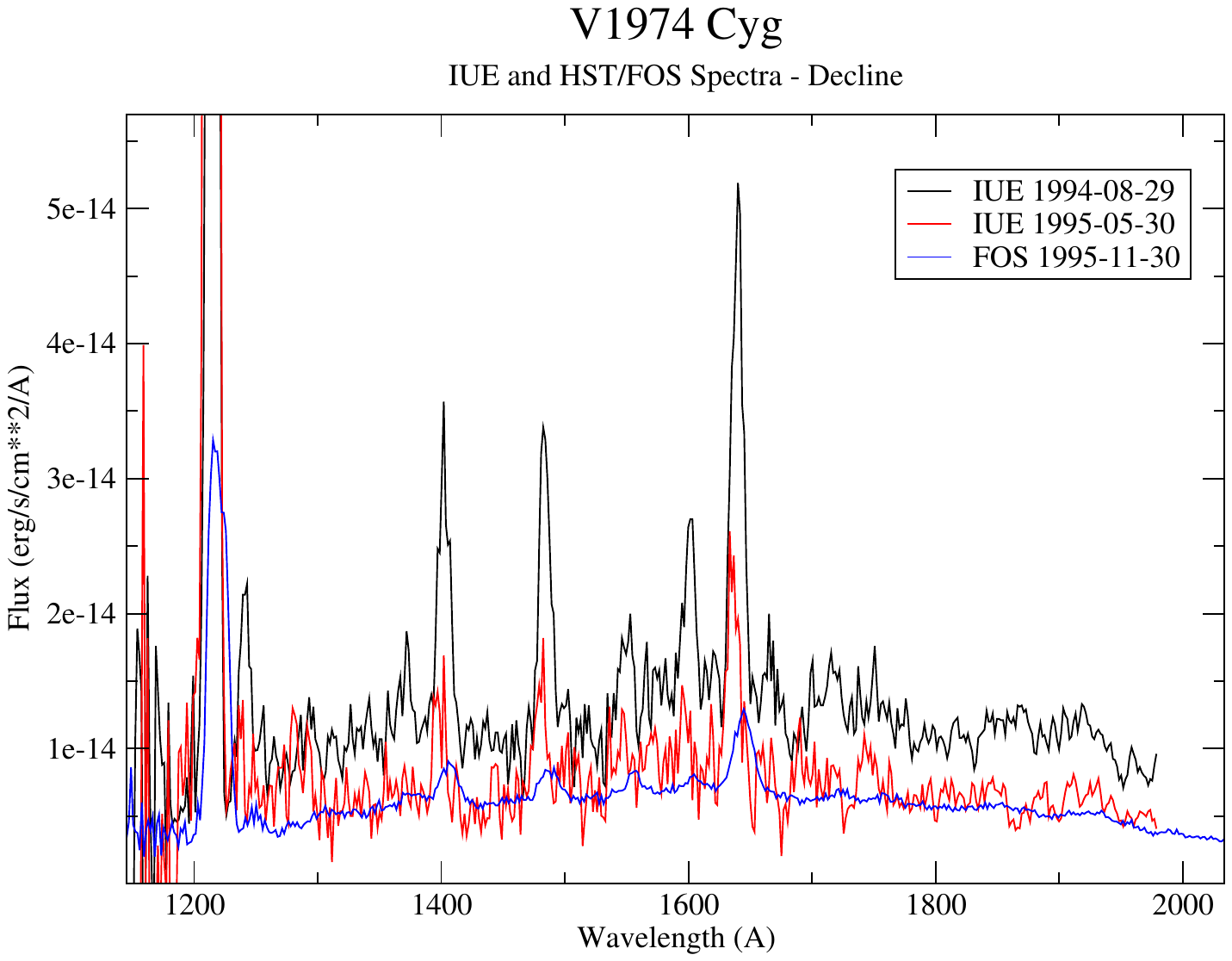} 
\caption{
The decline of the system in the years 1994-1995 as indicated 
from IUE and FOS spectra. 
The strength of the emission lines is decreasing with time and in the last spectrum 
the disk likely contributes most of the continuum flux. 
\label{1974-dec} 
} 
\end{figure} 

V1974 Cyg has a large number of IUE spectra obtained in 1992 and 1993, and only a few from 1994 and 1995. 
Similarly, it has HRS spectra obtained as early as 1992 following its outburst,  
and FOS spectra collected in 1994 and 1995. 
Its latest HST spectrum is from late November 1995, consisting of six FOS G160L (1150-2500~\AA) exposures covering the ISM bump.  

\begin{figure}[h!] 
\includegraphics[scale=0.32,trim= 0 30 0 30]{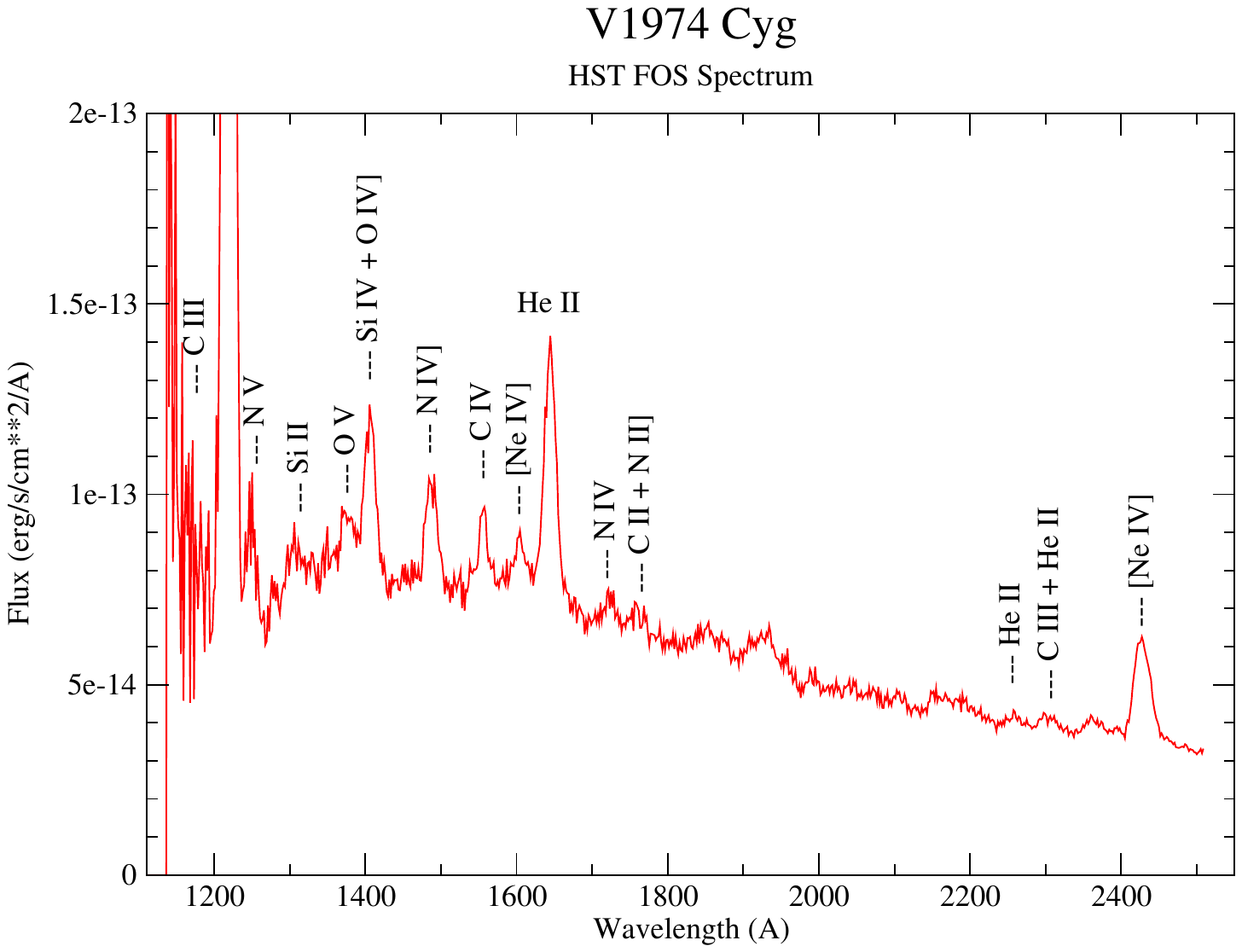} 
\caption{
The November 30, 1995 HST FOS spectrum of V1974 Cyg, dereddened assuming $E(B-V)=0.32$ with 
line identifications. Note the presence of Neon lines.  
\label{1974-lineID} 
} 
\end{figure}

In Fig.\ref{1974-dec} we display its last FOS spectrum compared to IUE spectra obtained $\sim$6 months and 
$\sim$1 year earlier. It seems that the FOS spectrum had reached a quiescent level with fewer and less pronounced
emission lines.  
In Fig.\ref{1974-lineID} we present the dereddened (last) FOS spectrum with line identifications    
based on \citet{fei88} and \citet{van05} who analyzed the emission lines.  

With the presence of many nebular lines in the FOS spectrum, 
it is clear that the nebular material was  
still affecting the spectrum (in late 1995), as also supported by its optical behavior 
\citep{mor01}. 

Its latest optical spectra are from 1994, they are dominated by emission lines, hence we decided to 
only use the November 30, 1995 FOS spectrum for our analysis.

\subsection{V842 Centauri - Nova Centauri 1986}

V842 Cen had a nova eruption toward the end of November 1986 and was observed with IUE
in early January 1987 with follow up observations through 1994. 
A coherent modulation with a period of $\sim 57$~s was detected in the optical, and,  
suspected to be the WD spin period, it was suggested \citep{wou09} 
that V842 Cen is likely an IP. However, the signal was never 
detected in the X-ray \citep{lun12} or UV \citep{sio13} bands, 
and since the system has a low inclination \citep{wou03,sch05}, 
its classification as an IP is still a matter of debate. 

The binary orbital period is uncertain, possibly between 3$\frac{1}{2}$ and 4~hr: 
\citet{wou09} first put it at 3.94~hr, which may have been
confirmed by the $3.64\pm0.46$~hr signal found by \citet{lun12}; 
the last attempt was made by \citet{rao25} who put it at 3.555~hr.  

\begin{figure}[h!] 
\includegraphics[scale=0.33,trim= 0 30 0 30]{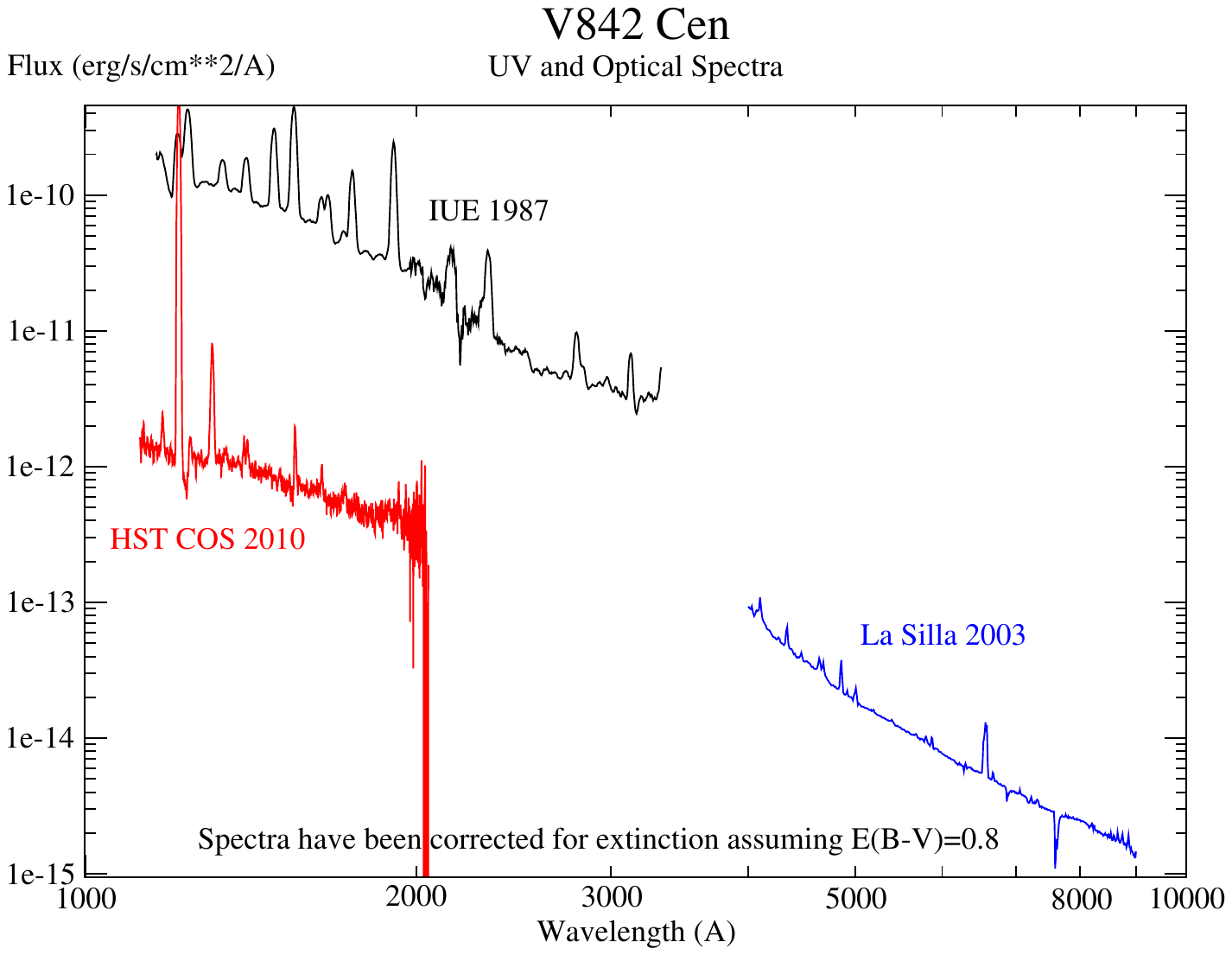}             
\caption{
The spectra of nova Centauri 1986 (V842 Cen): the IUE spectrum obtained in May 1987 
(6 months after the outburst; in black), the optical spectrum obtained in 2003 at La Silla
(in blue), and the HST COS spectrum obtained in 2010 (in red).  
The spectra have been corrected for extinction assuming a color excess $E(B-V)=0.8$. 
\label{v842cens} 
} 
\end{figure}

A color excess $E(B-V)=0.55\pm0.05$ was mentioned in the preliminary report by \citet{kra90} 
from an analysis of the 2175~\AA\ absorption feature in the May 21, 1987 IUE spectrum 
of V842 Cen, but the complete description of the results were never published elsewhere.
We used the same IUE SWP31020 (1987-05-21) and LWP10801 (1987-05-20) segments 
as \citet{kra90} to assess the reddening towards V842 Cen.
As mentioned, and explicitly shown in \S4.3, we find a color excess $E(B-V)=0.8\pm0.1$.
This is the value we use in our analysis. 
Note that the maximum Galactic reddening value \citep[e.g. from][]{sch98,sch11} 
is $\sim 1.5$ as V842 Cen lies in the line of sight of the Galactic disk. 
Its Gaia-derived distance is $1376^{+105}_{-72}$~pc.  
The system has a low inclination \citep{sch05,wou03,war15}, therefore we assume $i=18^\circ$.  
For our analysis we adopt a WD mass of $M_{\rm wd}=1.03M_\odot$ (from our grid of models) 
to agree with \citet{sha18}. We note that \citet{lun12} derived $M_{\rm wd}=0.88M_\odot$, 
and, therefore, we adopt a WD mass uncertainty $0.1M_\odot$.

\begin{figure}[h!] 
\includegraphics[scale=0.33,trim= 0 30 0 30]{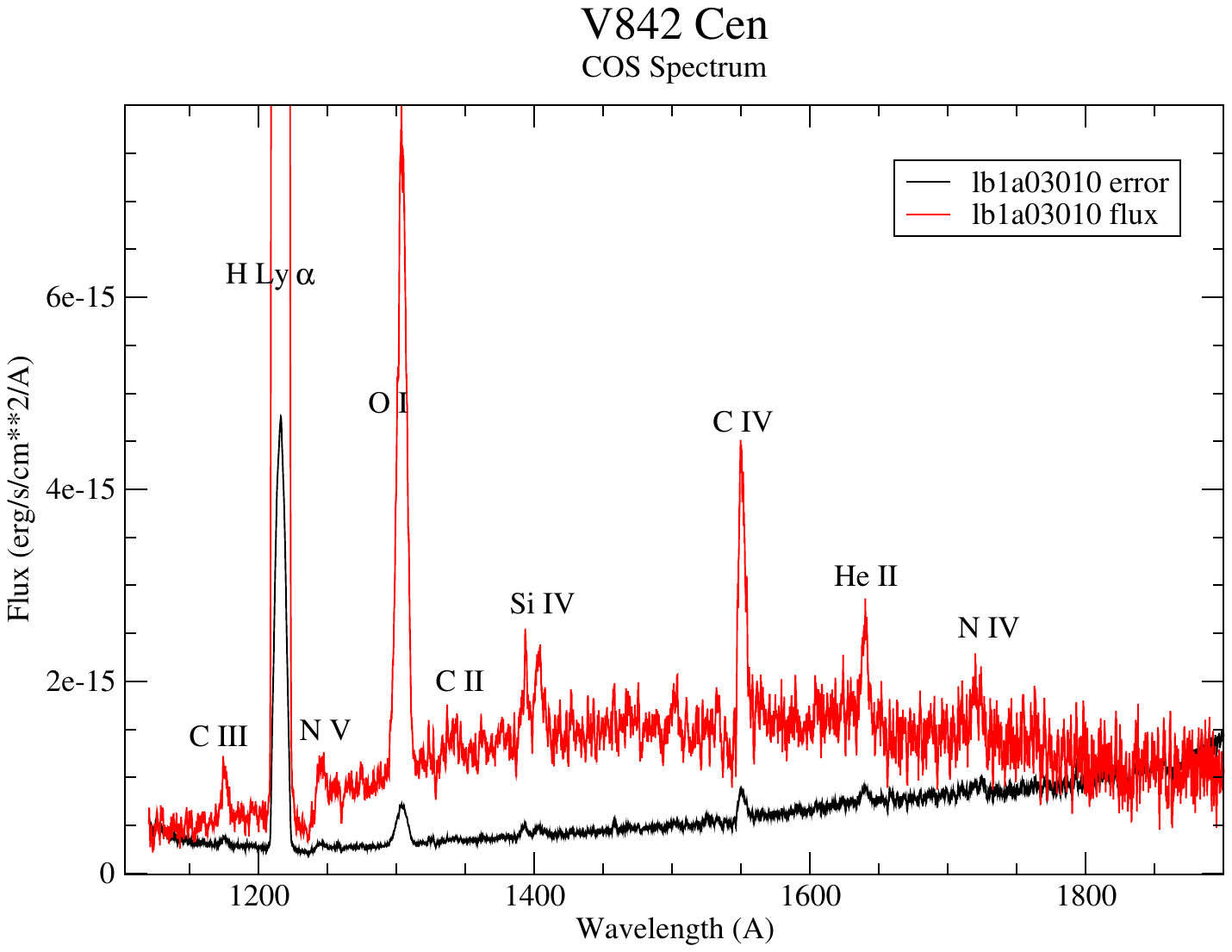}             
\caption{
The COS spectrum of V842 Cen with line identifications. 
The spectrum has not been corrected for extinction. 
Note that the error is as large as the signal for $\lambda > 1800$~\AA . 
\label{842-cos} 
} 
\end{figure}

V842 Cen has 22 IUE spectra from 1987 through 1994.  
However, it has a COS spectrum obtained as late as 2010 \citep{sio13}. 
V842 Cen has also an optical spectrum obtained in 2003 \citep{sch05} 
at La Silla Observatory with the European Southern Observatory (ESO) multi-mode instrument 
(EMMI) at the 3.5~m New Technology Telescope. 
This optical spectrum, the IUE 1987 and HST COS 2011 UV spectra are presented in Fig.\ref{v842cens}. 
Between 1987 and 2010 the UV flux has decreased by a factor of about 100, 
and the 2003 optical spectrum does not match the UV spectra.  
Hence, we only use the HST COS 2010 spectrum for our analysis,
it is presented in Fig.\ref{842-cos} with line identifications. 
\\ 


\section{Spectral Analysis Tools and Technique} 
We use the \texttt{Fortran} suite of codes \textsc{tlusty, synspec, rotin} and
\textsc{disksyn} \citep[or simply referred to as \textsc{tlusty},][]{hub88,hub91,hub94,hub95} 
to generate synthetic spectra of stellar atmosphere and disks. 
The code is publicly available online and well documented
(Hubeny \& Lanz 2017a, b, \& c).
\subsection{WD Stellar Spectra.} 
Novae are accreting at a relatively high rate and are dominated by 
emission from the accretion disk, so we do not expect any significant
contribution from the WD. However, the above suite of codes offers the capability 
to generate synthetic WD spectra.  
\textsc{tlusty} is first run (iteratively until convergence is achieved) 
to generate a one-dimensional (vertical)
stellar atmosphere structure for a given surface gravity ($Log(g)$),
effective surface temperature ($T_{\rm eff}$), and 
surface composition. It is followed by a run of \textsc{synspec} using the output
from \textsc{tlusty} as an input. \textsc{synspec} generates the detailed 
radiation and flux distribution of the continuum and lines as well as the 
stellar spectrum. The code includes the treatment of hydrogen quasi-molecular
satellite lines (low temperature), LTE and NLTE (high temperature) options.  
Rotational and instrumental broadening as well as limb darkening is 
applied with \textsc{rotin}.

\subsection{Accretion Disk Spectra.} 
The disk is assumed to be optically thick, vertically thin, 
axi-symmetric, and in steady state.  
It is integrated in the vertical dimension; namely the disk is 
one-dimensional and the mass transfer rate is the same at all radii. 
This is the standard ({\it alpha}) disk model \citep{sha73} with a nearly Keplerian velocity
and an analytical expression for the temperature profile $T(r)$ \citep{pri81}.  
The disk is divided in a set of concentric rings, 
each behaving like an independent radiating slab, analogous 
to a stellar atmosphere. \textsc{tlusty} is run for each 
independent ring: the input is the mass of the WD star, the inner
radius of the disk, the radius of the ring, and the mass accretion
rate. \textsc{synspec} is run for a given
set of inclinations for which the emerging intensities are 
computed. And last, \textsc{disksyn} is run to integrate all the 
rings outputs over a set of azimuthal segments with the rings 
projected Doppler velocity into a disk spectrum, taking into account
the projected area of the disk for each given inclination angles.   

We first set the inner disk radius to be the radius of the WD as in \citet{wad98}, namely
the radius for a nearly zero-temperature WD for a given mass (contrary to the stellar
spectra where the WD radius is given for a given temperature). 
For that reason, we also ran disks where the inner disk radius is larger than the
WD radius since the latter is expected to increase with temperature. 
Following \citet{wad98}, we consider WD masses $M_{\rm wd}/M_\odot=0.35, 0.55, 0.80, 1.03,$ and $1.21$,
inclinations $i=18, 41, 60,$ and $75$~deg (since the disk is assumed to 
be flat, its approximation is much less appropriate for nearly edge-on $\sim80^\circ$
disks). In the text and tables, we have in some instances rounded these numbers:   
$1 M_\odot$ for $1.03 M_\odot$, $1.2 M_\odot$ for $1.21 M_\odot$, $i=20^\circ$ for $i=18^\circ$, 
and $i=40^\circ$ for $i=41^\circ$. Due to the appearance of rather large uncertainties
in $E(B-V)$, $M_{\rm wd}$, and $i$, this has no effect on the results. 
The grid of disk models is in steps of $0.5$ in the log of the mass accretion rate, $\log(\dot{M})$, 
starting as low as $\dot{M}=10^{-10.5} M_\odot$/yr and going as high as $10^{-6.0}M_\odot$/yr. 
Since this is a discrete grid, we linearly interpolate between (two) discrete values to find
the mass accretion rate.

The non-zero temperature WD radius is taken from \citet{woo95}: it increases with
increasing temperature. 
In this manner, we create a grid of accretion disk models covering a wide range
of WD masses, accretion rates, and inclinations. 
We generated spectra covering the entire UV spectral
range and into the optical, namely, from 900~\AA\ to 7500~\AA .

In contrast to our previous spectral analyzes of accretion disks \citep[e.g.][]{god17,god18,god24},
where we ran \textsc{synspec} to generate emergent flux, in the present work we ran    
\textsc{synspec} to generate specific intensity evaluated at the given angles above. 
This has the advantage of not relying on limb-darkening correction factors \citep[that were computed for
the UV at 1448~\AA , see ][]{dia96},   
which are needed when using the emergent flux to generate the disk spectrum at a given inclination angle.  

We also have the possibility to generate disk rings of arbitrary temperature which can be used
to change the radial temperature profile of the disk if needed. 
The standard disk model has a temperature decreasing outward. We stop the standard disk model at a radius
where it reaches a temperature of about 12,000~K (which is similar but a little higher than in \citep{wad98}). 
Indeed, \textsc{tlusty} has difficulties converging below that temperature as convection becomes important.  
Beyond that radius, we keep the option to extend the disk as an isothermal ($\sim 12,000$~K) disk up to 
an outer radius $R_d$ which is taken as a free parameter. The size of $R_d$ is dictated by the fit;
namely, $R_d$ is increased until a best-fit is obtained. While this appears rather empirical,    
tidal interaction, spiral waves, the interaction of the stream with the disk, and irradiation of the
disk by the hot WD and hot inner disk all work to heat up the outer disk. 

We carried out such modeling in the past, e.g. \citet{god17}, and in the present analysis we also
considered such disks together with standard disk models. 
Further details on the disk modeling, for each nova, are given in the results section.  

\subsection{Dereddening of the Spectra.} 

Since novae are rarely nearby, they are often strongly affected by
ISM extinction which reddens their spectra.  
The ISM polycyclic aromatic hydrocarbon (PAH) grains \citep{li01}   
produce a strong broad absorption feature centered at 2175~\AA\   
in the longer wavelength segment of the IUE spectra (i.e. LWP or LWR), 
as well as in some HST spectra (e.g. STIS, FOS; depending on the instrument settings).  
Thanks to that absorption feature, one can assess and correct for the reddening of the spectra
\citep{ver87} in the following manner.

\begin{figure}[h!] 
\includegraphics[scale=0.35,trim= 0 0 0 10]{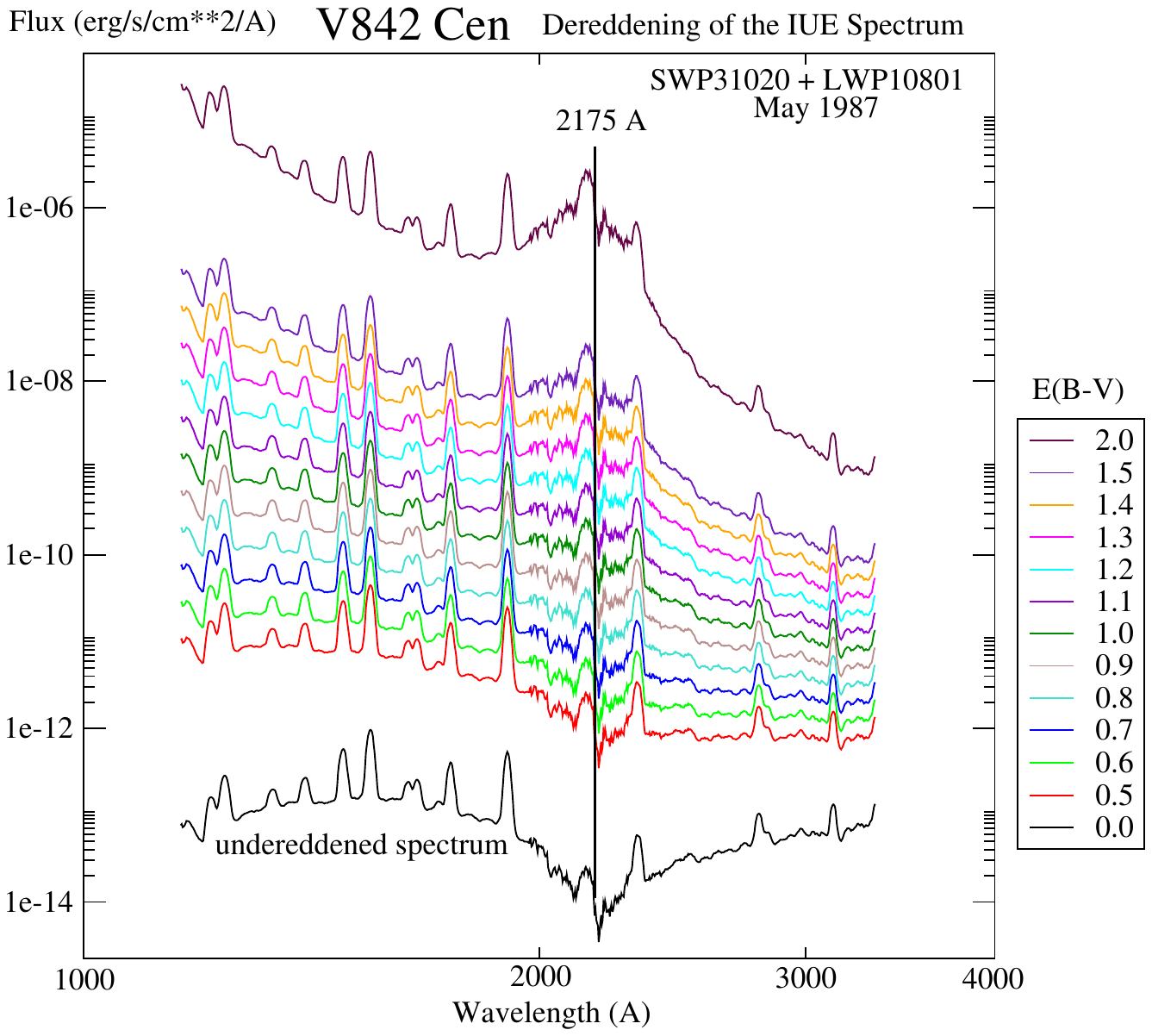} 
\caption{
The 1987 IUE spectrum of the nova V482 Cen corrected for
ISM extinction (``dereddened'') on a log-log scale. 
The original spectrum (not corrected) is shown in black 
(bottom line) and exhibits the well-known 2175~\AA\ absorption feature (ISM `bump'; \cite{li01}). 
To correct for extinction, we deredden the spectrum assuming increasing $E(B-V)$ values.  
The true color excess value $E(B-V)$ is that for which the 2175~\AA\ absorption feature vanishes. 
Values that are too large yield to an emission-like feature. 
\label{der1} 
}
\end{figure}

\begin{figure}[h!] 
\includegraphics[scale=0.35,trim= 40 50 0 10]{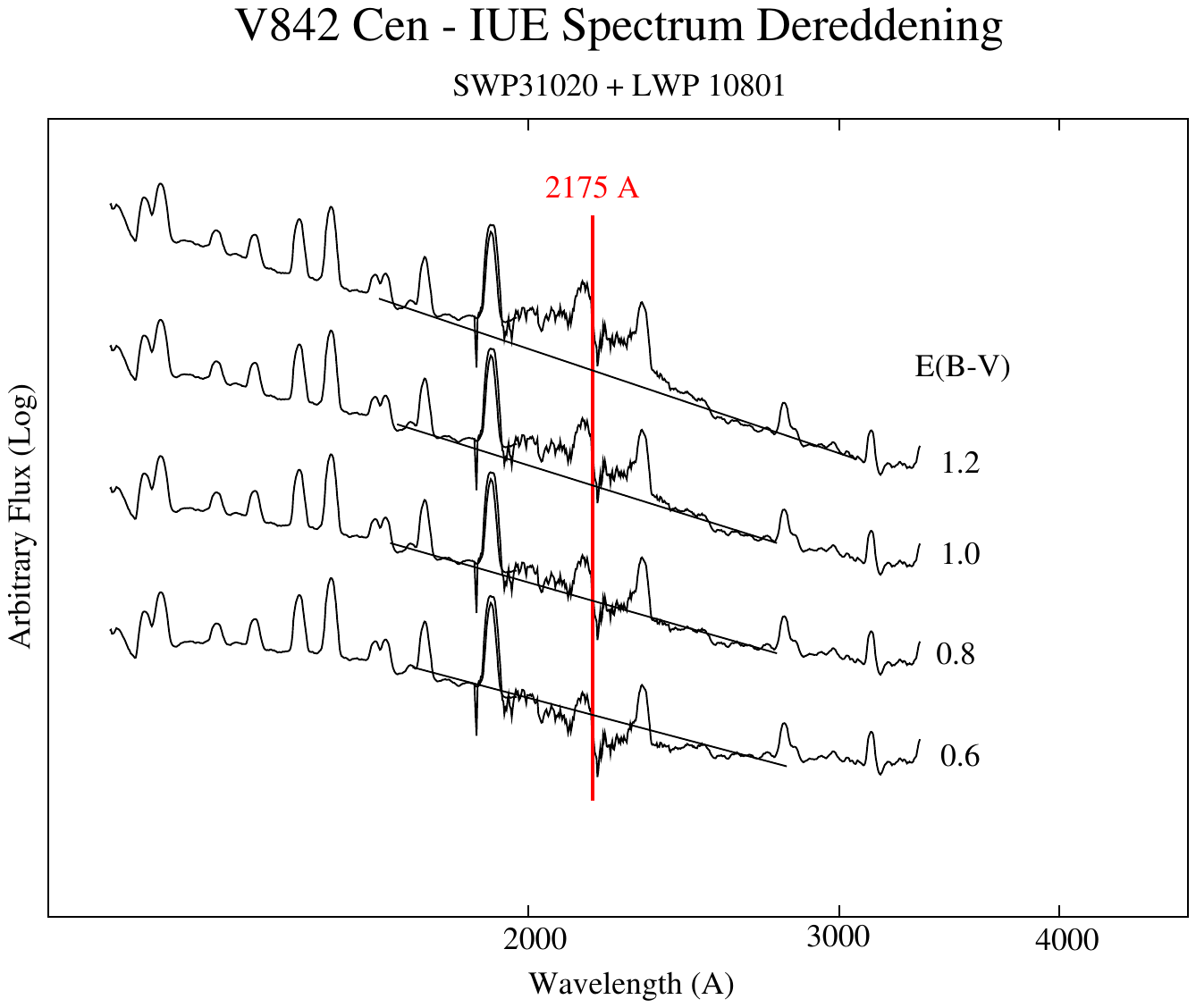} 
\caption{
Same as Fig.\ref{der1}. For clarity, only a few $E(B-V)$ values are shown, 
straight lines have been drawn, and we staggered the spectra vertically
on an arbitrary scale.
Fine tuning yields a value of $E(B-V)=0.8\pm0.1$ for V442 Cen. 
\label{der2} 
}
\end{figure}

For the extinction law, we adopt the analytical expression of \citet{fit07}.   
It is then used to deredden the observed spectra for increasing values of the 
color excess $E(B-V)$, the value for which the 2175~\AA\ absorption feature vanishes 
gives the desired value of $E(B-V)$ and the corrected spectrum.  
This is illustrated in Figs.\ref{der1} and \ref{der2}, where we present the dereddening of the
IUE spectrum of V842 Cen. 

It is to be noted that the 2175~\AA\ feature poorly correlates with the far UV extinction
\citep{gre83}, and nova ejecta can also contribute dust affecting the spectra of novae \citep{sho18}.
However, based on the size and composition of the nova ejecta larger grains, they do not contribute to the 2175~\AA\ 
absorption feature. This introduces an unknown amount of possible extinction that, unlike the ISM 
extinction, cannot be corrected using the 2175~\AA\ feature. 

\subsection{Fitting Technique.} 

Before fitting an observed spectrum with theoretical synthetic
spectra, we deredden the spectrum and 
mask any strong emission lines, absorption features
(e.g. ISM), and/or detector artifacts.

While we usually use a least $\chi^2_\nu$ method to assess the best fit, 
for a given WD mass and inclination, the scaling of the theoretical
disk spectrum to the observed spectrum provides a distance which is 
compared to the actual distance to the nova for validation. 
Not every least $\chi^2_\nu$ value provides the correct distance
and not every correct distance gives the best fit. 
For that reason, we do not use the $\chi^2_{\nu}$ value. 
We first use the correct distance then assess
the fit by visual inspection.

With the distance known from Gaia DR3, the main 
significant uncertainties are from the inclination $i$, the
WD mass, and color excess $E(B-V)$. 
It is to be remarked that the statistical error is much smaller than errors due to 
uncertainties in the Gaia distance $d$, $M_{\rm wd}$, inclination $i$, and $E(B-V)$. 
The propagation of the uncertainties in the above parameters 
on the resulting mass accretion rate $\dot{M}$ is estimated by fitting the spectrum to disk models 
while changing the values of the parameters $d$, $M_{\rm wd}$, $i$, and $E(B-V)$. 
For example, to assess the effect of a $\sim 0.2 M_\odot$ uncertainty in a $\sim 1.0 M_\odot$ WD model on
$\dot{M}$, we fit the observed spectrum with accretion disk models assuming $M_{\rm wd}=0.8 M_\odot$, $1.0 M_\odot$ and $1.21 M_\odot$, 
as shown explicitly in \S3.4 for CP Lac and in \S3.6 for V533 Her.  
A typical uncertainty of 20\% in $M_{\rm wd}$ gives an error of 28\% in $\dot{M}$. 
And similarly, for $i$ and $E(B-V)$, 
an uncertainty of 10$^\circ$ in the inclination $i$ (near 45$^\circ$) gives 30\% in $\dot{M}$, 
and an uncertainty of $0.05$ in $E(B-V)$ gives a 45\% error in the flux near 1500~\AA\ and only 12\% at 6000~\AA .
Since the flux scales like the inverse square of the distance, 
a relative uncertainty of 10\% in the distance $d$ gives an error of $\sim$21\% in $\dot{M}$. 
The errors are then added by simple quadrature. 
\\

\section{Spectral Analysis and Results}

\subsection{BK Lyn}

\begin{figure}[t!] 
\includegraphics[scale=0.36,trim=052 190 0 150,clip]{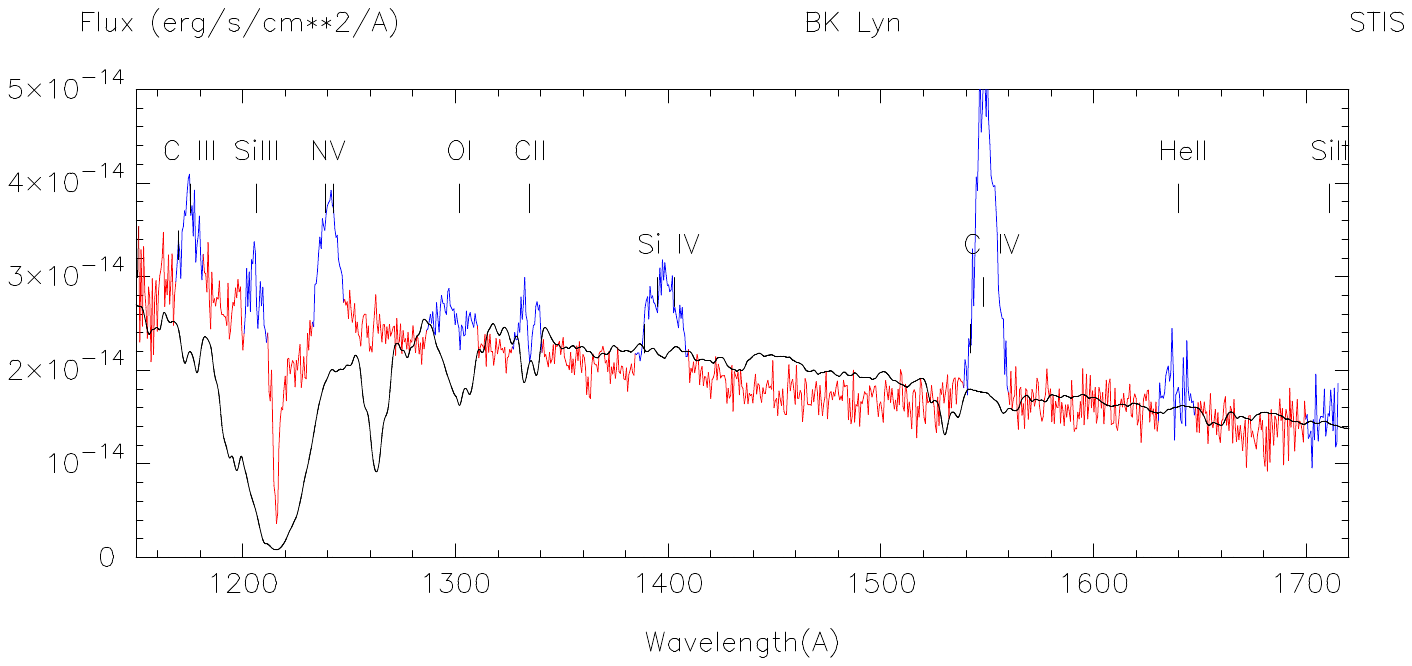}             
\includegraphics[scale=0.36,trim=052 190 0 185,clip]{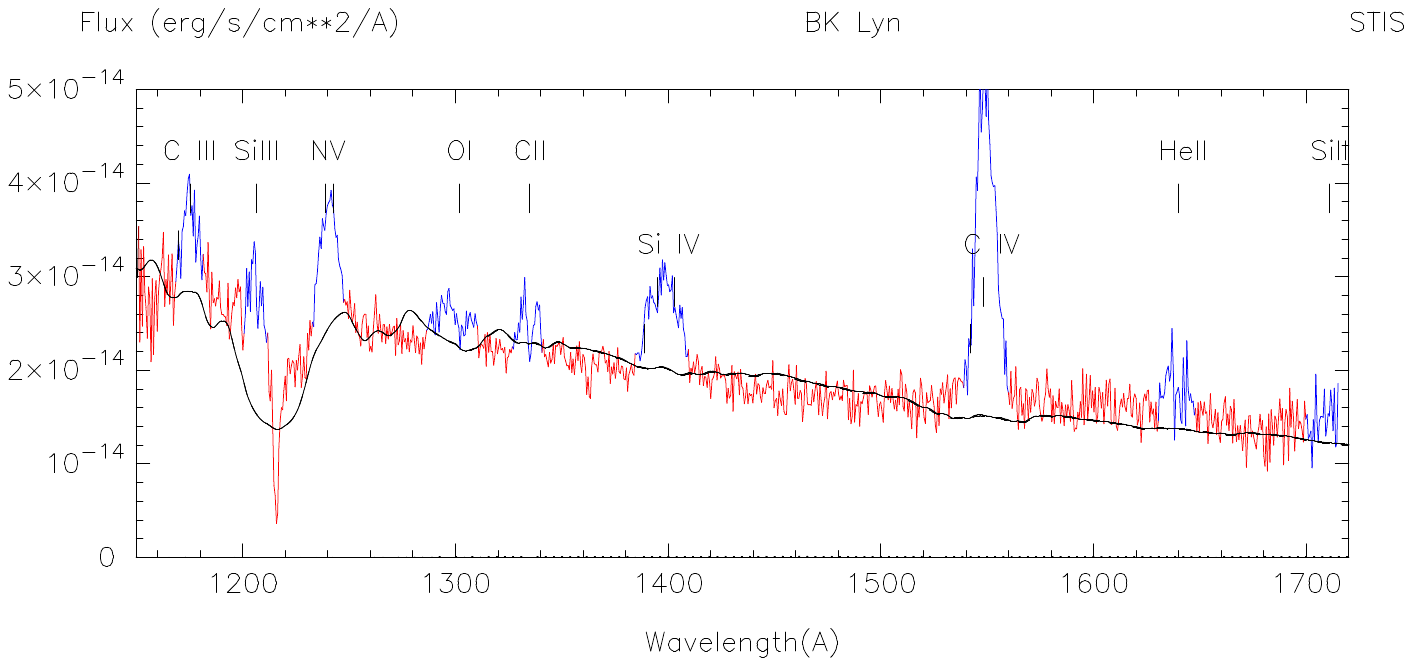}  
\includegraphics[scale=0.36,trim=052 120 0 185,clip]{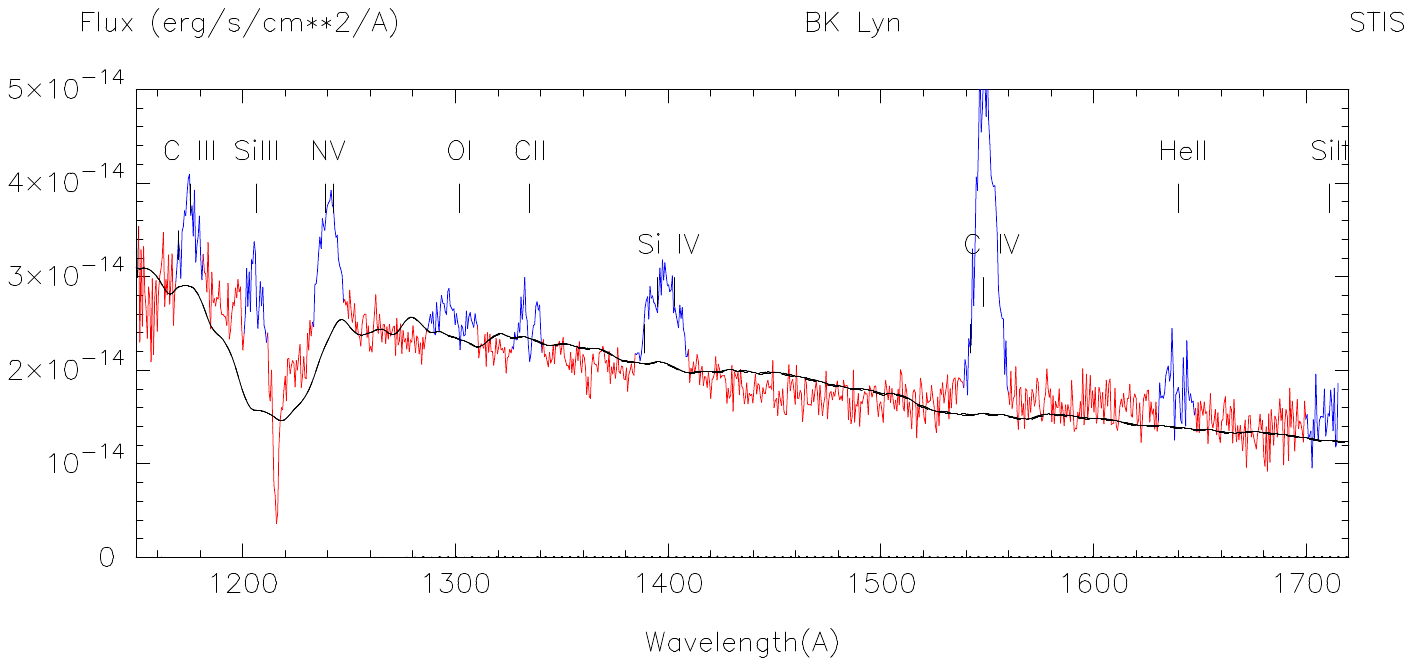} 
\caption{Accretion disk fits to the STIS spectrum of BK Lyn. Prominent emission lines 
have been annotated. The STIS spectrum is in red and the emission lines (in blue) 
have been masked for the fitting, the theoretical disk spectrum is in black. 
{\it (a) Top panel.} 
The accretion disk model  
has $M_{\rm wd}=0.35M_\odot$, $\dot{M}=1.1 \times 10^{-9}M_\odot$/yr, $i=40^\circ$.
Due to the relatively low disk temperature, the model provides a very poor fit in the short wavelength region. 
{\it (b) Middle panel.} 
We increased the WD mass to $M_{\rm wd}=1.0M_\odot$ and obtained a much better fit, 
with $\dot{M}=2.3 \times 10^{-10}M_\odot$/yr, and $i=60^\circ$.  
{\it (c) Bottom panel.} 
As the WD mass was increased to $M_{\rm wd}=1.2M_\odot$, 
the mass accretion rate decreased to $\dot{M}=1.45 \times 10^{-10.}M_\odot$/yr; here too $i=60^\circ$.  
\label{bkl} 
} 
\end{figure}

Since the mass of the WD in BK Lyn is unknown, we ran models ranging from $0.35M_\odot$
to $1.21M_\odot$. 
For disk models with $M_{\rm wd}=$0.35, 0.55, and $0.80M_\odot$, 
no good fit could be obtained and they had to be excluded. 
This is due to the broad Ly$\alpha$ feature in the disk model which doesn't fit
the observed spectrum, even when accounting for the presence of emission lines (see Fig.\ref{bkl}a). 
We note, however, that for the $0.35M_\odot$ WD mass \citep[as derived by][who also found $i=32^\circ$]{dob92} 
model, the fit yielded a mass accretion rate of $1.1 \times 10^{-9}M_\odot$/yr. 

For $M_{\rm wd}=1.03 M_\odot$, the inclination had to be set to $i=60^\circ$ to agree with the Ly$\alpha$ region,
giving $\dot{M}=2.3 \times 10^{-10}M_\odot$/yr (presented in Fig.\ref{bkl}b). 
The same model with $i=41^\circ$ didn't fit the Ly$\alpha$ region and, it too, had to be excluded. 
For $M_{\rm wd}=1.21 M_\odot$ a good fit was obtained for both $i=40^\circ$ and $i=60^\circ$, 
yielding $\dot{M}=7.2 \times 10^{-11}M_\odot$/yr and $1.45 \times 10^{-10}M_\odot$/yr
(see Fig.\ref{bkl}c).

\begin{figure}[t!]  
\includegraphics[scale=0.36,trim=052 190 0 150,clip]{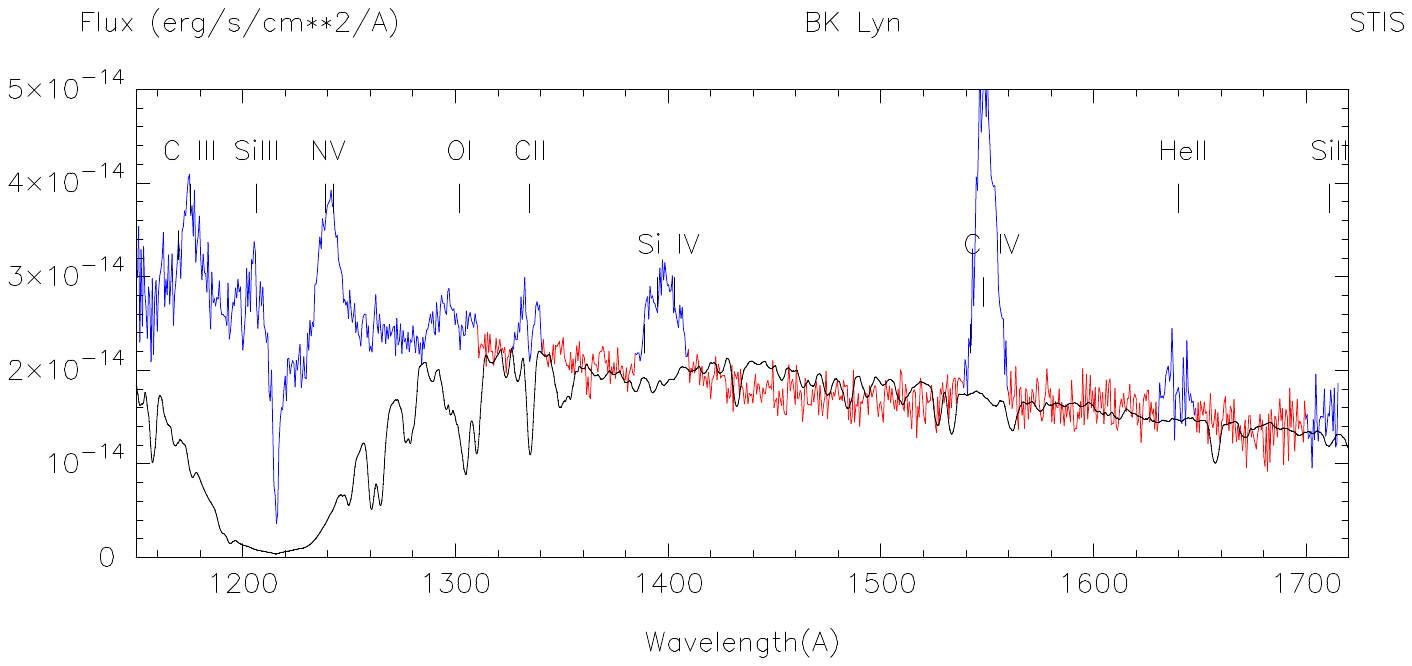}              
\includegraphics[scale=0.36,trim=052 170 0 185,clip]{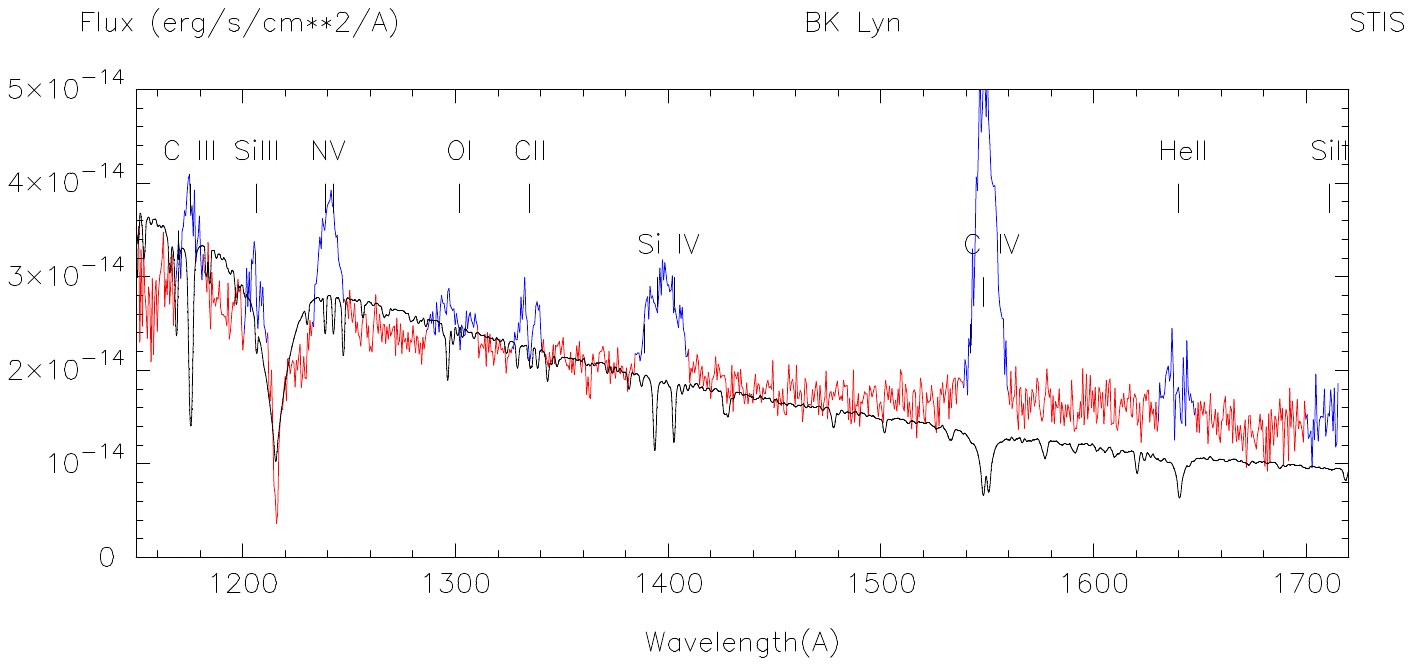}   
\includegraphics[scale=0.36,trim=052 120 0 185,clip]{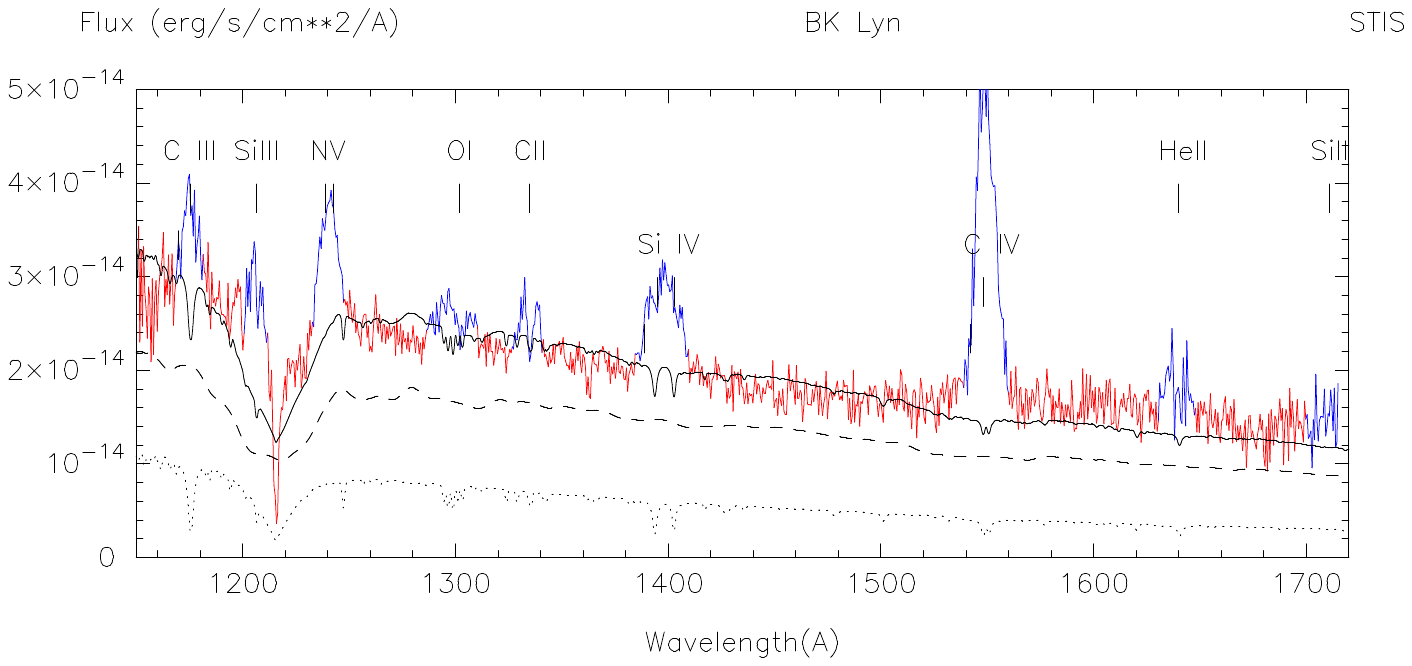} 
\caption{White Dwarf and Accretion disk fits to the STIS spectrum of BK Lyn. 
{\it (a) Top panel.} 
A $18,000$~K WD model with a mass $M_{\rm wd}=0.8M_\odot$ is fitted to the STIS spectrum
of BK Lyn. 
{\it (b) Middle panel.} A 55,000~K WD model with a mass of $1 M_\odot$ is fitted
to the STIS spectrum of BK Lyn. 
{\it (c) Bottom panel.} 
A combined accretion disk plus WD model fit to the STIS spectrum of BK Lyn. 
The accretion disk with $\dot{M}=10^{-10}M_\odot$/yr and $i=60^\circ$ provides 70\% of the flux 
(dashed line) and the 40,000~K WD contributes the remaining 30\% (dotted line). 
\label{bkl2} 
} 
\vspace{0.5cm} 
\end{figure}

For comparison, based on optical data alone and assuming $M_{\rm wd}= 0.8M_\odot$, $i=45^\circ$, and $d=800$~pc, 
\citet{pat13} derived a mass transfer of about $2 \times 10^{-9}M_\odot$/yr,
the limit between NL and DN behavior. 
For the same parameters (mass, inclination, distance) we found $\dot{M}=5.4 \times 10^{-10}M_\odot$/yr,
about four times lower than \citet{pat13}. However, the optical flux varies a lot, and, at the time
of the STIS observation, the system was in a lower brightness state (as stated earlier in \S3.2).

The STIS spectrum was previously modeled by \citet{zel09}, who  
for a $0.8M_\odot$ WD mass obtained $\dot{M}=3.2 \times 10^{-9}M_\odot$/yr
with $d=560$~pc, and for a $1.2M_\odot$ WD mass obtained 
$\dot{M}=1 \times 10^{-9}M_\odot$/yr with $d=300$~pc, both assuming $i=75^\circ-81^\circ$. 
The high $\dot{M}$ they obtained is due to the larger inclination; 
indeed, $i=75^\circ-81^\circ$ significantly reduces the flux, and to match the
observed continuum flux level, $\dot{M}$ has to be increased. 

Since the uncertainty in the results is mainly dictated by the unknown mass and inclination, 
we summarize the disk results as follows: $\dot{M}= 1.5 \pm 0.8 \times 10^{-10}M_\odot$/yr,
for $M_{\rm wd}=1.1 \pm 0.1M_\odot$ and $i=50^\circ \pm 10^\circ$. 

Since the flux (and therefore the derived accretion rate) is so low, we also checked  
a WD fit to the observed STIS spectrum. We find that the slope of the STIS spectrum agrees with 
a $\sim$18,000~K WD stellar atmosphere model, but such a low temperature model has a deep and broad 
Ly$\alpha$ absorption feature that doesn't fit the STIS spectrum (see Fig.\ref{bkl2}a); furthermore, 
the resulting distance is far too small (of the order of $\sim100$~pc).  
Higher temperature WD models ($T_{\rm wd} \sim 50,000$~K) that fit the distance have a continuum 
slope that is far too steep (see Fig.\ref{bkl2}b). 

We therefore tried combined accretion disk plus WD models. 
Since the WD model overall degrades the disk fit, we find that models where the WD contributes $\sim$1/3
or less of the flux can provide a reasonable fit to the STIS spectrum. 
A better fit is also obtained for the larger WD models and larger inclinations.
We present such a fit in Fig.\ref{bkl2}c: the WD, with $M_{\rm wd}=1.2 M_\odot$ and $T_{\rm wd}=40,000$~K, 
contributes 30\% of the flux, and the disk, with $\dot{M}=10^{-10}M_\odot$ and $i=60^\circ$, contributes
the remaining 70\%.

\subsection{HR Del}

For the analysis, we assumed both $M_{\rm wd}=0.55M_\odot$ and $0.8M_\odot$, 
and an inclination of $41^\circ$ to agree with \citet{bru82}, \citet{kue88} and \citet{sha18}.  
The luminosity from the central source (i.e. WD) was estimated to be $10^{36}$erg/s
with $T_{\rm eff}=65,000$~K \citep{mor09}. 
We note here that such a WD would contribute very little flux at such a distance
and can completely be neglected when compared to the contribution of the disk. 
However, the radius of a $0.55M_\odot$ WD increases significantly with the temperature 
and for $T_{\rm eff} \approx 65,000$~K and higher it increases by 50 to 100\%. 
Therefore, in the modeling we assumed an inner disk $R_0$ radius of $1.5$ and $2.0 \times R_{\rm wd}$,
where $R_{\rm wd}=9,050$~km is the radius of a $0.55 M_\odot$ {\it cold} WD.

We first modeled the IUE spectrum separately from the optical spectrum. 
For the above disk models, we found a mass accretion rate of $4.3$ and $5.4 \times 10^{-7} M_\odot$/yr
for $R_0=1.5$ and $2.0 R_{\rm wd}$, respectively. 

\begin{figure}[h!]  
\includegraphics[scale=0.38,trim= 60 120 0 130]{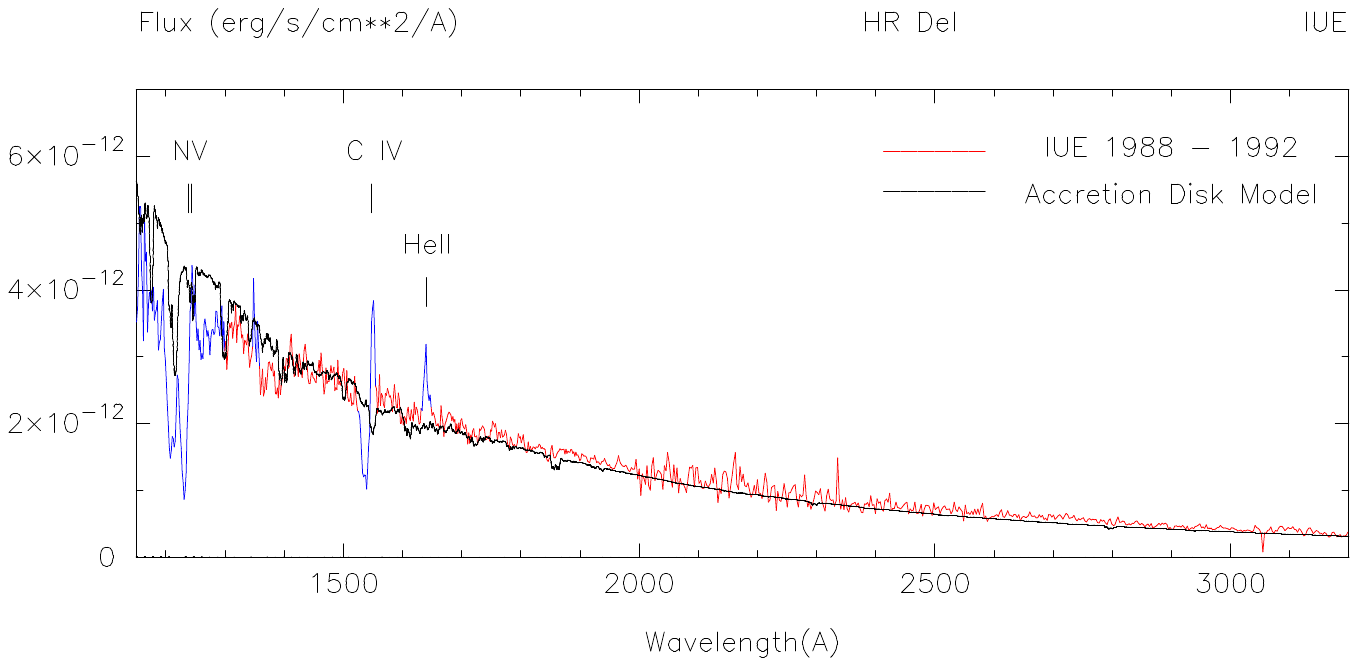} \\  
\caption{An accretion disk fit (in black) to the IUE spectrum of HR Del (in red, blue regions have been omitted). 
A $0.80 M_\odot$ WD mass is assumed, yielding to a mass accretion rate of $\dot{M}=3.04 \times 10^{-7}M_\odot$/yr.
At such a high mass accretion rate, all the disk models are undistinguishable: they all fit the observed spectrum 
equally well, except for the shorter wavelength region ($<$1300~\AA ), where they all have too much flux. 
\label{hrdeliuefit} 
} 
\end{figure}

For the $0.80M_\odot$ WD mass model, we set the inner radius of the disk $R_0=1.2R_{\rm wd}$, where here $R_{\rm wd}$ is 
the radius of a cold $0.8M_\odot$ WD, namely 6,990~km. The increase of the WD radius with temperature is much less
pronounced for larger WD masses. We obtained $\dot{M}=3.04 \times 10^{-7}M_\odot$/yr.
This model is presented in Fig.\ref{hrdeliuefit}. 

We note that even with a WD mass of $1.2 M_\odot$, we still obtained a very large mass
accretion rate: $1.5 \times 10^{-7}M_\odot$/yr. However, such a model had to be excluded 
because it does not agree with the (current estimate of the) WD mass. 

\begin{figure}[h!] 
\epsscale{1.3} 
\plotone{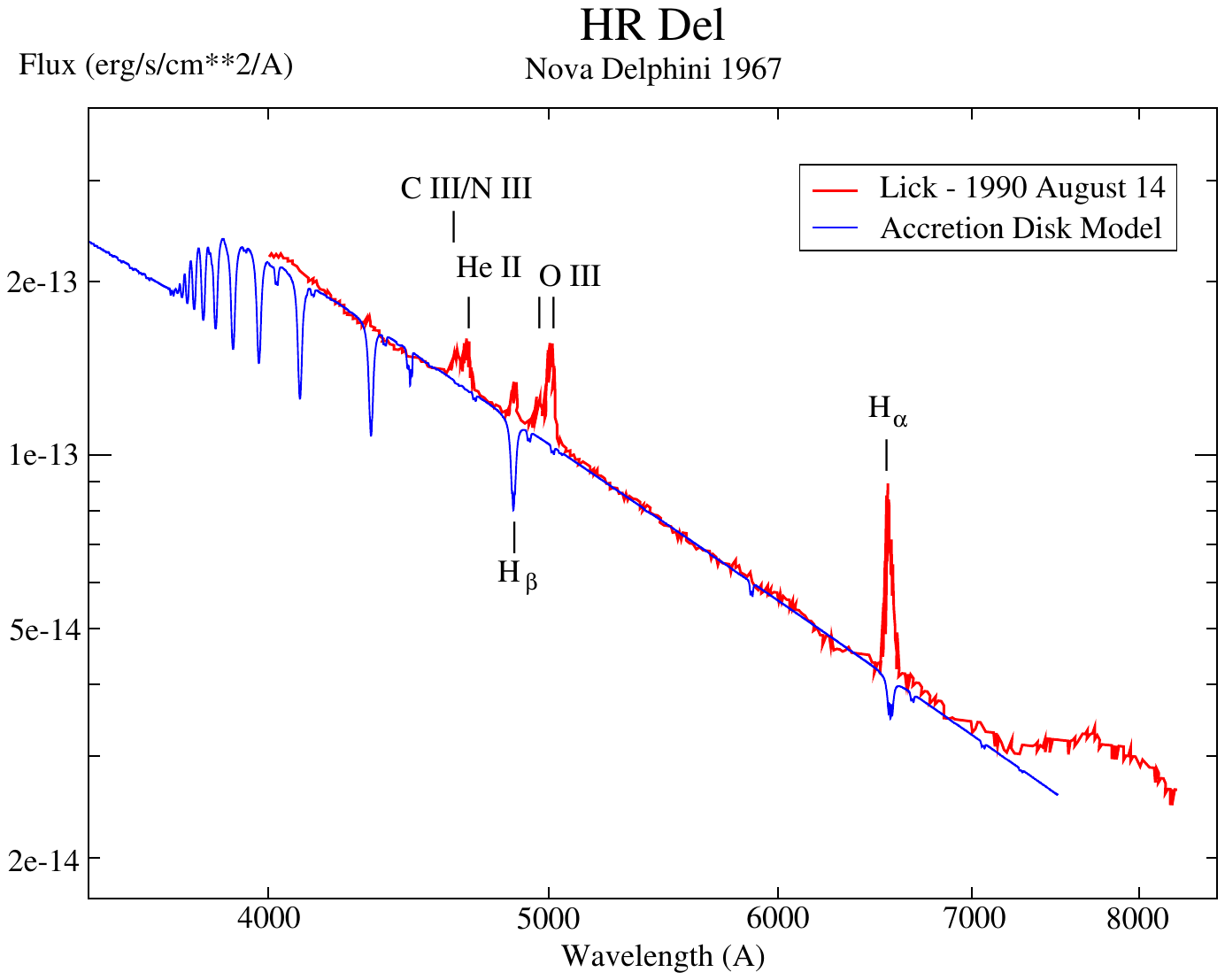} 
\caption{
The Lick Optical spectrum of HR Del (in red) obtained in 1990 is modeled with an accretion disk model (in blue). 
For a WD mass of $0.8M_\odot$, an inclination $i=41^\circ$, and a Gaia parallax-derived distance of 896~pc, the accretion rate obtained is $\dot{M}=3.04\times 10^{-7}M_\odot$/yr.  
Some nebular emission lines (e.g. O\,{\sc iii}) are still present 33 years after its nova eruption.  
Note that both axes (Flux \& Wavelength) are logarithmic for convenience. 
The excess flux in the longer wavelength range is an artifact (it can be seen in all the other Lick spectra of \citet{rin96}), and in any case it is extending beyond our limit of 7,500~\AA . 
\label{hrdeloptfit}
} 
\end{figure}

Next, we fitted the optical spectrum and found, quite surprisingly, that it can be fitted
with exactly the same disk models as the UV spectrum. 
We present in Fig.\ref{hrdeloptfit} the fitting of the optical spectrum with the disk
model with $M_{\rm wd}=0.8M_\odot$.  

\begin{figure} 
\epsscale{1.40} 
\plotone{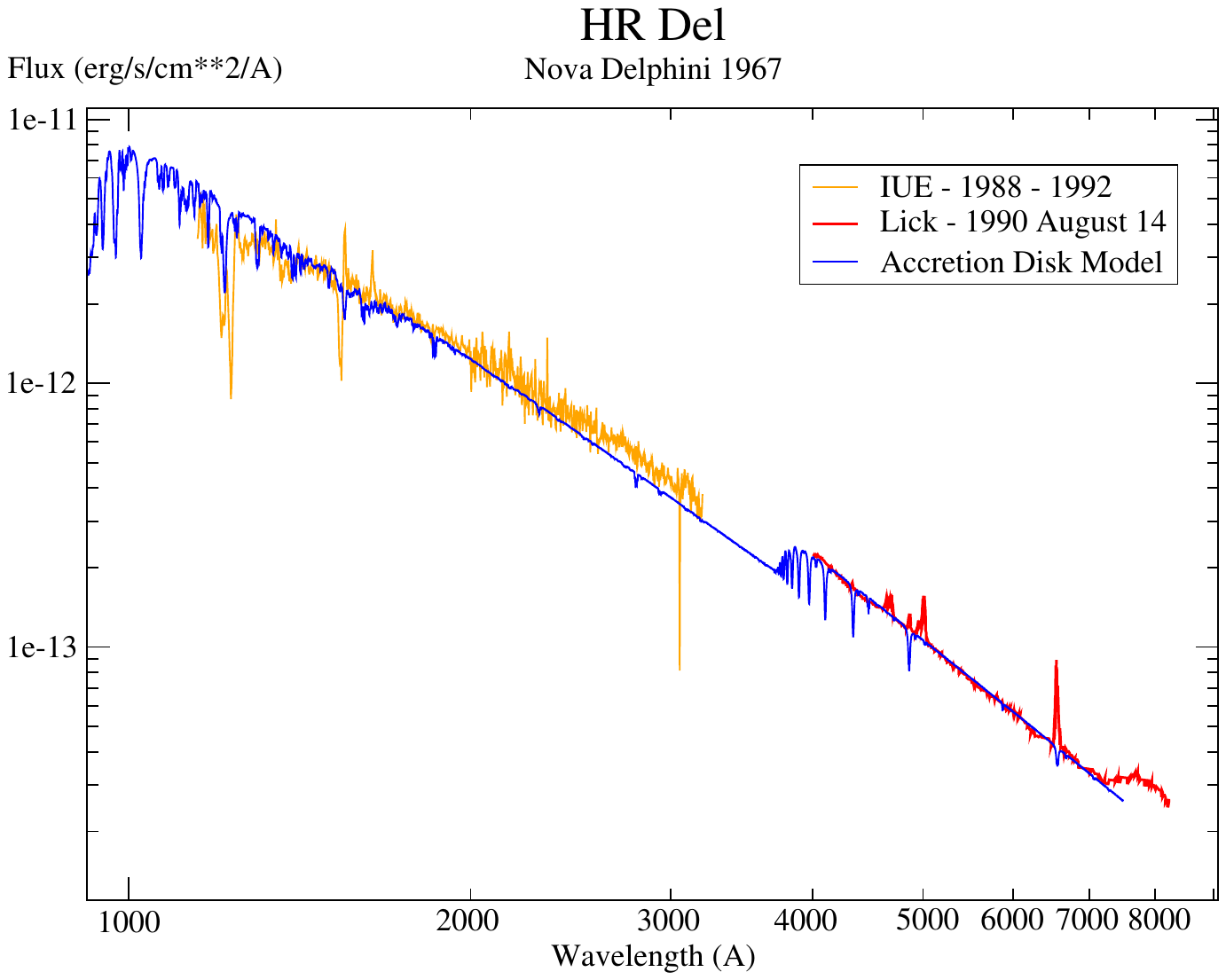} 
\caption{Accretion disk model fit to the UV and Optical spectra of HR Del. 
	The spectra agree relatively well with an accretion disk model with 
$M_{\rm wd}=0.80M_\odot$, $i=41^\circ$, and $\dot{M}=3 \times10^{-7}M_\odot$/yr.  
\label{hrdelfit}
} 
\end{figure}

In Fig.\ref{hrdelfit} we present the same model fit showing both the UV and optical spectral wavelength range.  
In spite of the fact that the optical and UV data were obtained with different telescopes and completely different calibrations, the spectral fit is remarkably robust and accurate.  
The fit is less accurate between 2000-3000~\AA , and deviates by as much as 17\%, where the IUE LWR/LWP segments have a relative error of 10\% to 20\%.  

Assuming a $0.6M_\odot$ WD mass, \citet{sel19} obtained $\dot{M}=1.5\times 10^{-7}M_\odot$/yr
in their analysis of the IUE spectra of HR Del.  
For comparison, we ran model fits with $M_{\rm wd}=0.55M_\odot$ and obtained a mass accretion rate 3 times larger. 
Some of the discrepancy with \citet{sel19} could be attributed to their distance of $d=$850~pc 
(instead of 900~pc), lower reddening (0.16 vs. 0.18), larger WD mass ($0.60M_\odot$ vs $0.55M_\odot$) and smaller 
WD radius (8,700~km vs 18,000~km). However, it was already noted before \citep{sel19} that UV spectral analyses
using accretion disk models \citep{pue07} yield higher mass accretion rates than simply integrating
the UV and optical fluxes.     

Since the WD mass is the main source of uncertainty in our analysis of HR Del, 
we summarize the results as $\dot{M}=4.2\pm1.2 \times 10^{-7}M_\odot$/yr, for a
WD mass $M_{\rm wd}=0.68^{+0.12}_{-0.13}M_\odot$ and $i=41^\circ$. 

\subsection{RR Pic} 

For the analysis of RR Pic, we assumed a WD mass of 
$0.55M_\odot$ and  $1.03M_\odot$, together with an inclination of $60^\circ$ and $75^\circ$.
For an inclination of $60^\circ$, the spectral fit gave a mass accretion of the order  
$\sim2$ and $\sim5 \times 10^{-8}M_\odot$/yr for a WD mass of $1.03M_\odot$ and $0.55M_\odot$, respectively. 
However, the fit is rather poor as it doesn't fit the slope of the continuum flux level
even when increasing the inclination to $75^\circ$. 

As mentioned in \S3.3, it is likely that the outer disk is heated up and for that reason 
we assumed an effective surface temperature of 12,000~K in the outer regions of the disk. 
We carried out spectral fits of such disks with a heated outer region, 
and found a mass accretion $\dot{M}=1.2 \times 10^{-8} M_\odot$/yr for  
$M_{\rm wd} = 1 M_\odot$ and $i=60^\circ$. 
The heated outer disk region extends from $25R_{\rm wd}$ ($\sim$140,000~km) to $75 R_{\rm wd}$
($\sim 0.45a$, where $a$ is the binary separation). 
For an inclination of $75^\circ$ the mass accretion rate had to be increased to 
$4.4 \times 10^{-8} M_\odot$/yr.

\begin{figure}[h!] 
\includegraphics[scale=0.37,trim= 50 30 0 140]{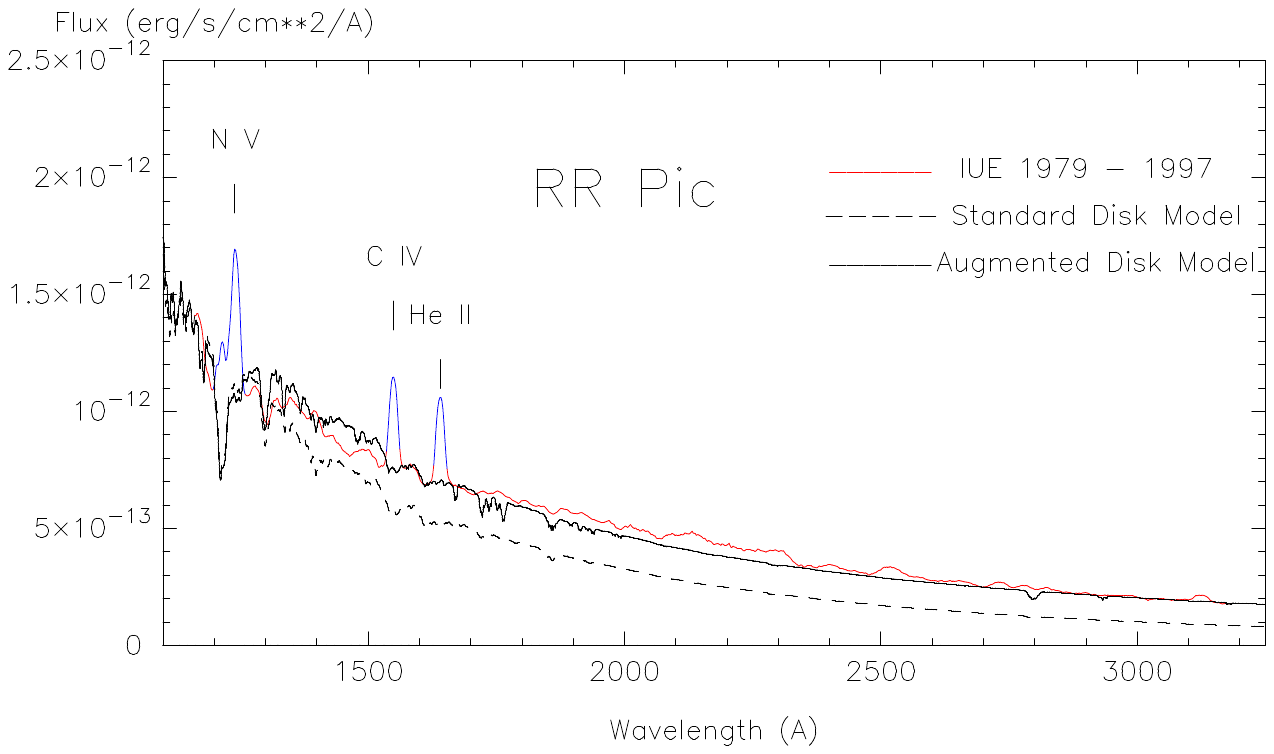}
\caption{
Modeling the IUE spectrum of RR Pic with an accretion disk. 
The observed spectrum is in red, regions that have been masked
(omitted) are in blue. A standard disk model (dashed black line) 
provides a continuum slope that is too steep. The addition of a 
12,000~K component in the outer disk provides a much better fit 
(solid black line).  See text for details. 
\label{rrpicfit}
}
\end{figure}

In Fig.\ref{rrpicfit} we present the fitting of the IUE spectrum of RR Pic  
assuming $M_{\rm wd}=1.03 M_\odot$ and $i=60^\circ$. The solid black line is the model with the heated 
outer disk, and the dotted line represents the model without a heated outer disk.

For the $1.03M_\odot$ WD mass the result can be summarized as 
$\dot{M}=2.8 \pm 1.6 \times 10^{-8} M_\odot$/yr.  
The large error is due mainly to the $\sim$17\% modulation in the continuum flux level as a function of the orbital phase
as shown in \S3.3 and to the uncertainty in the inclination  ($i\approx 67^\circ \pm 8^\circ$).   

A similar result is obtained assuming a $0.55M_\odot$ WD mass but with a higher mass accretion rate: 
$\dot{M}=7.0 \pm 4.0 \times 10^{-8} M_\odot$/yr for $i\approx 67^\circ \pm 8^\circ$.   

As to the unknown WD mass ($\sim 0.6M_\odot$ versus $\sim 1.0M_\odot$) 
it propagates as an uncertainty in the mass accretion of a factor of $\sim 2.5$. 
Altogether, the mass accretion rate could be as low as $\sim1.2 \times 10^{-8}M_\odot$/yr 
(for $M_{\rm wd}=1.12 M_\odot$, $i=60^\circ$) or as large as $1.1 \times 10^{-7}M_\odot$/yr
(for $M_{\rm wd}=0.45 M_\odot$, $i=75^\circ$).

\subsection{CP Lac} 

We ran models with $1.0 M_\odot$ and  $1.21 M_\odot$ WD masses, both with an inclination of $60^\circ$, 
and attempted to fit the UV spectrum and optical spectrum separately, as well as together. 
However, we found that no disk model could be fitted to the combined UV and optical spectrum.
This is not unexpected since the spectra were not obtained at the same time and the system
exhibits some small-amplitude outbursts. 

The spectral fit to the IUE UV spectrum, assuming a WD mass of $1.03 M_\odot$ and an inclination of $60^\circ$,
yielded a mass accretion rate of $8.2 \times 10^{-9}M_\odot$/yr for a Gaia distance
of 1163~pc. This standard disk model, i.e. similar to the \citet{wad98}, had a slope that is too
steep/blue. Therefore, we decided to model it in a manner similar to the way we modeled 
the disk of RR Pic, namely with a heated outer disk with a temperature of 12,000~K. 

\begin{figure}[h!]  
\includegraphics[scale=0.33,trim= 0 25 0 0]{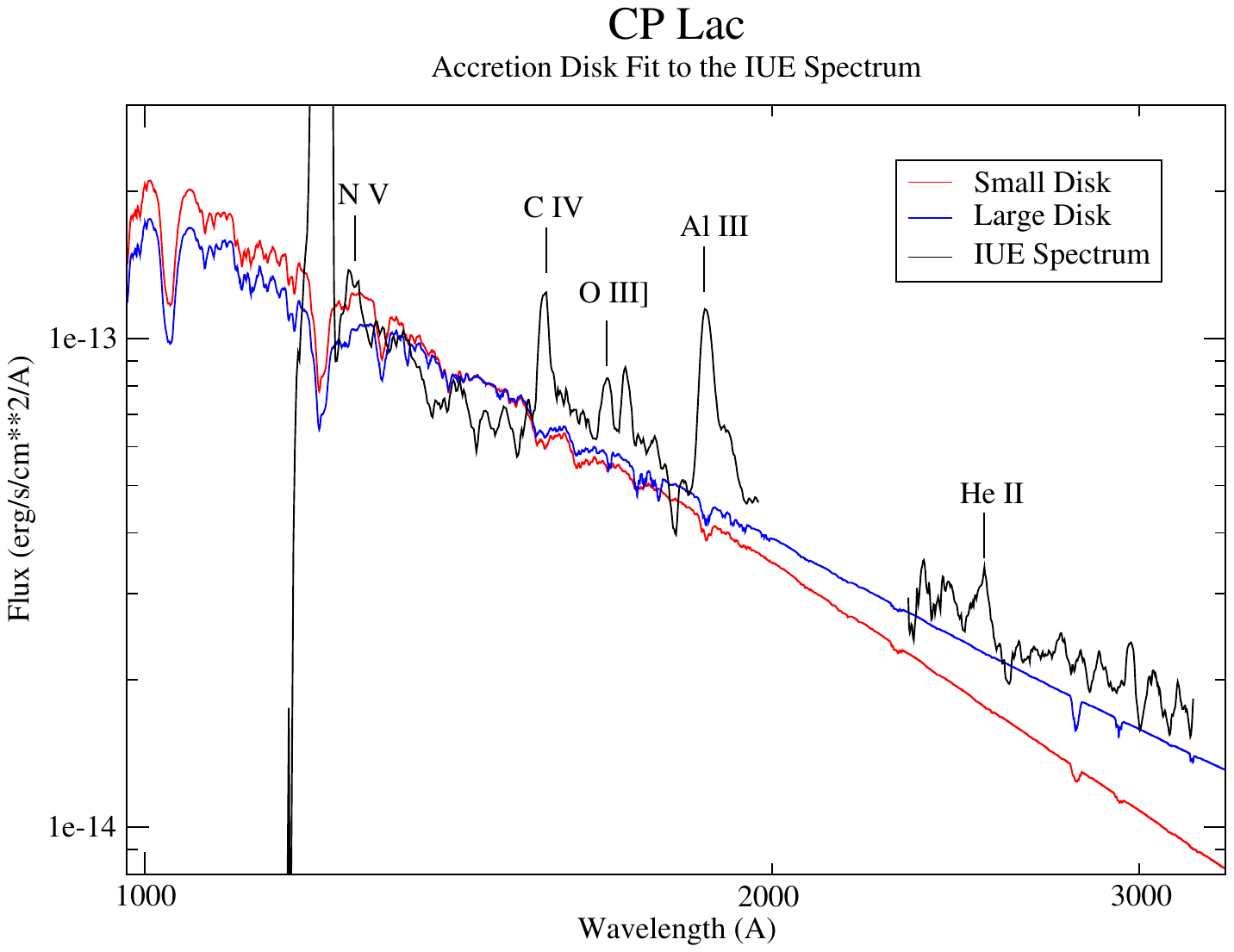} 
	\caption{Accretion disk fits to the IUE spectrum of CP Lac. 
The slope of the spectrum agrees with a standard disk model augmented
with an isothermal outer disk with a temperature of 12,000~K. The resulting larger disk model 
is shown in blue, a standard disk model fit alone is shown in red.
The mass accretion rate of the augmented disk model is $6.85\times 10^{-9}M_\odot$/yr.	
\label{cplaciuefit}	} 
\end{figure}

The standard disk model was augmented with an outer isothermal ($T=12,000$~K) region 
extending to $\sim 0.45a$. 
We refer to this model as the large disk model.
Assuming a WD mass of $1 M_\odot$ and an inclination of $60^\circ$, we found 
a mass accretion rate of $6.85 \times 10^{-9}M_\odot$/yr for a Gaia distance of 1163~pc.  
These two models, the smaller standard disk model of \citet{wad98} and this larger 
disk model augmented with an outer isothermal region, are presented in Fig.\ref{cplaciuefit}. 

\begin{figure}[h!]  
\includegraphics[scale=0.33,trim= 0 00 0 30]{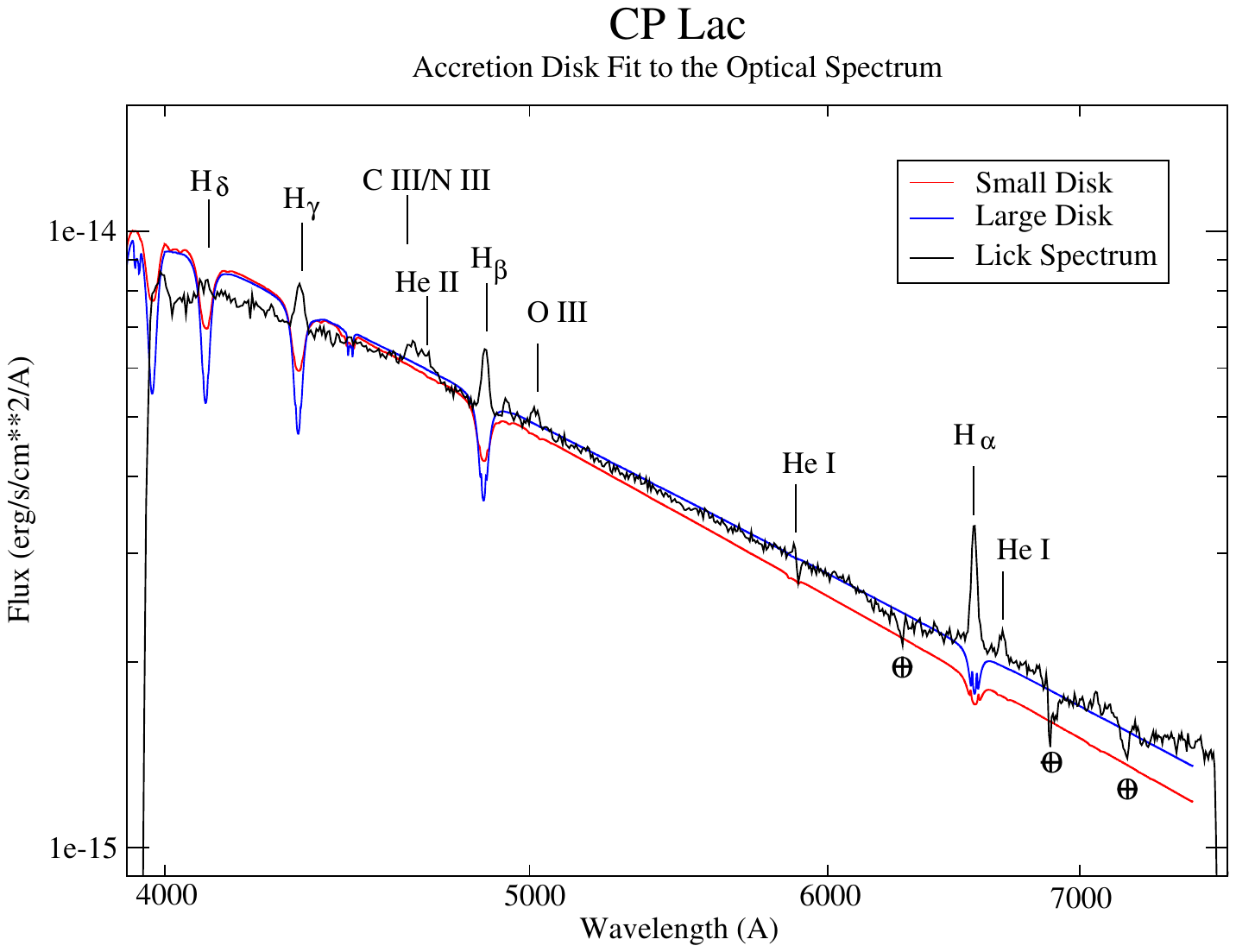} 
	\caption{Accretion disk fits to the optical spectrum of CP Lac. 
	Here too, the slope of the spectrum agrees with an augmented disk 
	model (in blue), but the mass accretion rate is $3.7\times10^{-9}M_\odot$/yr. 
        A standard disk model fit alone is shown in red.
\label{cplacoptfit} } 
	\vspace{0.5cm} 
\end{figure} 

We then carried out the spectral fit to the optical spectrum. Here too, we found that the
spectrum better agrees with a standard disk model augmented with an outer isothermal 
(12,000~K) region. The result is presented in Fig.\ref{cplacoptfit}.  
The large disk model has a mass accretion rate of $\dot{M}=3.7 \times 10^{-9}M_\odot$/yr, 
for a WD mass of $1.03 M_\odot$ and an inclination of $60^\circ$. 
For a WD mass of $1.21M_\odot$ we found a mass accretion rate about 30\% lower. 

Taking into account all the errors, we obtain a mass accretion rate 
$\dot{M} = 5.8 \pm 2.3 \times 10^{-9} M_\odot$/yr for the IUE UV spectrum obtained in 1991 
and $\dot{M} = 3.2 \pm 1.3 \times 10^{-9} M_\odot$/yr for the optical spectrum obtained in 1990, 
for a WD mass of $M_{\rm wd} = 1.1 \pm 0.1 M_\odot$, an inclination $i=60^\circ\pm5^\circ$
(though we only ran models with $i=60^\circ$, we did take an error of $\pm 5^\circ$ in the 
inclination into account), 
and a color excess of $E(B-V)=0.30\pm0.07$. The small error in the distance did not affect the results.  
\\

\subsection{DI Lac} 

\begin{figure}[h!] 
\includegraphics[scale=0.33,trim= 0 20 0 20]{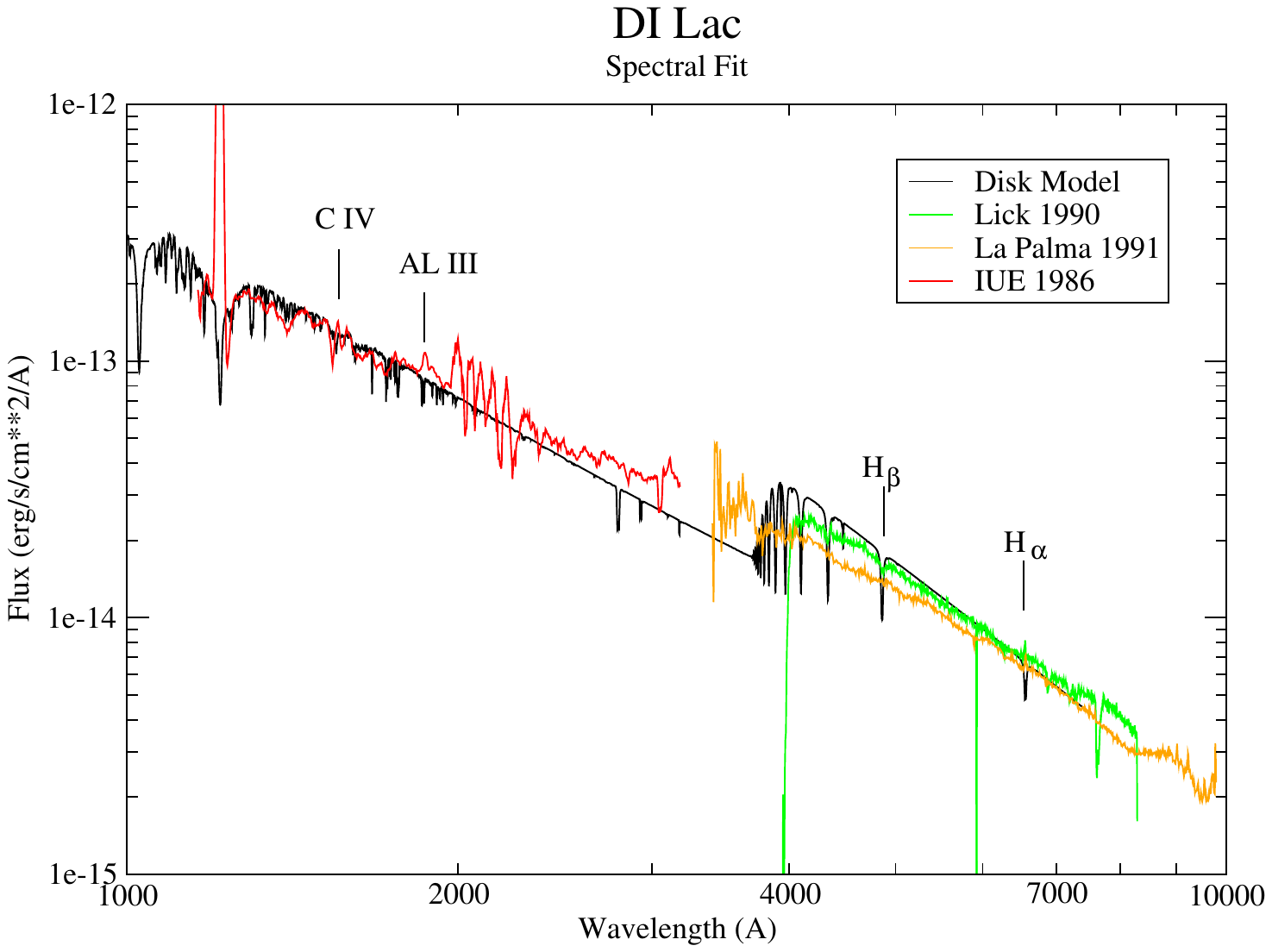} 
\caption{
An accretion disk model is fitted to the IUE + Optical spectra of DI Lac. 
The 1986 IUE spectrum is in red, the 1990 Lick optical spectrum is in green, the 1991 La Palma
optical spectrum is in orange, and the accretion disk model is in black. 
The derived mass accretion rate is $\dot{M}=9.2 \times 10^{-9}M_\odot$/yr.
See text for details. 
\label{dilacfit} }
\end{figure}

We followed the same procedure as for RR Pic and CP Lac, and we found that a
disk with a heated outer region extending to $R_d=0.27a$ gave a reasonable fit to both the UV
and optical spectra simultaneously. 
We found a mass accretion rate $\dot{M}=9.2 \pm 3.1 \times 10^{-9} M_\odot$/yr, 
for a WD mass of $1.0\pm0.1 M_\odot$, an inclination $i=18^\circ$, a color
excess of $E(B-V)=0.24\pm0.03$, and a Gaia distance of $1731\pm50$~pc. 
The spectral fit is presented in Fig.\ref{dilacfit}. 

\subsection{V533 Her} 

\begin{figure}[h!]  
\includegraphics[scale=0.37,trim= 50 100 0 130]{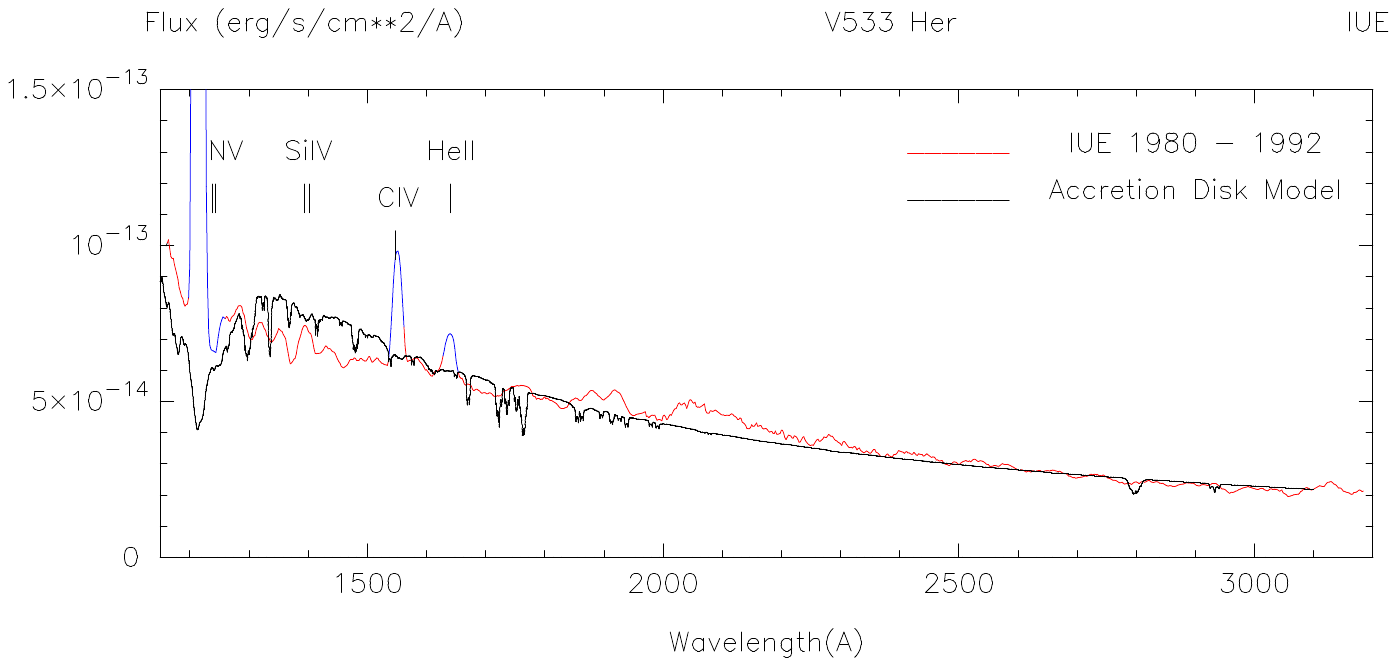}     
\caption{Accretion disk fit to the IUE spectrum of V533 Her. 
The observed spectrum is in red with masked emission lines in blue, and the
model spectrum is in black. 
Assuming a WD mass of $1.0M_\odot$, the mass accretion rate is $3.4\times10^{-9}M_\odot$/yr,
for an inclination of $60^\circ$. See text for details. 
\label{v533herfit} 
	} 
\end{figure}

For the analysis of V533 Her, we used the following system parameters: 
$i=60^\circ$, $E(B-V)=0.025$, a Gaia parallax-derived distance of 1248~pc, 
and the WD mass $M_{\rm wd}=1.03 M_\odot$ from our grid of disk models. 
Here too we found that a standard disk model cannot fit the relatively {\it flat} 
IUE spectrum and we augmented the disk with an outer isothermal ($T=12,000$~K) region. 
We found a mass accretion rate of $3.4\times10^{-9}M_\odot$/yr.
This provided a better fit, but the fit was still not as good as for the other systems,
as can be seen from Fig.\ref{v533herfit}.  
Note that we increased the outer radius of the 12,000~K region to fill in the Roche lobe,
$\approx 0.5a$. 

Taking into account the uncertainties in the parameters, we derived a final
value $\dot{M}=3.4 \pm 1.2 \times 10^{-9} M_\odot$/yr, for a WD mass of $1.03 \pm 0.10 M_\odot$, 
an inclination $i=60^\circ$, a color excess $E(B-V)=0.025\pm0.025$, and a distance $d=1248^{+130}_{-80}$~pc. 


As is the case with SW Sex stars \citep{hel94}, it is likely that the L1 stream of matter
partially overflowing the disk edge is seen projected against the brighter disk. Parts of the disk could be 
masked/veiled at given orbital phases generating a rather unusual UV spectrum that doesn't 
represent an accretion disk anymore. It is also possible that the emission from
the hot central region of the disk is being reprocessed by the overflowing stream. 
All these would explain the rather flat spectrum.

\subsection{V446 Her} 

We ran accretion disk models with $M_{\rm wd}=1.03$ and $1.21 M_\odot$, 
$i=18^\circ, 41^\circ,$ and $60^\circ$, and $E(B-V)=0.43\pm0.05$. 
We tried both standard disk models and models augmented with a 
heated outer disk region, and found that here too the heated outer 
disk models better fit the STIS spectrum. 
In the disk model the heated outer region extended to about 0.3a.  

The resulting mass accretion rate we obtained from the fitting of the augmented disk model
is $\dot{M} =1.26 \times 10^{-9.0} M_\odot$/yr  
for a WD mas of $1.1 M_\odot$ and an inclination $i=41^\circ$. 
The error was driven mainly by the uncertainty in the inclination, 
$\approx \pm 20^\circ$, giving an uncertainty corresponding to a factor
2 in $\dot{M}$, with a lower limit $\dot{M}_-=0.63 \times 10^{-9}M_\odot$/yr for $i=18^\circ$, 
and an upper limit $\dot{M}_+=2.52 \times 10^{-9}M_\odot$/yr for $i=60^\circ$ . 

\begin{figure}[h!] 
\includegraphics[scale=0.35,trim= 50 70 0 120]{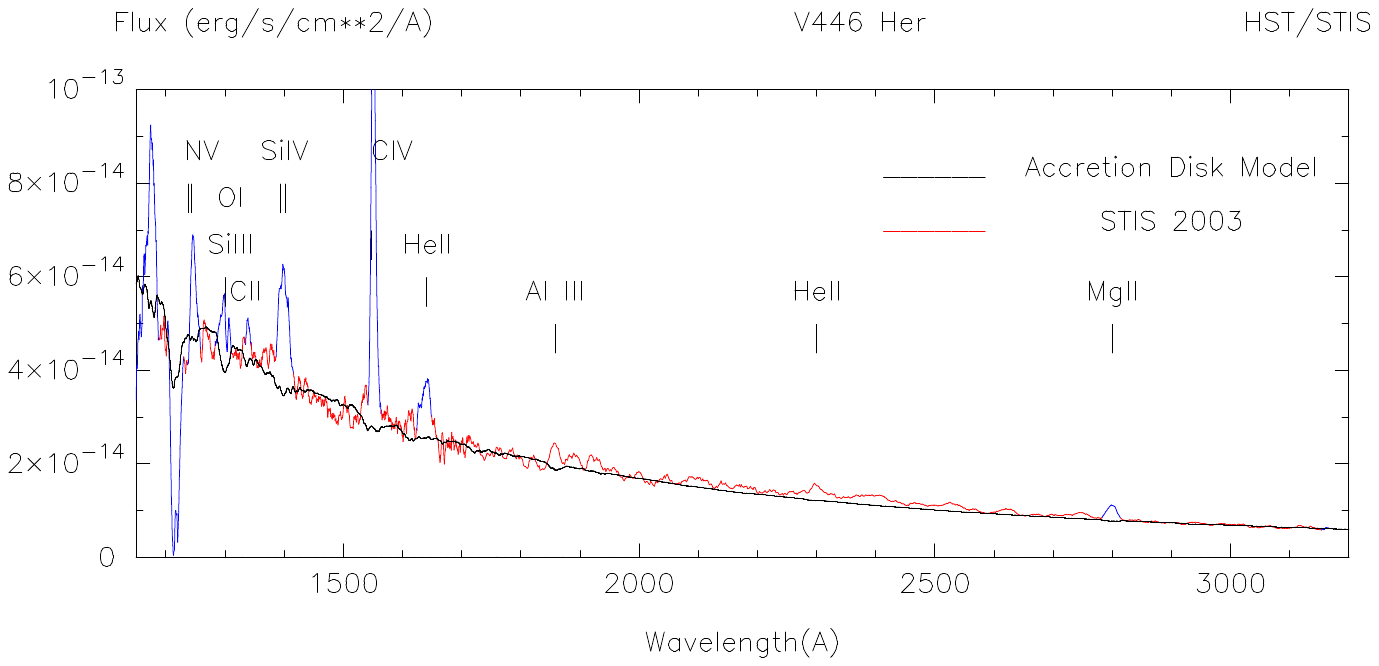} 
\caption{
An accretion disk model fit (in black) to the  
HST STIS (G140L/1425~\AA , G230L/2376~\AA ) spectrum of the nova V446 Her
(in red with masked emission lines in blue). 
The observed spectrum has been corrected for extinction assuming $E(B-V)=0.43$. 
\label{v446herfit} }
\end{figure}

The spectral fit to the STIS spectrum of V446 Her is displayed in Fig.\ref{v446herfit},
where we present the fit for our $1.03 M_\odot$ model with an inclination of $60^\circ$. 
For comparison, \cite{sel19} obtained $\dot{M}=1 \pm 0.5 \times 10^{-9}M_\odot$/yr,  
assuming an inclination $i=58\circ\pm7^\circ$.

Since the mass accretion rate we obtained was rather low, we also considered the possibility
that the WD contribution to the spectrum might not be negligible. 
We ran WD+disk models and found that 
the contribution of the WD, 
even if rather hot, was only few percent, it degraded the disk fit
and was therefore not taken into account.

\subsection{V1974 Cygni} 

\begin{figure}[h!] 
\includegraphics[scale=0.37,trim= 50 100 0 100]{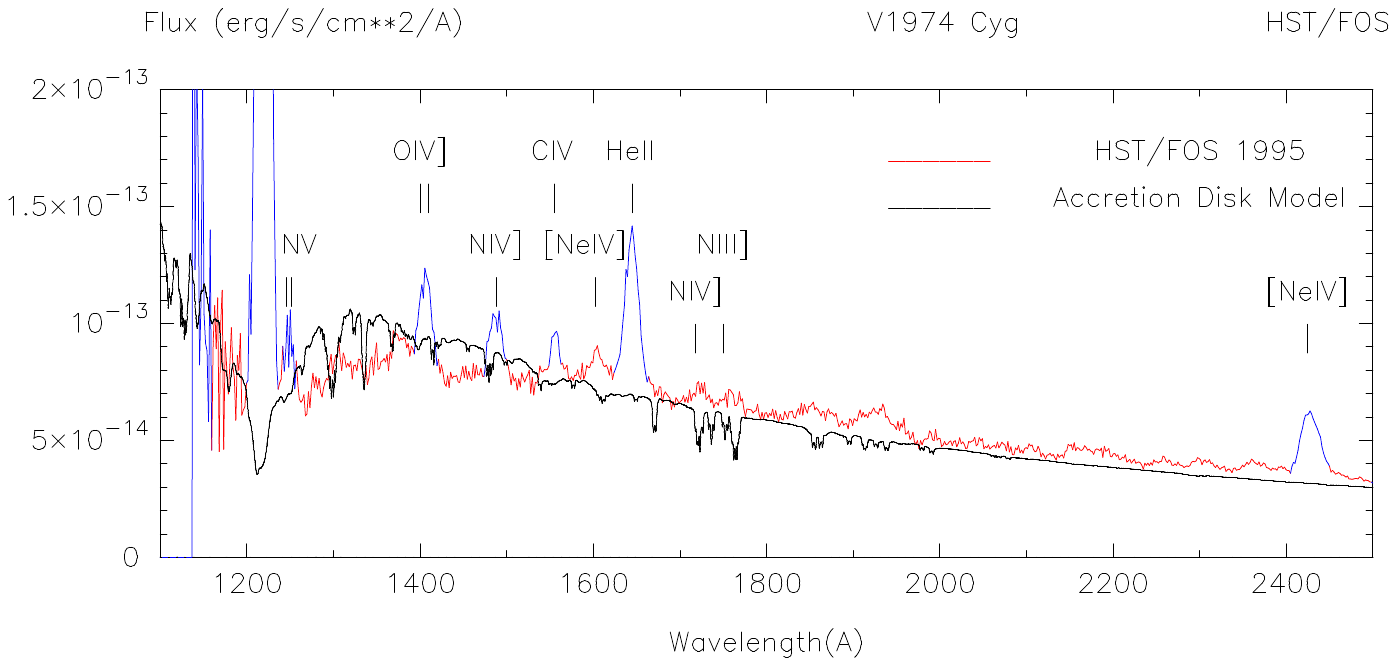} 
\caption{
Spectral fit to the FOS spectrum of V1974 Cyg. 
The FOS spectrum is in red and prominent emission lines, in blue, have been masked. 
The accretion disk model is in black. 
For an accreting WD mass of $1 M_\odot$, we find a mass accretion rate of 
$3.9 M_\odot$/yr. See text for details. 
\label{1974fit}	} 
\end{figure}

In the case of V1974 Cyg, the positive superhumps \citep{ret96,ole02,bru23a} are a signature that the disk reaches
the 3:1 resonance, and tidal shearing, spiral shocks and resonant eccentricity could contribute to the heating of the outer disk. 
Therefore, here too we ran both standard disk models and augmented disk models, we assumed  
WD masses of $1.03 M_\odot$ and $1.21 M_\odot$ to match the $1.1 M_\odot$ WD,
and we set $i=41^\circ$.  
We found for the standard disk models a mass accretion rate of $\dot{M}=7.2\pm1.4\times 10^{-9}M_\odot$/yr, 
while for the augmented disk models $\dot{M}$ decreased to  $\dot{M}=3.3\pm0.6\times 10^{-9}M_\odot$/yr, 
both for a WD mass of $M_{\rm wd}=1.1 \pm 0.1 M_\odot$ and a Gaia distance of $d=1617$~pc.  
The augmented disk model provided a better fit than the standard disk model. 
We present such a model in Fig.\ref{1974fit}, for our $M_{\rm wd}=1.03 M_\odot$ 
which gives in this particular case $\dot{M}=3.9 \times 10^{-9}M_\odot$/yr. 

As it was the case for V533 Her, the observed spectrum appears flatter between $\sim$1300~\AA\ and $\sim$2000~\AA\ 
than the augmented disk model. 
The radius of the 12,000~K outer disk was set to its 
maximum value, namely filling the Roche lobe of the WD: $\sim$0.55a. 
We checked if the addition of a WD improved the fit, 
but even if the WD was still very hot \citep[$\sim 100,000$~K;][]{mor01} in 1995 from its SSS phase \citep{sho96,bal98,cas00,hac05},   
it would contribute less than 1 percent of the UV flux at these wavelengths and distance.
However, a heated WD and/or inner disk are potential sources of ionization, and could irradiate the outer disk, the secondary and the nebular material.
We also note that if the WD was strongly magnetic \citep{cho97}, it could truncate the inner disk, further affecting the shape of the continuum.  
All these could contribute to the spectrum being rather flat. 

\subsection{V842 Cen}

\begin{figure}[h!] 
\includegraphics[scale=0.39,trim= 65 100 0 130]{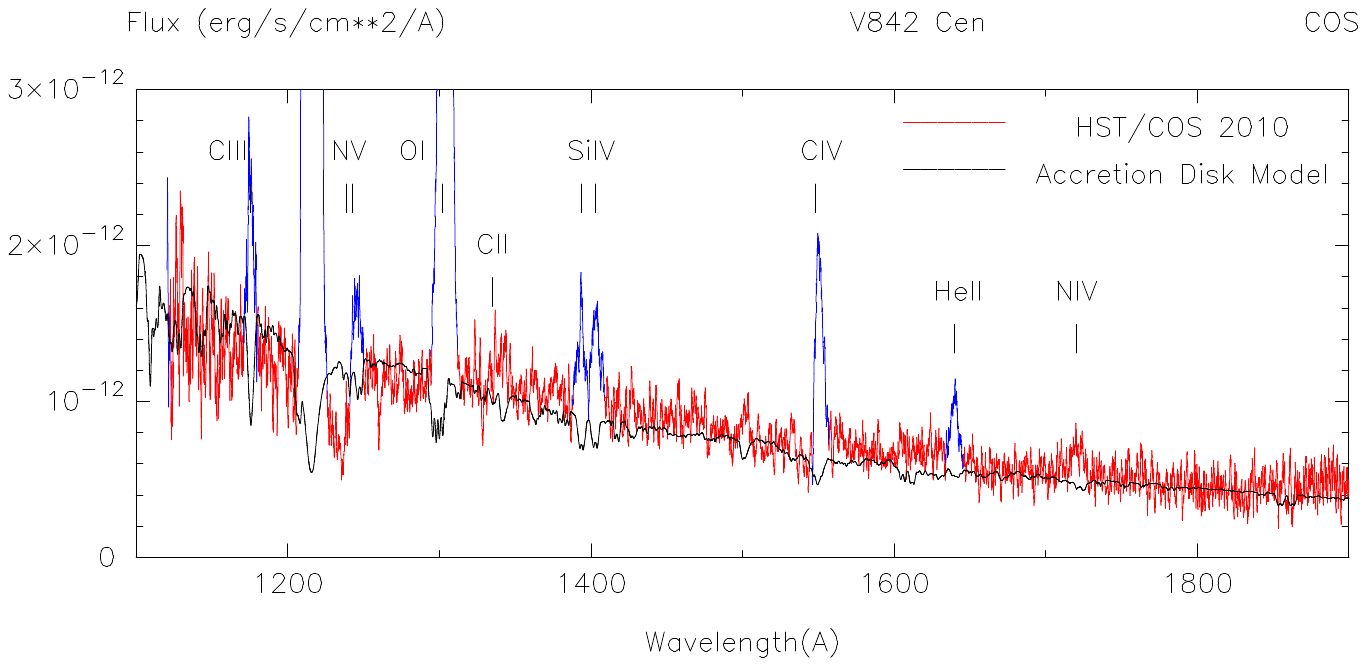}             
\caption{
Accretion disk fit (in black) to the COS spectrum (in red) of V842 Cen. 
For a $1.03 M_\odot$ WD mass, an inclination of $18^\circ$, a distance of $1376$~pc, and   
a color excess $E(B-V)=0.8\pm0.1$, the mass accretion rate 
$\dot{M} = 1_{-0.5}^{+1.0} \times 10^{-7} M_\odot$/yr. 
Emission lines (in blue) have been masked. 
\label{842-cos-fit3} 
} 
\end{figure}

For disk models with a $1.03 M_\odot$ WD mass, an inclination of $18^\circ$, a distance of $1376$~pc, and   
a color excess $E(B-V)=0.8\pm0.1$, we found a mass accretion rate of $\sim 10^{-7}M_\odot$/yr using a standard disk model
(see Fig.\ref{842-cos-fit3}).
The addition of a 12,000~K outer region did not noticeably affect the fit,
however, it slightly lowered the mass accretion rate. 
The main uncertainty came from the error in the reddening, dereddening assuming $E(B-V)=0.7$ 
gave a mass accretion rate smaller by a factor of two, and dereddening
assuming $E(B-V)=0.9$ gave a mass accretion larger by a factor of two, 
namely, $\dot{M} = 1_{-0.5}^{+1.0} \times 10^{-7} M_\odot$/yr, where we have rounded 
the values of the results since the error is so large. 

\begin{table*}[t!]  
\begin{center} 
{\bf   Table 4. Spectral Analysis Results} \\   [3pt]  
\begin{small} 
\begin{tabular}{lccccc}
\hline    
\noalign{\vskip 1.0ex}   
System & $M_{\rm wd}$ & $\dot{M}$               & $i$    & Outer  & $R_d$ \\ [1pt]  
Name   & $(M_\odot$)  & ($M_\odot/$yr)          & (deg)  & Disk   & $a$  \\ [1pt]       
\hline  
\noalign{\vskip 1.0ex}   
BK Lyn       & $1.1 \pm 0.1$          & $1.5 \pm 0.8 \times 10^{-10}$         &  $50 \pm 10$      &  N  & ---  \\     
HR Del       & $0.68^{+0.12}_{-0.13}$ &  $4.2\pm1.2 \times 10^{-7}$           & $41$              &  N  & ---  \\ 
RR Pic       & $1.0 \pm 0.1$          & $2.8 \pm 1.6 \times 10^{-8}$          & $67 \pm 8$        &  Y  & 0.45 \\    
RR Pic       & $0.55 \pm 0.1$         & $7.0 \pm 4.0 \times 10^{-8}$          & $67 \pm 8$        &  Y  & 0.45 \\    
CP Lac (UV)  & $1.1\pm0.1$            & $5.8\pm2.3 \times 10^{-9}$            & 60$\pm 5$         &  Y  & 0.45 \\ 
CP Lac (Opt) & $1.1\pm0.1$            & $3.2\pm1.3 \times 10^{-9}$            & 60$\pm 5$         &  Y  & 0.45 \\ 
DI Lac       & $1.0\pm0.1$            & $9.2\pm3.1 \times 10^{-9}$            & 18                &  Y  & 0.27 \\ 
V533 Her     & $1.0 \pm 0.1$          & $3.4 \pm 1.2 \times 10^{-9}$          & $60$              &  Y  & 0.50 \\ 
V446 Her     & $1.1\pm0.1$            & $1.26^{+1.26}_{-0.63} \times 10^{-9}$ & $41^{+19}_{-21}$  &  Y  & 0.30 \\ 
V1974 Cyg    & $1.1\pm0.1$            & $3.3\pm0.6\times 10^{-9}$             & 41                &  Y  & 0.55 \\  
V842 Cen     & $1.0\pm0.1$            & $1.0^{+1.0}_{-0.5} \times 10^{-7}$    & 18                &  N  & ---  \\ 
T Pyx        & $1.2\pm0.2$            & $2.0\pm1.0\times 10^{-7}$             & $50 \pm 5$        &  N  & ---  \\ 
\hline         
\noalign{\vskip 1.0ex}   
\end{tabular}   \\
Note: The WD masses and inclinations are the adopted values together with
their assumed uncertainties. For BK Lyn the inclination is a result
of the analysis. 
Where the inclination is known within a few degrees, no uncertainty in $i$ 
was assumed since the error is mainly due to uncertainties in the mass, color
excess, and distance. 
All the models listed agree with the Gaia parallax-derived distances.
CP Lac is listed twice: first from the analysis of its UV spectrum, 
then for the analysis of its optical spectrum. 
The column before last indicates whether an isothermal 12,000~K outer disk 
was added to the standard disk model, the last column lists the extend of the 
isothermal outer disk in units of the binary separation $a$. 
\end{small} 
\end{center}     
\end{table*} 

\citet{wou09} suggested V842 Cen is likely an IP,
but there is no strong evidence for it being an IP, except
for the signal detected in the optical. 
Therefore, we also tried disks with a truncated inner region, namely
with an inner radius twice as large as the WD radius. 
However, at such a high accretion rate, this made no difference in the fit 
and only yields to a very slightly higher mass accretion rate.

We do not compare our results with those of \citet{sio13}, since 
they assumed $E(B-V)=0.55$ and found the best fit to yield a very short distance (ten times closer
than the Gaia parallax-derived distance) and with a very small mass accretion rate.
\citet{sch05} also claimed V842 Cen is a low mass transfer rate system {\it because it is intrinsically very faint}, 
however, they used the observed spectrum without correcting for extinction.
It is clear that assuming a short distance and/or not correcting for color excess were the reasons for finding 
a lower mass accretion rate. 
It is worth noting that, for the optical spectrum, extinction [assuming E(B-V)=0.8] reduces the flux by a factor of two at 9000~\AA , 
and by a factor of 20 at 4000~\AA ; and, for the UV spectrum, extinction reduces the flux by up to 3 orders of magnitude 
(see Fig.\ref{der1}).  
\\ \\ 

\section{Discussion and Conclusions}

In Table 4 we list the results for all the systems, where for
clarity we have rounded up (or down) some of the values. 
We have added T Pyx to the table from the results of our previous analysis \citep{god24}.

\begin{figure}[h!] 
\includegraphics[scale=0.34,trim= 30 20 00 30]{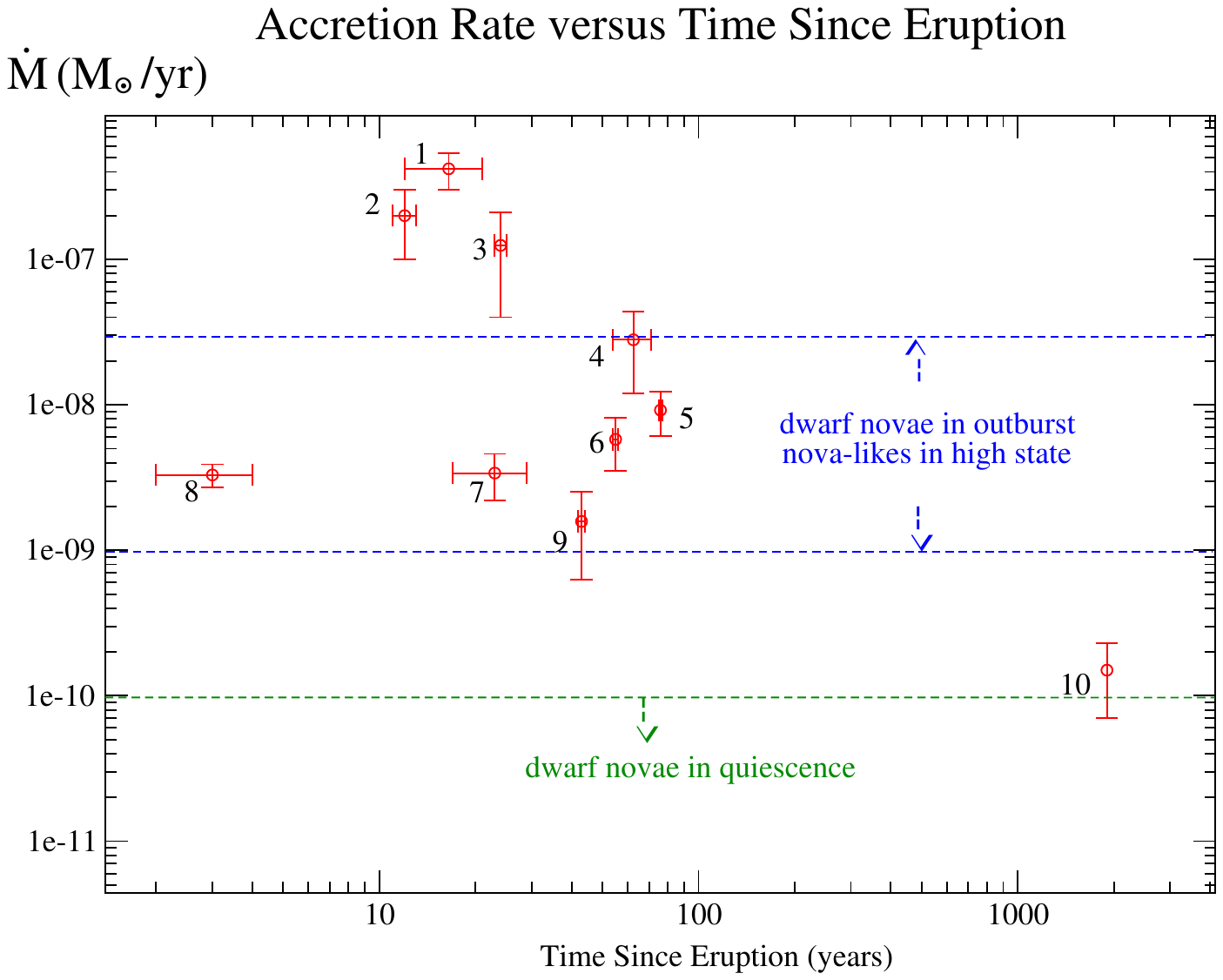} 
\caption{The mass accretion rate versus the time since eruption.
The data points are as follows: 01 - HR Del, 02 - T Pyx, 03 - V842 Cen, 
04 - RR Pic, 05 - DI Lac, 06 - CP Lac, 07 - V533 Her, 08 - V1974 Cyg, 
09 - V446 Her, 10 - BK Lyn.   
For some of the novae, the spectra obtained years apart were combined 
together, thereby generating error bars larger than the one-year default
for the time since eruption.  
For comparison, DNe in outburst and NLs in high state have 
accretion rates ranging from $10^{-9}M_\odot$/yr to a few $10^{-8}M_\odot$/yr
(between the two blue dashed lines), 
while DNe in quiescence have accretion rate $10^{-12}M_\odot$/yr to $10^{-10}M_\odot$/yr
(below the green dashed line).  
\label{mdotvstime}	} 
\end{figure}

Though results for 10 systems cannot be considered statistically significant, 
we still decided to draw a graph of the mass accretion rate versus the time
since eruption, which we present in Fig.\ref{mdotvstime}.

HR Del, T Pyx and V842 Cen, only 10-20 years post eruption,
have the highest mass accretion rate ($\sim 10^{-7}M_\odot$/yr),   
while BK Lyn, $\sim$2000 years after its eruption, has the lowest mass accretion rate
($\sim 10^{-10} M_\odot$/yr). Together with RR Pic, DI Lac and CP Lac, $\sim 70$ years
post-eruption, with an intermediate mass accretion ($\sim 10^{-8}M_\odot$/yr), these 
seven systems seem to hint at a declining $\dot{M}$ with time after nova eruption. 
This, however, certainly needs to be confirmed with many more data points, especially
that the STIS spectrum of BK Lyn was obtained when it was in a low state similar to 
a quiescent DN (see \S 3.1).

For HR Del, the large mass accretion we derived leads to stable burning on the WD surface
(self-sustained feedback loop SSS phase), 
and possibly at a rate slower than the matter is accreted. 
It must form a red giant-like structure or it must be carried away in a wind (see e.g. Fig.2 in \citet{cho21}).  
For T Pyx, the self-sustained feedback loop is shutting down as its accretion rate has been declining for 
the last $\sim$135 years \citep{sch10b,god24}. 
For V842 Cen, by evaluating the color excess from its IUE spectrum, we have now added it to the list 
of self-sustained high accreting novae.
It is clear that these three high accreting novae are distinct from novae like V533 Her 
and V1975 Cyg, which have a mass accretion of only a few $10^{-9}M_\odot$/yr less than 2 decades after
their nova eruption. 

These three novae are locked in the self-sustained irradiation-driven high mass accretion rate feedback loop
that prevents the secular decline predicted by the hibernation scenario.
Consequently, they should be excluded from the graph $\dot{M}(t)$, as should be BK Lyn and V446 Her 
as they were caught in a low state.   
This leaves only 5 novae (numbered 4,5,6,7,and 8) in Fig.\ref{mdotvstime}, all {\it comfortably}  
within the range of $\dot{M}$ characteristic of dwarf novae in outburst and nova-likes in high state. 
As such, our results (so far) are not different than the results of \citet{sel19}, namely they
do not reveal a correlation between $\dot{M}$ and time since eruption. 

\begin{figure} 
\includegraphics[scale=0.30,trim= 00 00 0 00]{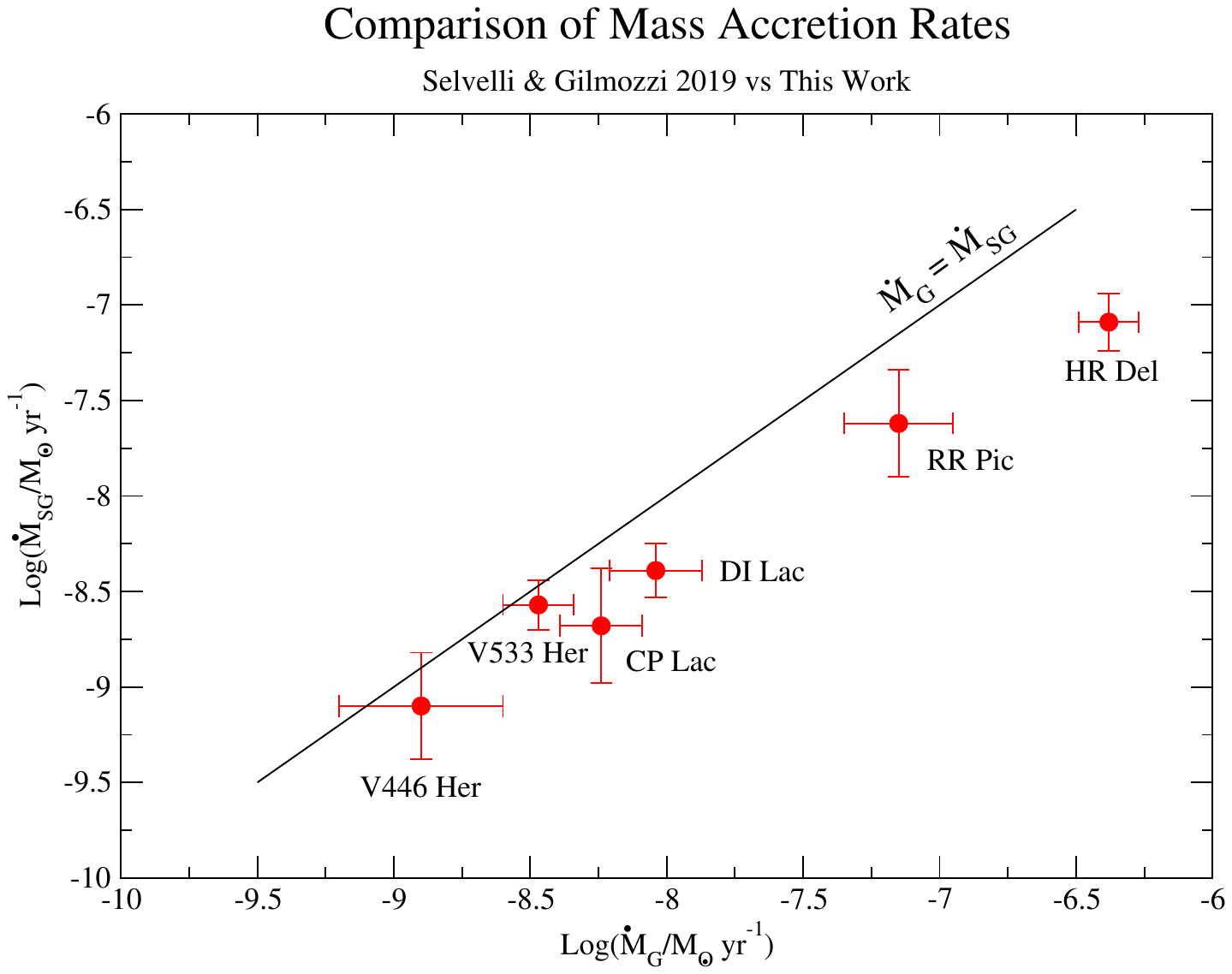}   
\caption{
The mass accretion rate obtained by integrating the UV and optical fluxes \citep{sel19}.  
versus the mass accretion rates we obtained in this work. 
Our results are systematically larger.
\label{gsg} } 
\end{figure}

A comparison with the previous analyses of \citet{moy03}, \citet{pue07}, and \citet{sio17} cannot be made
as they assumed different distances, $E(B-V)$ values, and WD masses. 
We are thus left to further compare our results with those of \citet{sel19} for the six systems
we have in common. 
For V533 Her and V446 Her, we obtained accretion rates that are only slightly larger (by $\sim20$\%). 
For DI Lac and CP Lac, we obtained accretion rates twice as large, however, the analysis
of the optical spectrum of CP Lac yielded the same accretion rate as \citet{sel19}. 
For RR Pic, we obtained an accretion rate three times larger. 
For HR Del, we adopted a WD mass very close to that of \citet{sel19} and obtained 
an accretion rate five times larger. The results are displayed in Fig.\ref{gsg}.    

As mentioned earlier in \S1, the mass accretion rates derived using the standard disk model
\citep[i.e. with {\textsc{tlusty};}][]{pue07} are on average 3 times larger than derived based on the observed
UV and optical luminosity \citep{gil24}. We notice that the largest discrepancy we obtained 
with \citet{sel19} is for HR Del, which has a spectrum that best agrees with 
a standard disk model. For the remaining five novae in common, the outer region of the disk was assumed
to have a temperature of $\sim 12,000$~K. That large outer region provided additional flux
and therefore, the resulting mass accretion rate was lower than for the standard disk model,
in better agreement with the results of \citet{sel19}.  

In the present work we adopted WD masses mostly from \citet{sha18}, 
which are larger than the WD masses derived by \citet{sel19} and \citet{hac19}.  
In particular, WD masses smaller than $\sim  1.0-1.1 M_\odot$ in \citet{sel19} and \citet{hac19} 
are on average $\sim 0.2M_\odot$ larger in \citet{sha18}, and the discrepancy
increases with decreasing WD mass.
The remaining nova WDs larger than $\sim 1.1 M_\odot$ in \citet{sel19} and \citet{hac19}
have on average the same masses in \citet{sha18}. 
This {\it bias} is explicitly documented and discussed in \citet{sch25}. 

The direct consequence of assuming a larger WD mass is a lower mass accretion rate.
Namely, had we assumed lower WD masses, we would have obtained even larger mass accretion rates 
(as seen explicitly in Table 4 for RR Pic). 

To further constrain the WD mass, it is important to consider the following. 
Positive superhumps are believed to be caused by the accretion disk becoming eccentric 
(due to tidal forces from the secondary) and precessing. 
Namely, superhumps are triggered when the outer disk reaches the 3:1 resonance radius, 
implying a small mass ratio $q=M_2/M_1 \le 0.3.$ 
This naturally occurs in systems with a smaller secondary mass below the period gap 
(where $M_2 \le 0.2 M_\odot$, such as for BK Lyn \& V1974 Cyg), 
but above the period gap it implies a large WD primary mass ($\sim 1 M_\odot$).   
RR Pic and V533 Her exhibit positive superhumps \citep{bru22,bru23a,bru23b}, they are therefore expected to have
a large WD mass. 

For four decades already, the validity of the standard $\alpha$-disk model has been questioned,
mainly because disk models, as those generated using \textsc{tlusty}, have systematically disagreed  
with the observations: the continuum of the observed spectra have a shallower slope, or alternatively,
the theoretical spectra are bluer \citep{wad84,wad88}, and the Balmer jump is either absent or much reduced. 
And while many explanations and solutions have
been advanced, this issue remains a matter of debate \citep[see][and references therein]{zsi24}. 
In the present work, we found that for systems accreting at a very high rate, $\sim 10^{-7}M_\odot$/yr
(as for V842 Cen, HR Del, and T Pyx), the observed spectra are bluer and agree with the standard disk model. 
As the mass accretion rate decreases the observed spectra seem to become flatter, though many other factors
can affect the spectra: e.g. highly inclined systems have flatter spectra; short spectral range spectra
(e.g. from 1200~\AA\ to 1700~\AA ) are easier to fit than large spectra range (e.g. covering UV + optical);
low S/N, and so on. Not unlike the nova-likes BZ Cam and V592 Cas \citep{god17}, we find that the addition of 
an outer disk with a temperature of the order of $\sim 12,000$~K improves the fit. 
If indeed, a $12,000$~K component with a large emitting area is contributing to the 
UV and optical, its relative contribution to the UV and optical flux decreases as $\dot{M}$ increases, 
in agreement with our disk modeling.  

Though our modeling is entirely empirical, there are observational evidence 
and physical processes at work to support the heating of the outer disk.  
The superhumps, mentioned above, through the 3:1 resonance, inject extra energy into the outer disk.  
Tidal interaction, even without the disk reaching the 3:1 resonance, heats up the outer region of the disk,
either through spiral shocks (at higher Mach number/low viscosity/low mass accretion rates) 
or through continuous background shearing due to the non-axisymmetric trajectories
of the gas elements (at low Mach number/high viscosity/high mass accretion rates), 
or through both \citep{ogi02,ju16}. 
The bright spot, that region at the edge of the disk being hit by the L1 stream, is also providing 
additional heating in the outer disk. Even material from the L1 stream overflowing the disk's edge 
will fall back onto the disk near orbital phase $\sim$0.5-0.6 \citep[e.g.][]{lub89,god19} and release 
some of its kinetic energy there. 
As the outer disk heats up, it extends vertically, thereby intercepting more ionizing radiation from the 
hot WD, boundary layer, and/or inner disk, which further increases its temperature. 
And while it is clear that heating of the outer disk takes place during the self-sustained enhanced
accretion phase (since the source irradiating the secondary also irradiates the disk), 
it is also very likely that it takes place at lower mass accretion rates. 
However, the emission from the hot inner (standard) disk itself completely dominates the UV and optical 
when the mass accretion rate is large, and only at lower accretion rates does the contribution of 
the outer disk becomes relatively significant. 

We therefore suggest that heating of the outer disk might be responsible, {\it at least in part}, for the shallower
slope of the observed spectra of accreting WDs.  
Geometrical effects, such as self-occultation of the disk, and large stream disk overflow can also 
affect the shape of the continuum making it flatter, as observed e.g. in the STIS spectra of EM Cyg 
\citep[see Figs.1 and 2, lower left panel, in][]{god19}.  

At this stage of our research, we did not check whether the mass accretion rate correlates with the time ($t_3$) it takes
for the nova to fade by 3 magnitudes after its eruption. Our current statistically small number  
of novae is further diminished by the fact that two novae, V446 Her and BK Lyn, were caught in a state of low mass 
accretion. Furthermore, for BK Lyn, the $t_3$ time and the characteristics of its eruption 
are entirely unknown. We have, however, assessed its WD mass as $1.0-1.2 M_\odot$ and have 
put a lower limit on its mass accretion rate $\approx 10^{-10}M_\odot$/yr (quiescent accretion rate). 
For these results, the buildup time to reach an envelope mass capable of triggering
the next nova eruption ranges roughly from $\approx$300,000~yr to $\approx$1 million years. 
Since the accretion rate is the quiescent one, these times are upper limit, assuming of course that
BK Lyn doesn't go into hibernation. Therefore, even 2000~years after its eruption, this might
not reflect its long-term accretion rate between eruptions. 

It is interesting to note that 
the mass accretion rate we found for V446 Her, $6.3 \times 10^{-10} < \dot{M}/(M_\odot$/yr)$ < 2.52 \times 10^{-9}$,  
does not reflect a DN quiescent state with $\dot{M} < 10^{-10}M_\odot$/yr, nor an optically thin disk.  
Instead, it is possible that V446 Her might not be reaching a true DN quiescent state, especially
that the `outburst' amplitude is really small, of the order of 1.5 magnitudes. 

And, while we emphasized that our results, just like those \citet{sel19}, might not reveal a correlation 
between $\dot{M}$ and time since eruption, the recent work of \citet{sal25} 
does show a decrease in the (unabsorbed) X-ray luminosity with time since eruption. 
Such a decrease seems at first to support the hibernation scenario. However, one has to keep in mind that the X-ray luminosity of 
non-magnetic white dwarfs accreting at a high rate is only about 1\% \citep[and often less; e.g.][]{bal26} of the expected 
X-ray luminosity from the boundary layer. With such a low efficiency factor ($\le$ 0.01) 
and a boundary layer that is still misunderstood \citep[e.g.][]{sca26}, one should be cautious 
when treating the X-ray luminosity as a simple direct indicator of the mass accretion rate. 
This clearly highlights the need for additional $\dot{M}$ data points for old novae
(such as e.g. V1315 Aql, Z Cam, BZ Cam) from UV and optical spectroscopy. 
It is the aim of our ongoing research to analyze UV and (mostly) optical spectra of old novae 
\citep[such as e.g. from][]{tap12,tap13b,tap13a,tap14,tap15,tap16},
the results of which will be presented in forthcoming  publications in this series.

\section*{acknowledgements}  

Three of the optical spectra were obtained with the 3-m Shane telescope at Lick observatory 
by Koji Mukai, and we are very grateful to him for generously sharing these archival 
datasets for our analysis. 
We are pleased to thank Linda Schmidtobreick for kindly agreeing that we display the spectrum 
of V842 Cen she collected at La Silla in 2003 \citep[from][]{sch05}. 

Support for this research is provided by the National Aeronautics and Space Administration 
Astrophysics Data Analysis Program (NASA/ADAP) grant number 80NSSC25K7579 
to Villanova University. 

We wish to thank the members of the AAVSO for their constant monitoring of cataclysmic
variables and novae, and making their data publicly available.  
Lick Observatory is operated by the University of California. The Isaac Newton telescope
is operated on the island of La Palma by the Royal Greenwich Observatory in the Spanish Observatorio
del Roque de los Michachos of the Instituto de Astrofisica de Canarias. 
To extract galactic coordinates, we used the {\sc simbad} data base, which is  
operated at the CDS, Strasbourg, France. 
This research has made use of the STilism 3D reddening map hosted
by the observatoire de Paris, as well as Galextin (galextin.org) online tools to
calculate interstellar reddening. 
Our research also used data from the European Space Agency (ESA) mission
{\it Gaia} (\url{https://esa.int}), processed by the {\it Gaia} Data Processing and 
Analyses Consortium (DPAC, \url{https//esa.int}). 
Part of this research is based on observations made with the NASA/ESA Hubble Space
Telescope, obtained from the Data Archive at the Space Telescope Science Institute,
which is operated by the Association of Universities for Research in Astronomy, 
Inc., under NASA contract NAS 5-26555. 
This work additionally utilizes archival data from the International Ultraviolet Explorer 
(IUE) satellite, a joint project of NASA, ESA, and the SERC, obtained
from the Mikulski Archive for Space Telescopes (MAST) at STScI. 






\begin{thebibliography}{}

\bibitem[Am\^ores et al.(2021)]{amo21} 
Am\^ores, E.B., Jesus, R.M., Mointinho, A. et al 2021, MNRAS, 508, 1788 

\bibitem[Am\^ores and L\'epine(2005)]{amo05} 
Am\^ores, E.B., \& L\'epine, J.R.D., 2005, \aj, 130, 659  

\bibitem[Austin et al.(1996)]{aus96} 
Austin, S.J., Wagner, R.M., Starrfield, S. et al. 1996, \aj, 111, 869 

\bibitem[Balman et al.(2026)]{bal26} 
Balman, \c{S}, Schlegel, E.M., Godon, P., Drake, J., and Sion, E.M. 2026, \apj, 1003, 59 

\bibitem[Balman et al.(1998)]{bal98} 
Balman, \c{S}, Krautter, J., \"Ogelman, H. 1998, \apj, 499, 395 


\bibitem[Bruch(1982)]{bru82}
Bruch, A. 1982, \pasp, 94 916 

\bibitem[Bruch(2022)]{bru22}
Bruch, A. 2022, \mnras, 514, 4718 

\bibitem[Bruch(2023a)]{bru23a}
Bruch, A. 2023a, \mnras, 519, 352  

\bibitem[Bruch(2023b)]{bru23b}
Bruch, A. 2023b, \mnras, 525, 1953 

\bibitem[Capitanio et al.(2017)]{cap17}
Capitanio, L., Lallement, R., Vergely, J.-L. et al. 2017, \aap, 606, 65 

\bibitem[Casalegno et al.(2000)]{cas00} 
Casalegno, R., Orio, M., Mathis, J. et al. 2000, \aap, 361, 725  

\bibitem[Cassatella et al.(2002)]{cas02}
Cassatella, A., Altamore, A., \& Gonz\'alez-Riestra, R. 
2002, A\&A, 384, 1023

\bibitem[Celed\'on et al.(2024)]{cel24}
Celed\'on, L., Schmidtobreick, L., Tappert, C., and Selman, F. 2024, \aap, 681, 106 

\bibitem[Chochol et al.(1997)]{cho97} 
Chochol, D., Grygar, J., Pribulla, T. et al. 1997, \aap, 318, 908 

\bibitem[Chomiuk et al.(2021)]{cho21} 
Chomiuk, L., Metzger, B.D., Shen, J.J. 2021, Annu.Rev.Astron.Astrophys., 59, 391 

\bibitem[Collins(1992)]{col92} 
Collins, P. 1992, IAU Circ.No.5454 

\bibitem[Darnley et al.(2014)]{dar14}    
Darnley, M.J., Williams, S.C., Bode, M.F., et al. 2014, A\&A, 563, L9

\bibitem[Darnley et al.(2017)]{dar17}
Darnley, M.H., Hounsell, R., Godon, P., et al. 2017, \apj, 849, 96 

\bibitem[Diaz et al.(1996)]{dia96}
Diaz, M.P., Wade, R.A., and Hubeny, I. 1996, \apj, 459, 236 

\bibitem[Dobrzycka \& Howell(1992)]{dob92}
Dobrzycka, D., \& Howell, S.B. 1992, \apj, 388, 614  

\bibitem[Feibelman et al.(1988)]{fei88} 
Feibelman, W.A., Oliversen, N.A., Nichols-Bohlin, J., and Garhart, M.P. 1988, 
International Ultraviolet Explorer Spectral Atlas of Planetary Nebulae, Central Stars, and
Related Objects, NASA Reference Publication 1203, Washington DC: NASA  

\bibitem[Fitzpatrick \& Massa(2007)]{fit07} 
Fitzpatrick, E.L, \& Massa, D. 2007, \apj, 663, 320 

\bibitem[Friedjung et al.(2010)]{fri10}
Friedjung, M., Dennefeld, M., Voloshina, I. 2010, \aap, 521, 84 

\bibitem[Fuentes-Morales et al.(2018)]{fue18} 
Fuentes-Morales, I., Vogt, N., Tappert, C. et a. 2018, \mnras, 474, 249 

\bibitem[Fujimoto(1982)]{fuj82} 
Fujimoto, M.Y. 1982, \apj, 257, 752 

\bibitem[Gaia Collaboration et al.(2021)]{gai21}
Gaia Collaboration, Brown, A.G.A., Vallenari, A., et al. 2021, \aap, 649, A1  

\bibitem[Gilmozzi \& Selvelli(2024)]{gil24} 
Gilmozzi, R., and Selvelli, P. 2024, \aap, 681, 83 

\bibitem[Ginzburg \& Quataert(2021)]{gin21} 
Ginzburg, S., \& Quataert, E. 2021, \mnras, 507, 475 

\bibitem[Godon(2019)]{god19}
Godon, P. 2019, \apj, 870, 112 

\bibitem[Godon et al.(2017)]{god17} 
Godon, P., Sion, E.M., Balman, \c{S}., and Blair, W.P. 2017, \apj, 846, 52 

\bibitem[Godon et al.(2018)]{god18}
Godon, P., Sion, E.M., Williams, R.E., \& Starrfield, S. 2018, \apj, 862, 89 

\bibitem[Godon et al.(2024)]{god24}
Godon, P., Sion, E.M., Williams, R.E. et al. 2024, \apj, 974, 202 

\bibitem[Greenberg \& Chlewicki(1983)]{gre83}
Greenberg, J.M., Chlewicki, G. 1983, \apj, 272, 563 

\bibitem[Guzman et al.(2019)]{guz19}
Guzman, G., Sion, E.M., \& Godon, P. 2019, \aj, 158, 99 

\bibitem[Hachisu \& Kato(2005)]{hac05}
Hachisu, I., \& Katro, M. 2005, \apj, 631, 1094 

\bibitem[Hachisu \& Kato(2016)]{hac16}
Hachisu, I., \& Katro, M. 2016, \apj, 816, 26    

\bibitem[Hachisu \& Kato(2019)]{hac19}
Hachisu, I., \& Katro, M. 2019, \apjs, 242, 18    

\bibitem[Haefner \& Betzenbichler(1991)]{hae91} 
Haefner, R., \& Betzenbichler, W. 1991, Inf.Bull.Variable Stars, 3665  

\bibitem[Haefner \& Metz(1982)]{hae82}
Haefner, R., \& Metz, K. 1982, \aap, 109, 171 

\bibitem[Hellier \& Robinson(1994)]{hel94}
Hellier, C., \& Robinson, E.L. 1994, \apj, 431, L107 

\bibitem[Hillman et al.(2016)]{hil16}    
Hillman, Y., Prialnik, D., Kovetz, A., \& Shara, M.M. 2016, \apj, 819, 168

\bibitem[Hillman et al.(2020)]{hil20}
Hillman, Y., Shara, M.M., Prialnik, D., Kovetz, A. 2020, Nature Astronomy, vol.4, p.886 


\bibitem[Hoard et al.(2000)]{hoa00} 
Hoard, D.W., Szkody, P., Honeycutt, R.K. et al. 2000, \pasp 

\bibitem[Honeycutt et al.(2011)]{hon11}
Honeycutt, R.K., Robertson, J.W., Kafka, S. 2011, \aj, 141, 121 

\bibitem[Honeycutt et al.(1995a)]{hon95b} 
Honeycutt, R.K., Robertson, J.W., Turner, G.W. 1995a, \apj, 446, 838  

\bibitem[Honeycutt et al.(1995b)]{hon95} 
Honeycutt, R.K., Robertson, J.W., Turner, G.W. 1995b, in Cataclysmic Variables, 
ed.A.Bianchini, M., Della Valle, \& M.Orio (Dordrecth:Kluwer), 
Astrophysics \& Space Science Library, Vol.205, p.75  

\bibitem[Honeycutt et al.(1998)]{hon98} 
Honeycutt, R.K., Robertson, J.W., \& Turner, G.W. 1998, AJ, 115, 2527 

\bibitem[Hubeny(1988)]{hub88} 
Hubeny, I. 1988, Comput.Phys.Comm., 52, 103

\bibitem[Hubeny(1991)]{hub91} 
Hubeny, I. 1991, in Structure and Emission Properties of 
Accretion Disks, 
Proceedings of IAU Colloq. 129, p.227

\bibitem[Hubeny \& Lanz(1995)]{hub95} 
Hubeny, I., \& Lanz, T. 1995, \apj, 439, 875

\bibitem[Hubeny \& Lanz(2017a)]{hub17a} 
Hubeny, I., \& Lanz, T. 2017a, 
A Brief Introductory Guide to {\textsc tlusty} and {\textsc synspec}, 
arXiv:1706.01859 (not published elsewhere) 

\bibitem[Hubeny \& Lanz(2017b)]{hub17b} 
Hubeny, I., \& Lanz, T. 2017b, 
{\textsc tlusty} User's GUide II: Reference Manual,                
arXiv:1706.01935 (not published elsewhere) 

\bibitem[Hubeny \& Lanz(2017c)]{hub17c} 
Hubeny, I., \& Lanz, T. 2017c, 
{\textsc tlusty} User's GUide III: Operational Manual,                
arXiv:1706.01937 (not published elsewhere) 

\bibitem[Hubeny et al.(1994)]{hub94} 
Hubeny, I., Lanz, T., \& Jeffrey, S. 1994, 
St. Andrews Univ. Newsletter on Analysis of Astronomical Spectra, 20, 30


\bibitem[Ju et al.(2016)]{ju16}
Ju, W., Stone, J.M., Zhu, Z. 2016, \apj, 823, 81 

\bibitem[Kato et al.(2014)]{kat14}  
Kato, M., Saio, H., Hachisu, I., \& Nomoto, K. 2014, \apj, 793, 136

\bibitem[King(1988)]{kin88}
King, A.R. 1988, QJRAS, 29, 1  

\bibitem[Knigge et al.(2000)]{kni00}
Knigge, C., King, A.R., Patterson, J. 2000, \aap, 364, L75 

\bibitem[Kraft(1964)]{kra64} 
Kraft, R. 1964, \apj, 139, 457 

\bibitem[Krautter \& Snijders(1990)]{kra90}
Krautter, J., \& Snijders, M.A.J. 1990, in Cataclysmic Variables and Low-Max X-ray Binaries,
Proceedings of the 11th North American Workshop, 9-13 October 1989, Santa-Fe, Ed. C.W. Mauche,
Cambridge, UK (Cambridge University Press), 387 

\bibitem[Kuerster \& Barwig(1988)]{kue88}
Kuerster, M., Barwig, H. 1988, \aap, 199, 201  

\bibitem[Kutter \& Sparks(1980)]{kut80} 
Kutter, G.S., \& Sparks, W.M. 1980, \apj, 239, 988 

\bibitem[Lallement et al.(2018)]{lal18} 
Lallement, R., Capitano, L., Ruiz-Dern, L., et al. 2018, \aap, 616, A132 

\bibitem[Lallement et al.(2014)]{lal14} 
Lallement, R., Vergely, J.-L., Vallette, B., et al. 2014, \aap, 561, A91  

\bibitem[Leichty et al.(2022)]{lei22} 
Leichty, M., Garnavich, P., Littlefield, C. et al. 2022, Res.Notes AAS, 6, 91 

\bibitem[Li \& Draine(2001)]{li01}
Li, A., \& Draine, B.T. 2001, \apj, 554, 778 

\bibitem[Livio \& Pringle(2011)]{liv11} 
Livio, M., \& Pringle, J.E. 2011, \apjl, 740, L18  

\bibitem[Lubow(1989)]{lub89} 
Lubow, S.H. 1989, \apj, 340, 1064 

\bibitem[Luna et al.(2012)]{lun12} 
Luna, G., Diaz, M., Brickhouse, N., Morales, M. 2012, MNRAS, 423, 75 

\bibitem[Luri et al.(2018)]{lur18}
Luri, X., Brown, A.G.A., \& Sarro, L.M. 2018, \aap, Special Gaia issue, 616, A9   

\bibitem[Moraes \& Diaz(2009)]{mor09}
Moraes, M., Diaz, M. 2009, \aj, 138, 1541 

\bibitem[Moro-Mart\'in et al.(2001)]{mor01} 
Moro-Mart\'in, A., Garnavich, P.M., \& Noriega-Crespo, A. 2001, \aj, 121, 1636 

\bibitem[Moyer et al.(2003)]{moy03} 
Moyer, E., Sion, E.M., Szkody, P., G\"ansicke, B.T., \& Howell, S. 2003, \aj, 125, 288 

\bibitem[Mukai \& Orio(2005)]{muk05} 
Mukai, K. \& Orio, M., 2005, \apj, 622, 602


\bibitem[Naylor et al.(1992)]{nay92}
Naylor, T., Charles, P.A., Mukai, K., and Evans, A. 1992, \mnras, 258, 449 

\bibitem[Ogilvie(2002)]{ogi02}
Ogilvie, G.I. 2002, \mnras, 330, 937 

\bibitem[Olech(2002)]{ole02}
Olech, A. 2002, Acta Astronomica, 52, 273 

\bibitem[Paczy\'nski(1967)]{pac67} 
Paczy\'nski, B. 1967, Acta Astron., 17, 287 

\bibitem[Paczy\'nski \& Sienkiewicz(1981)]{pac81} 
Paczy\'nski, B., \& Sienkiewicz, R. 1981, \apjl, 248, L27       

\bibitem[Pagnotta(2015)]{pag15} 
Pagnotta, A. 2015, Acta Polytechnica, 2, 199

\bibitem[Pagnotta et al.(2009)]{pag09}     
Pagnotta, A., Schaefer, B. E., Xiao, L., Collazzi, A. C., \& Kroll, P.
2009, \aj, 138, 1230

\bibitem[Pagnotta \& Schaefer(2014)]{pag14}                  
Pagnotta, A., \& Schaefer, B.E. 2014, \apj, 788, 164

\bibitem[Patterson(1979)]{pat79}
Patterson, J. 1979, \apj, 233, L13 

\bibitem[Patterson(1984)]{pat84}
Patterson, J. 1984, \apjs, 54, 443 

\bibitem[Patterson(1994)]{pat94} 
Patterson, J. 1994, \pasp, 106, 209 

\bibitem[Patterson et al.(2013)]{pat13} 
Patterson, J., Uthas, H., Kemp, J. et al. 2013, \mnras, 434, 1902 

\bibitem[Patterson et al.(2017)]{pat17}
Patterson, J., Oksanen, A., Kemp, J. et al. 2017, \mnras, 466, 581  

\bibitem[Patterson et al.(2022)]{pat22} 
Patterson, J., Kemp, J., Monard, B., et al. 2022, \apj, 924, 27 

\bibitem[Peters \& Thorstensen(2006)]{pet06} 
Peters, C.S., \& Thorstensen, J.R. 2006, \pasp, 118, 687   

\bibitem[Prialnik et al.(1982)]{pri82} 
Prialnik, D., Livio, M., Shaviv, G., \& Kovetz, A. 1982, \apj, 257, 312 

\bibitem[Pringle(1981)]{pri81}
Pringle, J.E. 1981, ARA\&A, 19, 137 
 
\bibitem[Puebla et al.(2007)]{pue07} 
Puebla, R.E., Diaz, M.P., Hubeny, I. 2007, \apj, 134, 1923 

\bibitem[Rao et al.(2025)]{rao25} 
Rao, S.M., Paney, J.C., Rawat, N., Joshi, A., Singh, A.K. 2025, A\&A, in print (2025arXiv250604371R) 

\bibitem[Rappaport et al.(1982)]{rap82}
Rappaport, S., Joss, P.C., \& Webbink, R.F. 1982, \apj, 254, 616 

\bibitem[Retter et al.(1996)]{ret96} 
Retter, A., Leibowitz, E.M., Ofek, E.O. 1996, in Evans H., Wood J.H., eds, 
Proc. IAU Colloq.158, Cataclysmic Variables and Related Objects, Kluwer, Dordrecth, p.321 

\bibitem[Ringwald et al.(1996a)]{rin96}
Ringwald, F.A., Naylor, T., \& Mukai, T. 1996, \mnras, 281, 192 

\bibitem[Ringwald et al.(1996b)]{rin96a} 
Ringwald, F.A., Thorstensen, J.R., Honeycutt, R.K., \& Robertson, J.W. 1996, \mnras, 278, 125 

\bibitem[Ritter \& Kolb(1998)]{rit98}
Ritter, H., \& Kolb, U. 1998, \aaps, 129, 83 

\bibitem[Rodr\'iguez-Gil \& Mart\'inez-Pais(2002)]{rod02}  
Rodr\'iguez-Gil, P. \& Mart\'inez-Pais, I.G. 2002,  Classical Nova Explosions, 637, 558  

\bibitem[Rodr\'iguez-Gil \& Torres(2005)]{rod05} 
Rodr\'iguez-Gil, P., \& Torres, M.A.P. 2005, A\&A, 431, 289   

\bibitem[Sala \& Hernanz(2005)]{sal05} 
Sala, G., \& Hernanz, M. 2005, \aap, 1057, 2005 

\bibitem[Sala et al.(2025)]{sal25} 
Sala, G., Haberl, F., Schwope, A. et al. 2025, \aap, 704, A72 



\bibitem[Scaringi et al.(2026)]{sca26} 
Scaringi, S., Knigge, C., de Martino, D. 2026, Space Science Review, 222, 32 

\bibitem[Schaefer(2010)]{sch10} 
Schaefer, B.E. 2010, \apjs, 187, 275

\bibitem[Schaefer(2018)]{sch18} 
Schaefer, B.E. 2018, \mnras, 481, 3033 

\bibitem[Schaefer(2023)]{sch23} 
Schaefer, B.E. 2023, \mnras, 525, 785     

\bibitem[Schaefer(2025)]{sch25} 
Schaefer, B.E. 2025, \apj, 993, 232       

\bibitem[Schaefer et al.(2010)]{sch10b} 
Schaefer, B.E., Pagnotta, A., \& Shara, M.M. 2010, \apj, 708, 381  

\bibitem[Schlafly \& Finkbeiner(2011)]{sch11}
Schlafly, E.F., \& Finkbeiner, D.P. 2011, \apj, 737 

\bibitem[Schlegel et al.(1998)]{sch98} 
Schlegel, D.J., Finkbeiner, D.P., Davis, M. 1998, \apj, 500, 525  

\bibitem[Schmidtobreick et al.(2005)]{sch05} 
Schmidtobreick, L., Tappert, C., Bianchini, A., Mennickent, R. 2005, A\&A, 432, 199 
                 
\bibitem[Schmidtobreick et al.(2003)Schmidtobreick, Tappert, \& Saviane]{sch03} 
Schmidtobreick, L., Tappert, C., \& Saviane, I. 2003, \mnras, 342, 145  

\bibitem[Selvelli \& Friedjung(2003)]{sel03}
Selvelli, P., \& Friedjung, M. 2003, \aap, 401, 297 

\bibitem[Selvelli \& Gilmozzi(2013)]{sel13}
Selvelli, P., \& Gilmozzi, R. 2013, \aap, 560, 49 

\bibitem[Selvelli \& Gilmozzi(2019)]{sel19} 
Selvelli, P., Gilmozzi, R. 2019, A\&A, 622, 186 

\bibitem[Shafter(2017)]{sha17}   
Shafter, A.W. 2017, \apj, 834, 196

\bibitem[Shafter et al.(2009)]{sha09}  
Shafter, A.W., Rau, A., Quimby, R.M., et al. 2009, \apj, 690, 1148

\bibitem[Shafter et al.(2015)]{sha15}           
Shafter, A.W., Henze, M., Rector, T.A., et al. 2015, \apjs, 216, 34

\bibitem[Shakura \& Sunyaev(1973)]{sha73}
Shakura, N.I., \& Sunyaev, R.A. 1973, \aap, 24, 337 

\bibitem[Shara et al.(1986)]{sha86} 
Shara, M.M., Livio, M., Moffat, A.F.M., \& Orio, M. 1986, \apj, 311, 163

\bibitem[Shara et al.(2018)]{sha18} 
Shara, M.M., Prialnik, D., Hillman, Y., \& Kovetz, A. 2018, \apj, 860, 110 

\bibitem[Shore et al.(2018)]{sho18}
Shore, S.M., Kuin, N.P., Mason, E., De Gennaro Aquino, I. 2018, \aap, 619, 104 

\bibitem[Shore et al.(1993)]{sho93}
Shore, S.N., Sonneborn, G., Starrfield, S., Gonzalez-Riestra, R., \& Ake, T.B. 1993, \aj, 106, 2408 

\bibitem[Shore et al.(1996)]{sho96} 
Shore, S.N., Starrfield, S., \& Sonnerborn, G. 1996, \apj, 463, L21 

\bibitem[Shugarov et al.(2002)]{shu02}
Shugarov, S.Yu., Goranskij, V.P., Pavlenko, E.P. 2002, in: 
{\it Classical Nova Explosions}, M. Hernanz, J. Jose (eds.), AIP Conf. Proc. Vol. 637, 323 

\bibitem[Sion et al.(2017)Sion, Godon, \& Jones]{sio17}
Sion, E.M., Godon, P., \& Jones, L. 2017, \aj, 153, 109 

\bibitem[Sion et al.(2013)]{sio13}
Sion, E.M., Szkody, P., Mukadama, A. et al. 2013, \apj, 772, 116 

\bibitem[Spruit \& Ritter(1983)]{spr83} 
Spruit, H.C., \& Ritter, H. 1983, \aap, 124, 267 

\bibitem[Starrfield et al.(2020)]{sta20}
Starrfield, S., Bose, M., Iliadis, C. et al. 2020, \apj, 895, 70 

\bibitem[Starrfield et al.(1976)]{sta76} 
Starrfield, S., Sparks, W.M., \& Truran, J. W. 1976, in IAU Symp. 73,
Structure and Evolution of Close Binary Systems, ed. P. Eggleton,
S. Mitton, \& J. Whelan (Cambridge: Cambridge Univ. Press), 155

\bibitem[Szkody \& Howell(1992)]{szk92} 
Szkody, P., Howell, S.B. 1992, \apjs, 78, 537  

\bibitem[Tappert et al.(2016)]{tap16} 
Tappert, C., Barria, D., Fuentes Morales, I. et al. 2016, \mnras, 562, 1371 

\bibitem[Tappert et al.(2012)]{tap12} 
Tappert, C., Ederoclite, A., Mennickent, R.E., Schmidtobreick, L., \& Vogt, N. 2012, \mnras, 423, 2476 

\bibitem[Tappert et al.(2013a)]{tap13b} 
Tappert, C., Schmidtobreick, L., Vogt, N., \& Ederoclite, A. 2013a, \mnras, 436, 2412 

\bibitem[Tappert et al.(2014)]{tap14} 
Tappert, C., Vogt, N.,  Della Valle, M., Schmidtobreick, L., \& Ederoclite, A. 2014, \mnras, 442, 565 

\bibitem[Tappert et al.(2013b)]{tap13a} 
Tappert, C., Vogt, N.,  Schmidtobreick, L., Ederoclite, A., \& Vanderbeke, J. 2013b, \mnras, 431, 92 

\bibitem[Tappert et al.(2015)]{tap15} 
Tappert, C., Vogt, N.,  Schmidtobreick, L., Ederoclite, A., \& Vanderbeke, J. 2015, \mnras, 450, 943 

\bibitem[Thorstensen \& Taylor(2000)]{tho00}
Thorstensen, J.R., \& Taylor, C.J. 2000, \mnras, 312, 629 


\bibitem[Vanlandingham et al.(2005)]{van05} 
Vanlandingham, K.M., Schwartz, G.J., Shore, S.N., Starrfield, S., Wagner, R.M. 2005, \apj, 130, 734 

\bibitem[Verbunt(1987)]{ver87}
Verbunt, F. 1987, \aaps, 71, 339 

\bibitem[Verbunt \& Zwaan(1981)]{ver81}
Verbunt, F., \& Zwaan, C. 1981, \aap, 100, L7 

\bibitem[Vogt et al.(2017)]{vog17}
Vogt, N., Schreiber, M.R., Hambsch, F.-J., Retamales, G., Tappert, C. et al.
2017, \pasp, 129, 4201 

\bibitem[Wade(1984)]{wad84} 
Wade, R.A. 1984, \mnras, 208, 381  

\bibitem[Wade(1988)]{wad88} 
Wade, R.A. 1988, \apj, 335, 394  

\bibitem[Wade \& Hubeny(1998)]{wad98}
Wade, R.A., \& Hubeny, I. 1998, \apj, 509, 350 

\bibitem[Warner(1995)]{war95} 
Warner, B. 1995, {\it Cataclysmic Variables}, Cambridge University 
Press: Cambridge

\bibitem[Warner(1996)]{war96}
Warner, B. 1996, \mnras, 219, 751 

\bibitem[Warner \& Woudt(2015)]{war15} 
Warner, B., Woudt, P. 2015, Mem.S.A.It. Vol.86, 108 

\bibitem[Williams et al.(2016)]{wil16} 
Williams, S.C., Darnley, M.M., Bode, M.F., \& Shafter, A.W. 2016, \apj, 817, 143 

\bibitem[Wood(1995)]{woo95}
Wood, M.A. 1995, LNP Vol.443: White Dwarfs, 41 

\bibitem[Woudt \& Warner(2003)]{wou03} 
Woudt, P.A., \& Warner, B. 2003, \mnras, 340, 1011       

\bibitem[Woudt et al.(2009)]{wou09} 
Woudt, P.A., Warner, B., Osborne, J., Page, K. 2009, MNRAS, 395, 2177 

\bibitem[Zellem et al.(2009)]{zel09} 
Zellem, R., Hollon, N., Ballouz, R.-L., Sion, E.M., Godon, P., 
G\"ansicke, B.T., \& Long, K.S. 2009, PASP, 121, 942  

\bibitem[Zsidi et al.(2024)]{zsi24}
Zsidi, G., Nixon, C.J., Naylor, T., \& Pringle, J.W. 2024, \mnras, 532, 592 

\end{thebibliography}
\end{document}